\documentclass[letterpaper,12pt]{article}
\pdfoutput=1
\usepackage{jheppub}
\usepackage{epsfig}
\usepackage{amsmath}
\usepackage{graphicx}
\usepackage{subcaption}
\usepackage{tikz-cd}

\DeclareSymbolFont{AMSa}{U}{msa}{m}{n}
\DeclareSymbolFont{AMSb}{U}{msb}{m}{n}
\let\Box\relax
\DeclareMathSymbol{\Box}{\mathord}{AMSa}{"03}
\newcommand{\f}{\frac}
\newcommand{\Sum}{\displaystyle\sum\limits}

\newcommand{\Poincare}{Poincar\'e}
\allowdisplaybreaks

\newcommand{\ti}[1]{\textit{#1}}

\newcommand{\half}{{\frac12}}

\newcommand{\cH}{{\mathcal H}}
\newcommand{\cI}{{\mathcal I}}
\newcommand{\cV}{{\mathcal V}}
\newcommand{\cX}{{\mathcal X}}
\newcommand{\cL}{{\mathcal L}}
\newcommand{\cA}{{\mathcal A}}
\newcommand{\N}{{\mathcal N}}
\newcommand{\cN}{{\mathcal N}}

\newcommand{\cS}{{\mathcal S}}
\newcommand{\cO}{{\mathcal O}}
\newcommand{\cF}{{\mathcal F}}

\newcommand{\fg}{{\mathfrak g}}
\newcommand{\fh}{{\mathfrak h}}

\newcommand{\Z}{{\mathbb Z}}
\newcommand{\R}{{\mathbb R}}
\newcommand{\C}{{\mathbb C}}
\newcommand{\PP}{{\mathbb P}}

\newcommand{\fro}{{\overline{\underline\Omega}}}

\newcommand{\abs}[1]{\lvert#1\rvert}

\newcommand{\de}{{\mathrm d}}
\newcommand{\e}{{\mathrm e}}
\newcommand{\I}{{\mathrm i}}

\DeclareMathOperator{\Tr}{Tr}

\title{Line defect Schur indices, Verlinde algebras and \texorpdfstring{$U(1)_r$}{U(1)r} fixed points}

\author[1]{Andrew Neitzke}
\author[2]{and Fei Yan}
\affiliation[1]{Department of Mathematics, The University of Texas at Austin, TX 78712.}
\affiliation[2]{Department of Physics, The University of Texas at Austin, TX 78712.}
\emailAdd{neitzke@math.utexas.edu}
\emailAdd{ fei.yan@utexas.edu}

\abstract{Given an $\cN=2$ superconformal field theory, we reconsider
the Schur index $\cI_L(q)$ in the presence of a half line defect $L$. Recently
Cordova-Gaiotto-Shao found that $\cI_L(q)$ admits
an expansion in terms of characters of the chiral algebra $\cA$
introduced by Beem et al., with simple coefficients $v_{L,\beta}(q)$.
We report a puzzling new feature of this expansion:
the $q \to 1$ limit of the coefficients $v_{L_,\beta}(q)$
is linearly related to the vacuum expectation values $\langle L \rangle$
in $U(1)_r$-invariant vacua of the theory compactified on $S^1$.
This relation can be expressed algebraically as a commutative diagram
involving three algebras: the algebra generated by line defects, the
algebra of functions on $U(1)_r$-invariant vacua, and a
Verlinde-like algebra associated to $\cA$.
Our evidence is experimental, by direct computation in
the Argyres-Douglas theories of type $(A_1,A_2)$, $(A_1,A_4)$,
$(A_1, A_6)$, $(A_1, D_3)$ and $(A_1, D_5)$.
In the latter two theories, which have
flavor symmetries, the Verlinde-like algebra which appears is a new deformation
of algebras previously considered.}

\preprint{\text{UTTG-05-17}}

\begin{document}
\maketitle
\flushbottom

\section{Introduction}

This paper describes a puzzling new feature of the \ti{line defect Schur index}
in $\cN=2$ theories, introduced in \cite{Dimofte:2011py} and recently
reconsidered in \cite{Cordova:2016uwk}.
In short, there is an unexpectedly close relation between:
\begin{itemize}
\item the Schur index in the presence of a supersymmetric (half) line defect $L$,
\item the vevs $\langle L \rangle$ in $U(1)_r$-invariant vacua of the theory compactified
on $S^1$.
\end{itemize}
The precise statements and some discussion appear in
\S\ref{sec:commutative-diagram}-\S\ref{sec:comments} below;
the intervening sections provide the necessary notation and background.

\subsection{Schur indices and chiral algebras}

In \cite{Beem:2013sza} a novel correspondence between 4d $\mathcal{N}=2$ SCFT and 2d
chiral algebras was discovered:
given an $\N=2$ SCFT, there is a corresponding chiral algebra $\cA$. The operators in the vacuum module of the chiral algebra $\cA$ correspond to local operators in the original $\cN=2$
theory which contribute to the Schur index $\cI(q)$ (and Macdonald index\footnote{Macdonald index and its relation to chiral algebra was studied in \cite{Song:2016yfd}.}).

The algebras $\cA$ corresponding to Argyres-Douglas theories have been intensively studied in e.g. \cite{Beem:2013sza,Beem:2014zpa,Buican:2015ina,Cordova:2015nma,Creutzig:2017qyf,Song:2017oew,Xie:2016evu,Buican:2017uka}. In particular, the chiral algebra for the
$(A_1,A_{2N})$ Argyres-Douglas theory\footnote{Here
and below we use the taxonomy of Argyres-Douglas theories from \cite{Cecotti:2010fi}, in which
they are labeled by pairs of ADE type Lie algebras. Argyres-Douglas theories were first discovered in \cite{Argyres:1995jj,Argyres:1995xn}.} was conjectured to be the Virasoro minimal model
with $(p,q) = (2,2N+3)$, and the chiral algebra for $(A_1,D_{2N+1})$ Argyres-Douglas theories was conjectured to be $\widehat{\mathfrak{sl}(2)}_{k}$ at level $k=-4N/(2N+1)$. The Schur indices for certain Argyres-Douglas theories have been computed and indeed match the vacuum characters of the corresponding 2d chiral algebra \cite{Buican:2015ina,Cordova:2015nma,Cordova:2016uwk,Buican:2017uka}.

\subsection{Schur indices with half line defects and Verlinde algebra}\label{sec:1.2}

In \cite{Cordova:2016uwk} this story was extended to include the \ti{non-vacuum} characters of the
chiral algebra $\cA$, by considering a new Schur index $\cI_L(q)$,
which counts operators
of the $\N=2$ SCFT which sit at the endpoint of a supersymmetric ``half line defect'' $L$.
In various examples, \cite{Cordova:2016uwk}
found that $\cI_L(q)$
can be expressed as a linear combination of characters
associated to modules for the algebra $\cA$:
\begin{equation} \label{eq:schur-defect-expansion}
	\cI_L(q) = \sum_\beta v_{L,\beta}(q) \chi_\beta(q)
\end{equation}
where
$\chi_\beta(q)$ are the characters, and $v_{L,\beta}(q)$ are some
simple Laurent polynomials in $q$, with integer coefficients.

In the expansion
\eqref{eq:schur-defect-expansion}, the index $\beta$ is running over
some finite collection of modules, which moreover are closed under
a canonical action of the modular $S$ matrix. This being so,
we can use the Verlinde formula to define a commutative and
associative algebra $\cV$, generated by the ``primaries'' $\Phi_\beta$
corresponding to the modules with characters $\chi_\beta(q)$,
with product laws of the form
\begin{equation}
	[\Phi_\beta] \times [\Phi_\alpha] = c_{\beta\alpha}^\gamma [\Phi_\gamma].
\end{equation}
In $(A_1,A_{2N})$ Argyres-Douglas theories this commutative product corresponds to the true fusion operation in the $(2,2N+3)$ Virasoro minimal model. More generally though, we do not claim to interpret this product as any kind of fusion operation: we just use the formal rule provided by the Verlinde formula. In the following we will often refer to these product laws as \ti{modular fusion rules}\footnote{We thank Christopher Beem for suggesting us to make a distinction from the true fusion rules.} of the Verlinde-like algebra $\cV$.

Now, let us return to the expansion \eqref{eq:schur-defect-expansion}
and specialize the coefficients $v_{L,\beta}(q)$ to $q = 1$, defining
\begin{equation}
V_{L,\beta} = v_{L,\beta}(q=1).
\end{equation}
Then for every line defect $L$ we get an element $f(L) \in \cV$ by
\begin{equation}
	f(L) = \sum_\beta V_{L,\beta} [\Phi_\beta].
\end{equation}
Remarkably, \cite{Cordova:2016uwk}
found evidence that this map is actually  a \ti{homomorphism} of commutative algebras,
\begin{equation} \label{eq:hom-1}
	f: \cL \to \cV
\end{equation}
where
$\cL$ is the commutative OPE algebra of line defects in the original $\N=2$ theory.

$f$ always maps the trivial line defect to the vacuum module, since the
Schur index without any line defect insertions is
the vacuum character of $\cA$.
Thus the fact that the trivial line defect is the identity in the OPE
algebra gets mapped to the fact that the vacuum module is the identity
in the Verlinde algebra $\cV$.

Evidence for the homomorphism property of the line defect Schur index was observed in \cite{Cordova:2016uwk}
in the $(A_1, A_2)$ and $(A_1,A_4)$ theories. In \S\ref{sec:A1A6} below we give evidence
that the same is true in the $(A_1,A_6)$ theory. We also extend to the $(A_1,D_3)$ and $(A_1,D_5)$
theories, in \S\ref{sec:A1D3} and \S\ref{sec:A1D5},
but this involves a little twist: see \S\ref{sec:flavor} below.

\subsection{A simple example} \label{sec:intro-example}

Just to fix ideas, we quickly review here the case of
the Argyres-Douglas theory of type $(A_1, A_2)$.
The basic data are:
\begin{itemize}
	\item
There are five distinguished nontrivial line defects $L_1, \dots, L_5$ in the theory,
which generate all the rest by operator products. In fact one only needs products
involving \ti{consecutive} $L_i$: the most general
simple line defect can be written \cite{Gaiotto:2010be}
\begin{equation} \label{eq:general-defect}
	L = L_i^m L_{i+1}^n
\end{equation}
for $i \in \{1, \dots, 5\}$ and $m, n \ge 0$ (letting $L_6 = L_1$).
We also have the trivial line defect which we write as $1$.

\item The chiral algebra $\cA$
is the $(2,5)$ Virasoro minimal model, with $c = -22/5$.
The corresponding Verlinde algebra $\cV$ has two generators $[\Phi_{1,1}]$, $[\Phi_{1,2}]$ corresponding
to the two primaries. $[\Phi_{1,1}]$ is the identity
element, so the only nontrivial product is $[\Phi_{1,2}] \times [\Phi_{1,2}]$, which is
\begin{equation} \label{eq:verlinde-A1A2}
	[\Phi_{1,2}] \times [\Phi_{1,2}] = [\Phi_{1,1}] + [\Phi_{1,2}].
\end{equation}
\end{itemize}
The line defect Schur indices come out to \cite{Cordova:2016uwk}
\begin{equation}
\cI_1(q) = \chi_{1,1}(q), \qquad \cI_{L_i}(q) = q^{-\frac{1}{2}}\big(\chi_{1,1}(q)-\chi_{1,2}(q)\big).
\end{equation}
Thus the homomorphism $f$ in this case is
\begin{equation} \label{eq:hom-A1A2}
 f(1) = [\Phi_{1,1}], \qquad f(L_i) = [\Phi_{1,1}] - [\Phi_{1,2}].
\end{equation}
In particular, $f$ forgets the index $i$, so it identifies the
$5$ generators $L_i$.\footnote{We will give a derivation of this property of $f$
in \S \ref{sec:schur-index-defect-ir}.} Moreover, $f$ collapses the infinite-dimensional
algebra $\cL$, spanned by the operators \eqref{eq:general-defect}, down to
the two-dimensional algebra $\cV$.

\subsection{Diagonalizing the Verlinde algebra} \label{sec:ver-diag}

To explain the main new results of this paper, we need a brief digression
to recall a structural fact about the Verlinde algebra
$\cV$: the modular $S$ operator gives a canonical diagonalization of $\cV$
\cite{Verlinde:1988sn}.
Concretely, if we choose an ordering of the $n$ primaries, then we can represent
the operation of fusion with $\Phi_i$
by an $n \times n$ matrix $N_{\Phi_i}$, and likewise $S$ by an $n \times n$
matrix; then the statement is that the matrices
\begin{equation}
	\hat N_{\Phi} = S N_{\Phi} S^{-1}
\end{equation}
are all diagonal.

For example, in the $(2,5)$ Virasoro minimal model, if we choose the ordering
of the primaries $(\Phi_{1,1}, \Phi_{1,2})$, then we have \cite{D.Francesco}
\begin{equation}
	N_{\Phi_{1,1}} = \begin{pmatrix} 1 & 0 \\ 0 & 1 \end{pmatrix}, \quad N_{\Phi_{1,2}} = \begin{pmatrix} 0 & 1 \\ 1 & 1 \end{pmatrix}, \quad
S=\frac{2}{\sqrt{5}}\begin{pmatrix}
-\sin\frac{2\pi}{5} & \sin\frac{4\pi}{5}\\
\sin\frac{4\pi}{5} & \sin\frac{2\pi}{5}
\end{pmatrix},
\end{equation}
from which we can compute
\begin{equation}
\hat N_{\Phi_{1,1}} = \begin{pmatrix}
1 & 0\\
0 & 1
\end{pmatrix},
\quad
\hat N_{\Phi_{1,2}} = \begin{pmatrix}
\frac{1-\sqrt{5}}{2} & 0\\
0 & \frac{1+\sqrt{5}}{2}
\end{pmatrix}.
\end{equation}

The representation of $\cV$ by the diagonal matrices $\hat N_\Phi$
shows that $\cV$ is naturally isomorphic
to a direct sum of copies of $\C$. Moreover these copies correspond
canonically to the primaries themselves, using the
ordering of the primaries we have chosen.
Another way of saying this is: \ti{$\cV$ is canonically isomorphic
to the algebra of functions on the set of primaries of $\cA$.}
We will use the statement in this form,
in \S\ref{sec:verlinde-fixed} below.

\subsection{Verlinde algebra and \texorpdfstring{$U(1)_r$}{U(1)r}-fixed points in three dimensions} \label{sec:verlinde-fixed}

Now we recall another place where the Verlinde algebra of $\cA$ has
recently appeared.

We consider the compactification of our
superconformal $\N=2$ theory to three dimensions on $S^1$.
As is well known, beginning with \cite{Seiberg:1996nz}, the Coulomb branch of the compactified theory
is a hyperk\"ahler space $\cN$. For example, if our theory
is a theory of class $\cS$, say $\cS[\fg,C]$, then $\cN$ is a moduli space
of solutions of Hitchin equations on $C$ with gauge algebra $\fg$
\cite{Gaiotto:2009hg,MR89a:32021}.

The $U(1)_r$
symmetry of the theory acts geometrically on $\cN$.
This action is an important tool in the
study of this space. For example, it can be
used to compute the Betti numbers of the Hitchin moduli spaces,
as was noted already in \cite{MR89a:32021}. More recently \cite{Gukov:2015sna,Gukov:2016lki} this $U(1)_r$ action has been used to define and compute a new
``$U(1)_r$-equivariant index'' for $\cN$, related to a Coulomb branch index
in the $\N=2$ theory. In both computations the starring role is
played by the \ti{fixed locus} $F \subset \cN$ of the $U(1)_r$ symmetry.
The points of $F$ are the $U(1)_r$-invariant vacua of the compactified theory.

For our purposes the key fact about $F$ is the following
recent observation: \ti{the points of $F$
are naturally in 1-1 correspondence with the primaries of $\cA$
\cite{Cecotti:2010fi,Fredrickson-Neitzke,Fredrickson-talk,Fredrickson:2017yka}.}\footnote{Some early hints of this appeared in \cite{Cecotti:2010fi}, and a
precise correspondence of this sort in
the case of $(A_m, A_n)$ Argyres-Douglas theories with
$(m+1,n+1) = 1$ is
developed in \cite{Fredrickson-Neitzke}, first
reported in \cite{Fredrickson-talk}. This correspondence was used extensively
in \cite{Fredrickson:2017yka}, where the $U(1)_R$ weights at the fixed points
were also worked out; that work also substantially broadened
the scope of the correspondence, well beyond
the class of $(A_m, A_n)$ theories.
Despite all this, as far as we know, nobody has yet provided a
first-principles explanation of \ti{why} the correspondence
between points of $F$ and primaries of $\cA$ exists.
In this paper we just take this correspondence as a given.}
Combining this correspondence with the picture of $\cV$
reviewed in \S\ref{sec:ver-diag}, we conclude that
there is a canonical isomorphism
\begin{equation}
	h: \cV \to \cO(F),
\end{equation}
where $\cO(F)$ means the algebra of functions on $F$.
Concretely, $h$ maps $[\Phi]$
to the vector of diagonal entries of $\hat{N}_\Phi$,
using the correspondence above to match up
the points of $F$ with the positions along the diagonal.

\subsection{Fixed points and vevs} \label{sec:fixed-points-and-vevs}

We consider the vacuum expectation values of $\half$-BPS line defects
wrapped around $S^1$ in $S^1 \times \R^3$. These vevs are functions on the vacuum moduli space
$\cN$; the process of taking vevs gives a homomorphism of commutative algebras
\begin{equation}
\cL \to \cO(\cN)
\end{equation}
from the OPE algebra of $\half$-BPS line defects to the algebra $\cO(\cN)$
of holomorphic functions on $\cN$.\footnote{In fact, in all examples we know,
this is an isomorphism $\cL \simeq \cO(\cN)$, though we do not need
this fact in anything that follows.} Now consider the \ti{restriction}
of these vevs to the $U(1)_r$-fixed locus $F \subset \cN$: this gives
another homomorphism of commutative algebras,
\begin{equation}
g: \cL \to \cO(F).
\end{equation}

In Argyres-Douglas theories, the map $g$ is very far from being an isomorphism:
it forgets most of the details of a line
defect, remembering only its vevs at the finitely many $U(1)_r$-invariant vacua.
This is reminiscent of the fact that the map $f$, built from line defect Schur indices
$\cI_L$, likewise
forgets most of the details of the line defects $L$. In the next section we flesh
this out into a precise sense in which $f$ and $g$ are ``the same.''

Before we state our main result, we would like to point out that the $\half$-BPS line defects that we are talking about in this section are \ti{full line defects}, which are by definition different from the \ti{half line defects} in \ref{sec:1.2}. However, away from the endpoints of the half line defects they are ``locally" the same object. In particular the OPE algebra of half line defects is isomorphic to the OPE algebra of full line defects, both of which we denote as $\cL$.

\subsection{The commutative diagram} \label{sec:commutative-diagram}

So far in this introduction we have described three \ti{a priori} unrelated
commutative algebras associated to an $\N=2$ SCFT:
\begin{itemize}
\item The OPE algebra $\cL$ of $\half$-BPS line defects,
\item The Verlinde algebra $\cV$ associated to the chiral algebra $\cA$,
\item The algebra $\cO(F)$ of functions on the set of $U(1)_r$-invariant
vacua of the theory compactified on $S^1$.
\end{itemize}
We also described three \ti{a priori} unrelated maps between these
algebras:
\begin{itemize}
	\item The map $f: \cL \to \cV$ obtained by computing Schur indices
	in the presence of half line defects and expanding them in terms of
	characters of $\cA$,

	\item The isomorphism $h: \cV \to \cO(F)$, constructed using
	the mysterious identification between $U(1)_r$-invariant vacua
	and chiral primaries, and using also the modular $S$ matrix,

	\item The map $g: \cL \to \cO(F)$ obtained by compactifying
	the theory on $S^1$ and evaluating line defect vevs in $U(1)_r$-invariant
	vacua of the reduced theory.
\end{itemize}
These ingredients can be naturally assembled into a diagram:
\begin{center}
\begin{tikzcd}
\cL \arrow{r}{f} \arrow[swap]{dr}{g} & \cV \arrow{d}{h} \\
& \cO(F)
\end{tikzcd}
\end{center}
This raises the natural question of whether the diagram \ti{commutes}, i.e. whether
\begin{equation} \label{eq:commutativity}
h \circ f = g.
\end{equation}
In \S\ref{sec:examples-even} below, we verify by direct computation that
\eqref{eq:commutativity} indeed holds,
in the Argyres-Douglas theories of type $(A_1, A_2)$, $(A_1, A_4)$, and
$(A_1, A_6)$.
In \S\ref{sec:examples-odd} we verify a similar statement in $(A_1, D_3)$
and $(A_1, D_5)$ theories: see \S\ref{sec:flavor} below for more on this.

The commutativity \eqref{eq:commutativity} is the main new result of this paper.
In a sense it is not surprising --- once you realize
that this diagram exists, it is hard to imagine that it would not commute --- but on the other
hand its physical meaning is not at all transparent, at least to us. It should be interesting
to unravel. We comment
a bit further on this question in \S\ref{sec:comments} below.

\subsection{Flavor symmetries} \label{sec:flavor}

In $\N=2$ theories with flavor symmetries the story described above
can be enriched.
The Schur index, rather than being a function $\cI_L(q)$,
is promoted to $\cI_L(q,z)$ where $z$ stands for the flavor fugacities.
The chiral algebra $\cA$ also contains currents for the flavor symmetry group,
and thus its characters are promoted to $\chi_i(q,z)$.
It is natural to ask whether there are analogues of the homomorphisms $f$, $g$, $h$
in such theories with the extra parameters $z$ included.\footnote{In
\cite{Cordova:2016uwk} the case of $(A_1,D_3)$ was considered,
after specializing to $z\to 1$ to ``forget'' the flavor symmetry. Though this limit is very special in the sense that characters of the two non-vacuum admissible representations diverge in this limit and only one linear combination of the two characters is well-defined. This linear combination and the vacuum character transform into each other under modular transformations \cite{Beem:2017ooy}.}

In \S\ref{sec:examples-odd} below we consider this question for
the $(A_1, D_3)$ and $(A_1, D_5)$ Argyres-Douglas theories, which have
flavor symmetry $SU(2)$. The Cartan subgroup of $SU(2)$
consists of matrices $diag(z, z^{-1})$ for $\abs{z} = 1$;
thus the fugacity in this case is just a single number
$z$.
The chiral algebras in these theories are
$\cA = \widehat{\mathfrak{sl}(2)}_{-\frac{4}{3}}$
and $\cA = \widehat{\mathfrak{sl}(2)}_{-\frac{8}{5}}$
respectively.

In the compactification of the theory on $S^1$, turning on the fugacity
$z$, with $\abs{z} = 1$,
corresponds to switching on a ``flavor Wilson line'' around the $S^1$.
Such a Wilson line leads to a deformation of $\cN$ which does not break
the $U(1)_r$ symmetry. Thus for any fixed $z$ we can consider the fixed locus
$F_z \subset \cN_z$, which turns out to be discrete, just as in the $(A_1, A_{2n})$
theories we considered above. Evaluating line defect vevs at
$F_z$ we get a homomorphism
\begin{equation}
	g_z: \cL \to \cO(F_z).
\end{equation}

Now we would like to repeat the story of \S\ref{sec:commutative-diagram}
here, i.e. to construct maps $f_z$ and $h_z$, and to verify
\eqref{eq:commutativity}.
A key question arises: what should we use as ``Verlinde algebra''?
There are no conventional two-dimensional conformal field theories with
$\cA$ as symmetry algebras; the usual candidate with symmetry $\widehat{\mathfrak{sl}(2)}_k$ would be the
WZW model, but that only makes sense for positive
integer $k$. Thus there is no clear physically-defined
notion of Verlinde algebra. Still, it was realized in \cite{V.G.Kac} that at \ti{admissible} levels there is a finite set of \ti{admissible} representations of $\cA$ whose characters span a representation of the modular group $SL(2,\Z)$.  A Verlinde-like algebra built from the \ti{admissible} representations $\cV_1$ was constructed in \cite{Koh-Sorba} where the fusion rules were given by naive application of the Verlinde formula \cite{V.G.Kac}. $\cV_1$ has the odd
feature that some of the structure constants are equal to $-1$.\footnote{Fusion rules of $\widehat{\mathfrak{sl}(2)}_k$ at admissible negative fractional level have been studied intensively over the years and have been completely solved and understood recently in \cite{Creutzig:2012sd,Creutzig:2013yca} (see also references therein). From this point of view, the negative structure constants
have to do with the fact that admissible representations are not closed under fusion. In any case, in
in our context we are simply considering a Verlinde-like algebra $\cV_1$ defined by naive application of the Verlinde formula,
and not worrying too much about whether it has a fusion interpretation.}

Nevertheless, we could try to construct $f_z$ and $h_z$, and verify \eqref{eq:commutativity}, using this algebra $\cV_1$.
What we find experimentally in \S\ref{sec:examples-odd} below is that this
does not quite work: we need to use a deformed Verlinde-like algebra
$\cV_z$. $\cV_z$ is obtained from $\cV_1$ by
replacing each structure constant $-1$ by $-z^2$.
Once we make this modification, the whole story goes through as in \S\ref{sec:commutative-diagram}
above.

\subsection{Interpretations and comments} \label{sec:comments}

\begin{itemize}

\item The main new result of this paper is the commutative diagram in \S\ref{sec:commutative-diagram}.
What is the physical interpretation of this commutative diagram? One tempting
possibility is that there is a new \ti{localization} computation of the Schur index.
Indeed, if we think of the Schur index as a kind of partition function on $S^3 \times S^1$,
we could imagine computing it by first reducing on $S^1$ and then making some computation
in the resulting effective theory on $S^3$. After this reduction the line defects become
local operators, which are determined by their vevs on $\cN$. In a localization computation
using $U(1)_r$,
they could get further reduced to just their vevs in the $U(1)_r$-invariant vacua.
This would match our observation that the object
$f(L)$ --- which contains much\footnote{Though not
quite all, because of the need to take $q \to 1$ in the coefficients
$v$} of the information of the Schur index $\cI_L$
--- is linearly related to $g(L)$, i.e. to the vevs of $L$ in the $U(1)_r$-invariant vacua.

\item Our verification of the commutativity \eqref{eq:commutativity}
requires us to evaluate explicitly the vacuum expectation values of $\half$-BPS line defects
at the fixed points of the $U(1)_r$ action on $\cN$. In the language
of the Hitchin system, this amounts to solving an instance of the \ti{nonabelian
Hodge correspondence}: for some specific Higgs bundles, we determine the
corresponding complex flat connections up to equivalence.
It would be very interesting to see how far one can push these
ideas: can we compute the vevs in every case where the vacua are isolated?
Can we extend beyond the fixed points, say to get some information about their
infinitesimal neighborhoods? Can we say anything
about non-isolated fixed points?

\item It is natural to ask how broadly the commutative diagram of \S\ref{sec:commutative-diagram} exists; so far
we have checked it only in five theories.
We conjecture that it exists more generally whenever it makes sense,
i.e. whenever the $U(1)_r$-invariant vacua of the theory reduced
on $S^1$ are all isolated. The $U(1)_r$-invariant vacua are isolated in
all Argyres-Douglas theories where the question has been investigated, e.g.
the $(A_m, A_n)$ theories for $gcd(m+1, n+1) = 1$, but more generally
they are usually not isolated.

\item One of the simplest examples where the $U(1)_r$-invariant
vacua are \ti{not} isolated is $\N=2$ super Yang-Mills with $G = SU(2)$
and $N_f=4$, compactified on $S^1$ with generic flavor Wilson lines.
In this theory it appears that there are $4$ isolated $U(1)_r$-invariant
vacua, but also an $S^2$ of $U(1)_r$-invariant vacua, as explained e.g. in
\cite{hauselthesis}. In this theory \cite{Fredrickson:2017yka} argued that nevertheless there
is a correspondence between \ti{connected components} of the space of
$U(1)_r$-invariant vacua and chiral primaries.
It would be very interesting to understand how the diagram \eqref{eq:commutativity}
can be extended to this case. (An obstacle to the most naive extension
is that the line defect vevs are not constant on the $S^2$ of invariant vacua.
Perhaps one needs instead to take the \ti{average} over this $S^2$.)

\item In this paper one of the main players is the homomorphism
$f: \cL \to \cV$. The observation that there is some relation
between algebras of line defects and Verlinde algebras
was made already in \cite{Cecotti:2010fi}. Indeed, that paper
described a map $f': \cL \to \cV$ in the $(A_1, A_{2N})$ theories,
constructed in a different
way, by mapping certain distinguished line defects directly
to minimal model primaries.\footnote{The distinguished line defects
in question actually coincide with the generators $A_i, B_i, \dots$
which we use in \S\ref{sec:examples-even}.}
To forestall a possible confusion,
we emphasize that $f$ and $f'$ are \ti{not} the same.
For example, in the $(A_1,A_2)$ theory we have $f'(L_i) = [\Phi_{1,2}]$,
while \eqref{eq:hom-A1A2} says $f(L_i) = [\Phi_{1,1}] - [\Phi_{1,2}]$.

\item Beyond line defects
one could also consider surface defects and interfaces between surface
defects. The Schur index in the
presence of surface defects, and its relation to 2d chiral algebra,
were studied quite recently in \cite{Cordova:2017ohl,Cordova:2017mhb}
and also featured in the ongoing work \cite{Beem-Peelaers-Rastelli}. It might
be interesting to incorporate surface defects into the story
of this paper.

\item
In this paper we focused on examples of $(A_1,A_{2N})$ and $(A_1,D_{2N+1})$ Argyres-Douglas theories, mainly because their chiral algebras have been relatively well understood and computation of line defect generators is not too complicated.
What about other $(A_1, \fg)$ Argyres-Douglas theories? There is one more example
which we expect should be relatively straightforward, namely
$(A_1, D_4)$, for which the chiral algebra
is $\widehat{{\mathfrak{sl}}(3)}_{-3/2}$ \cite{Beem,Rastelli,Cordova:2015nma,Buican:2015ina}.
Beyond this:
\begin{itemize}
\item The chiral algebra for $(A_1,A_{2N-1})$ Argyres-Douglas theories with $N>2$ is
conjectured to be the $\mathcal{B}_{N+1}$ algebra, the subregular quantum Hamiltonian
reduction of $\widehat{{\mathfrak{sl}}(N)}_{-N^2/(N+1)}$ \cite{Creutzig:2017qyf,Beem:2017ooy}\footnote{Chiral algebra for $(A_1,A_{2N-1})$ and $(A_1,D_{2N})$ Argyres-Douglas theories were reproduced in \cite{Song:2017oew} along with new results for generalized
Argyres-Douglas theories in the sense of \cite{Xie:2012hs,Wang:2015mra}.}. As pointed out in \cite{Fredrickson:2017yka}, the relevant modules associated with the $U(1)_r$ fixed points depend
on the parity of $N$, and for even $N$, the relevant modules are suitable representatives of local
modules which are closed under modular transformation \cite{Auger-Creutzig-Kanade-Rupert,Creutzig:2017qyf,Beem:2017ooy,Beem-Peelaers}. For odd $N$, $S$-transformation turns local modules into twisted modules \cite{Auger-Creutzig-Kanade-Rupert,Creutzig:2017qyf,Beem:2017ooy,Beem-Peelaers}, which makes the matching of $U(1)_r$ fixed points with relevant modules very subtle \cite{Fredrickson:2017yka}. These local and twisted modules and their modular properties are studied in \cite{Beem-Peelaers,Beem:2017ooy,Auger-Creutzig-Kanade-Rupert}.

\item The situation is similar for $(A_1,D_{2N})$ Argyres-Douglas
theories with $N>2$. Here the
chiral algebra has been conjectured to be the $\mathcal{W}_N$ algebra
coming from a non-regular quantum Hamiltonian reduction
of $\widehat{{\mathfrak{sl}}(N+1)}_{-(N^2-1)/N}$ \cite{Creutzig:2017qyf}. For even $N$, \cite{Fredrickson:2017yka} confirmed that the relevant modules are suitable representatives of local
modules listed in \cite{Creutzig:2017qyf},
while for odd $N$ the situation becomes subtle again \cite{Fredrickson:2017yka} since $S$-transformation
turns local modules into twisted modules \cite{Creutzig:2017qyf}.

 \item Chiral algebras for $(A_1, E_{6,7,8})$ Argyres-Douglas theories were conjectured in \cite{Cordova:2015nma,Song:2017oew},
 and at least for $(A_1, E_6)$ and $(A_1, E_8)$ there is a natural guess for the relevant class of modules.
 However, in these theories the computation of line defect generators and their
 framed BPS spectra has not been worked out; it would be interesting
 to develop it.
\end{itemize}

\end{itemize}

\section*{Acknowledgments}

We thank Christopher Beem, Clay C\'ordova, Jacques Distler, Davide Gaiotto, Pietro Longhi, Wolfger Peelaers,
Leonardo Rastelli, Shu-Heng Shao, and Jaewon Song for helpful discussions.
The work of AN is supported by National Science Foundation grant
1151693. The work of FY is supported in part by National Science Foundation grant PHY-1620610. FY would like to thank the organizers of ``Superconformal Field Theories in d $\geq$ 4" at the Aspen Center for Physics, ``Great Lakes String 2017" at the University of Cincinnati, and Pollica Summer Workshop 2017 for hospitality during various stages of this work. FY was also partly supported by the ERC STG grant 306260 during the Pollica Summer Workshop.

\section{Schur indices and their IR formulas}

In this section
we review the definition and IR formula for the ordinary Schur index and
the Schur index with half line defects inserted.

\subsection{The Schur index} \label{sec:schur-index}

The superconformal index of a four-dimensional $\mathcal{N}=2$ SCFT is defined as \cite{Kinney:2005ej,Gadde:2011uv}
\begin{equation}
\cI(p,q,t,a_i)=\text{Tr}(-1)^Fp^{j_2-j_1-r}q^{j_2+j_1-r}t^{R+r}\prod_{i}a_i^{f_i}e^{-\beta\delta_{2\dot{-}}},
\end{equation}
where
\begin{equation}
2\delta_{2\dot{-}}=\{\widetilde{Q}_{2\dot{-}},\widetilde{Q}^{\dagger}_{2\dot{-}}\}=E-2j_2-2R+r.
\end{equation}
Here $p,q,t$ are three superconformal fugacities, $a_i$ are flavor symmetry fugacities, $E$ is the scaling dimension, $j_1$ and $j_2$ are Cartan generators of $SU(2)_1\times SU(2)_2$, $R$ and $r$ are the Cartan generators of the $SU(2)_R\times U(1)_r$ $R$-symmetry group. The trace is taken over the Hilbert space on $S^3$ in radial quantization.

The Schur index is obtained by taking the $q=t$ limit \cite{Gadde:2011ik,Gadde:2011uv},
\begin{equation}\label{Schur}
\cI(q,a_i) = \text{Tr} (-1)^Fq^{E-R}\prod_{i}a_i^{f_i}.
\end{equation}
Here the contributing states are $\frac14$-BPS, annihilated by four supercharges: $Q^1_{-}$, $\tilde{Q}_{2\dot{-}}$, $S_1^{-}$ and $\tilde{S}^{2\dot{-}}$. Their quantum numbers satisfy
\begin{equation}\label{OrdSchurCondition}
E-j_1-j_2-2R=0,\quad j_1-j_2+r=0.
\end{equation}

\subsection{The IR formula for the Schur index} \label{sec:schur-index-ir}

Recently an IR formula for the Schur index was conjectured in \cite{Cordova:2015nma},\footnote{We follow the convention
of \cite{Cordova:2015nma,Cordova:2016uwk} for fermion number, $(-1)^F = \e^{2\pi \I R}$.} relating the Schur index to the trace of the ``quantum monodromy'' operator, a
$q$-series introduced in \cite{Cecotti:2010fi}:
\begin{equation}\label{eq:ir-formula}
\cI(q)=(q)_\infty^{2r}\text{Tr}[M(q)], \quad (q)_\infty:=\prod_{j=0}^\infty(1-q^{j+1}).
\end{equation}
In this section we review the mechanics of this formula.

To write down the operator $M(q)$,
we need to perturb to a point of the Coulomb branch of the theory, where
the only massless fields are those of abelian $\N=2$ gauge theory.
$M(q)$ will be built out of the massive BPS spectrum
of the theory.

Recall that massive BPS states in an $\cN=2$ theory
lie in representations of $SU(2)_J\times SU(2)_R$,
where $SU(2)_J$ is the little group.
The one-particle Hilbert space is graded by the IR charge lattice $\Gamma$, consisting of electromagnetic and flavor
charges:\footnote{The lattice $\Gamma$ strictly speaking
is the fiber of a local system, depending on the point
$u$ of the Coulomb branch, so we should really write it
as $\Gamma_u$; we will suppress this in the notation.} thus
$\cH = \oplus_{\gamma \in \Gamma} \cH_\gamma$.
Factoring out the center-of-mass degrees of freedom, we have:
\begin{equation}
\mathcal{H}_\gamma=[(2,1)\oplus(1,2)]\otimes h_\gamma.
\end{equation}
To count BPS particles refined by representations of $SU(2)_J\times SU(2)_R$, one
consider the protected spin character \cite{Gaiotto:2008cd}
\begin{equation}
\Tr_{h_\gamma}[y^J(-y)^R] = \Sum_{n\in\mathbb{Z}}\Omega_n(\gamma)y^n,
\end{equation}
with integers $\Omega_n(\gamma) \in \Z$,
and packages the $\Omega_n(\gamma)$ into the
``Kontsevich-Soibelman factor'':
\begin{equation} \label{eq:ks-factor}
K(q;X_\gamma;\Omega_i(\gamma)):=\prod_{n\in\mathbb{Z}}E_q((-1)^nq^{n/2}X_\gamma)^{(-1)^n\Omega_n(\gamma)}.
\end{equation}
$K$ is a $q$-series valued in the algebra of
formal variables $X_\gamma$; these variables
themselves are valued in the ``quantum torus'' algebra, obeying
the relations
\begin{equation}\label{IROPE}
X_\gamma X_{\gamma'}=q^{\langle\gamma',\gamma\rangle}X_{\gamma'}X_{\gamma}=q^{\f{1}{2}\langle\gamma,\gamma'\rangle}X_{\gamma+\gamma'},
\end{equation}
where $\langle,\rangle$ is the Dirac pairing on $\Gamma$.
$E_q(z)$ is the quantum dilogarithm defined as
\begin{equation}
E_q(z)=\prod_{j=0}^\infty(1+q^{j+\f{1}{2}}z)^{-1}=\Sum_{n=0}^\infty\f{(-q^{\f{1}{2}}z)^n}{(q)_n}.
\end{equation}
The quantum monodromy operator $M(q)$ is defined as
\begin{equation} \label{eq:quantum-monodromy}
M(q)=\prod_{\gamma\in\Gamma}^{\curvearrowleft}K(q;X_\gamma;\Omega_i(\gamma)).
\end{equation}
The ordering in this product is based on the central charges $Z_\gamma$: if $\text{arg}(Z_{\gamma_1})>\text{arg}(Z_{\gamma_2})$ then $K(X_{\gamma_1})$ is to the right of $K(X_{\gamma_2})$.
The flavor charges --- which have zero Dirac pairing with other charges --- form a sublattice $\Gamma_f \subset \Gamma$. The trace operation is defined by a
truncation to this sublattice:
\begin{equation}
\text{Tr}(X_\gamma) = \begin{cases} 0 & \text{if}\quad \gamma \notin \Gamma_f, \\
X_{\gamma} & \text{otherwise}. \end{cases}
\end{equation}
If we denote a basis for $\Gamma_f$ by
$(\gamma_{f_a})$, then the trace is a function of the
$X_{\gamma_{f_a}}$, which are related to the flavor fugacities $a_i$ in the UV definition of the Schur index \cite{Cordova:2015nma,Cordova:2016uwk}.

$\Tr M(q)$ is invariant when crossing walls of marginal stability in the Coulomb branch \cite{Gaiotto:2008cd,Kontsevich:2008fj,Gaiotto:2009hg,Dimofte:2009tm}.
Of course this is a necessity for \eqref{eq:ir-formula} to make sense,
since $\cI(q)$ is defined directly in the UV and does not depend on
a point of the Coulomb branch.

As pointed out in \cite{Cordova:2015nma,Cordova:2016uwk}, \eqref{eq:ir-formula} is only a formal definition: in principle, in evaluating it, we could meet infinitely many terms contributing to the same power of $q$. In practice we may
hope that these infinitely many terms will come with alternating signs so that they leave
a well-defined Laurent series in $q$, but at least we need to have some definite prescription for how we will order the terms. In \cite{Cordova:2016uwk} the authors propose a prescription to tackle this problem. First they rewrite (\ref{eq:ir-formula}) as
\begin{equation}\label{eq:ir-formula2}
\cI(q)=(q)_\infty^{2r}\text{Tr}[S(q)\overline{S}(q)],
\end{equation}
where $S(q)$ is the ``quantum spectrum generator'' (so called because it contains
enough information to reconstruct the full BPS spectrum),
\begin{equation}
S(q)=\prod_{\text{arg}(Z_\gamma)\in[0,\pi)}^{\curvearrowleft}K(q;X_\gamma;\Omega_i(\gamma)), \quad
\overline{S}(q)=\prod_{\text{arg}(Z_\gamma)\in[\pi,2\pi)}^{\curvearrowleft}K(q;X_\gamma;\Omega_i(\gamma)).
\end{equation}
Next, they conjecture that $S(q)$ and $\overline{S}(q)$ can be expanded as Taylor series in $q$,
with no negative powers of $q$ appearing. If this is so, then
one can try to compute the coefficient of $q^k$ in $\Tr M(q)$ by expanding $S(q)$ and $\overline{S}(q)$ up to some large finite
order $q^N$. The conjecture is that for large enough $N$
the coefficient of $q^k$ will stabilize
to some limiting value (in the examples investigated in \cite{Cordova:2016uwk}
it is sufficient to take $N$ larger than some theory-dependent linear function of $k$.)
In the examples we consider in this paper, we find that the necessary stabilization
does indeed occur, and thus we can use the prescription of \cite{Cordova:2016uwk}.

\subsection{The Schur index with half line defects} \label{sec:schur-index-defect}

Supersymmetric line defects in $\cN=2$ theories have been studied extensively:
a small sampling of references is \cite{Drukker:2009id,Drukker:2009tz,Gaiotto:2010be,Cordova:2013bza,Cordova:2016uwk}.

The line defects which have been studied most extensively are \ti{full} line defects. These are $\half$-BPS objects extended along a straight line in some fixed direction $n^\mu\in\mathbb{R}^4$.
For example, there are $\half$-BPS line defects that extend along the time direction and sits at a point in $\mathbb{R}^3$, preserving four {\Poincare} supercharges, time translation, $SU(2)_J$ rotation around the defect in $\mathbb{R}^3$,
and $SU(2)_R$ $R$-symmetry. The choices of half-BPS subalgebra which can be preserved by such a line defect
are parameterized by $\zeta\in\mathbb{C}^\times$.
When $\abs{\zeta}=1$, so that
$\zeta=\e^{-\I \theta}$, the line defect can be interpreted as
a heavy external BPS source particle, whose central charge has phase $\theta$.

In this section, following \cite{Cordova:2016uwk},
we will be interested in \ti{half} line defects in
superconformal $\cN=2$ theories. A half line defect extends along a ray in $\mathbb{R}^4$ and terminates at a point, say
the origin.
The half line defect looks like a full line defect except near its endpoint; in particular, the
indexing set labeling half line defects is the same as that for full line defects, and it will sometimes be
convenient to let the symbol $L$ stand simultaneously for a half line defect and for its corresponding full
line defect.
The endpoint, however, only preserves two {\Poincare} supercharges, and breaks all translation symmetry.
Moreover the endpoint supports a variety of local endpoint operators; these are the operators
which will be counted by the line defect Schur index.

More generally we can consider a junction of multiple half line defects $L_i$. To preserve some common supersymmetry, these half line
defects must lie in a common spatial plane $\R^2 \subset \R^3$.
Each $L_i$ ends at the origin and has orientation
\begin{equation}
n_i^\mu=(\cos \theta_i, \sin \theta_i, 0, 0),
\end{equation}
where $\theta_i$ is the phase of the central charge of $L_i$.
After conformal mapping to $S^3\times S^1$, each half line defect
wraps $S^1$ and sits at a point on a common great circle on $S^3$.
This configuration preserves one {\Poincare} supercharge and one conformal supercharge,
\begin{equation}\label{DefectSchurQ}
Q=Q^1_{-}+\tilde{Q}_{2\dot{-}}, \quad S=S_1^{-}+\tilde{S}^{2\dot{-}}.
\end{equation}
Recall from \cite{Gadde:2011uv} that $Q^1_{-}$, $\tilde{Q}_{2\dot{-}}$, $S_1^{-}$ and $\tilde{S}^{2\dot{-}}$ are exactly the four supercharges that annihilate Schur operators. Thus the definition of Schur index can be extended to include these half line defect insertions \cite{Dimofte:2011py,Cordova:2016uwk}:
\begin{equation}
\cI_{L_1(\theta_1)L_2(\theta_2)\cdots L_n(\theta_n)}(q) = \Tr_{\mathcal{H}'}[\e^{2\pi \I R}q^{E-R}].
\end{equation}
Here the trace is over the Hilbert space $\mathcal{H}'$ on $S^3$ with half line defects $L_i$ inserted along the great circle at angles $\theta_i$. $\mathcal{H}'$ consists of states annihilated by $Q$ and $S$ in (\ref{DefectSchurQ}).

For Lagrangian gauge theories with 't Hooft-Wilson half line defects, one could use a localization formula to compute the Schur index, as formulated in \cite{Dimofte:2011py,Cordova:2016uwk}. In this paper we consider half line defects in Argyres-Douglas theories, for which we do not have a Lagrangian description available.
Instead, we will use the IR formula conjectured by \cite{Cordova:2016uwk},
which we describe next.

\subsection{The IR formula for the line defect Schur index}
\label{sec:schur-index-defect-ir}

Suppose we fix a full line defect $L$ in $\R^4$ and go to a point $u$ in the Coulomb branch. Let $\mathcal{H}_{L,u}$ denote the Hilbert space of the theory with line defect $L$ inserted. In
this setting there is a new class of BPS states, called \ti{framed BPS states} \cite{Gaiotto:2010be}, which saturate the bound
\begin{equation}
M\geq\text{Re}(Z/\zeta), \qquad \zeta = \e^{\I \theta}.
\end{equation}
Framed BPS states form a subspace $\mathcal{H}_{L,u}^{\text{BPS}} \subset \mathcal{H}_{L,u}$. As usual $\mathcal{H}_{L,u}^{\text{BPS}}$ has a decomposition into sectors labeled by electromagnetic and flavor charges,
\begin{equation}
\mathcal{H}_{L,u}^{\text{BPS}}=\bigoplus_{\gamma\in\Gamma}\mathcal{H}_{L,u,\gamma}^{\text{BPS}}.
\end{equation}
The degeneracies of framed BPS states are counted by the ``framed protected spin character'' defined in \cite{Gaiotto:2010be}:
\begin{equation}\label{framedBPScount}
\fro(L,\gamma,u,q)=\text{Tr}_{\mathcal{H}_{L,u}^{\text{BPS}}}[q^J(-q)^R].
\end{equation}
In the infrared the line defect $L$ has a description as a sum of IR line defects,
which can be thought of as infinitely heavy dyons with charges $\gamma \in \Gamma$. These
IR line defects are represented by formal quantum torus
variables $X_\gamma$ with OPE given by (\ref{IROPE}). Then, for each $L$ one can
define a generating function counting the framed BPS states:
\begin{equation}
F(L(\theta)) = \Sum_{\gamma\in\Gamma}\fro(L,\gamma,u,q)X_\gamma.
\end{equation}
These generating functions are different in different chambers of the Coulomb branch, undergoing framed wall-crossing at the BPS walls \cite{Gaiotto:2010be}.

The IR formula of \cite{Cordova:2016uwk} for the Schur index with insertion of a half line defect $L$ with phase $\theta$ is:
\begin{equation}\label{eq:ir-formulaLD}
\cI_{L(\theta)}(q)=(q)_\infty^{2r}\text{Tr}[F(L(\theta))S_\theta(q)S_{\theta+\pi}(q)],
\end{equation}
where
\begin{equation}
S_\theta(q)=\prod_{\text{arg}(Z_\gamma)\in[\theta,\theta+\pi)}^{\curvearrowleft}K(q;X_\gamma;\Omega_i(\gamma)).
\end{equation}
As demonstrated in \cite{Cordova:2016uwk}, the right side of (\ref{eq:ir-formulaLD}) is invariant under framed wall-crossing,
as is needed since the left side manifestly
does not depend on a point of the Coulomb branch.  When computing half line defect Schur index we often choose $\theta=0$, in which case $S_{\theta}(q)$ and $S_{\theta+\pi}(q)$ reduce to $S(q)$ and $\overline{S}(q)$ respectively.\par

More generally, for multiple half line defects $L_i$, $i=1, \dots, k$, with phase relations $\theta_1<\theta_2<\cdots<\theta_k$, where there are no ordinary BPS
particles with phases in the interval $[\theta_1,\theta_k]$,
the IR formula of \cite{Cordova:2016uwk} for the Schur index is
\begin{equation} \label{eq:ir-formula-multiple}
\cI_{L_1(\theta_1)\cdots L_k(\theta_k)}=(q)_\infty^{2r}\text{Tr}[F(L_1(\theta_1))\dots F(L_k(\theta_k))S_{\theta_k}(q)S_{\theta_k+\pi}(q)].
\end{equation}

We note that this formula is ``compatible with operator products'', in the following sense. The Schur index with two half line defects inserted, $\cI_{L_1(\theta)L_2(\theta+\delta\theta)}$ with $\delta\theta$ small, only depends on $\text{sgn}(\delta\theta)$. In particular, in the limit of $\delta\theta\to 0$ this looks like taking the non-commutative OPE of two parallel half line defects with phase $\theta$. Therefore computing $\cI_{L_1(\theta)L_2(\theta+\delta\theta)}$ and taking the $q\to1$ limit in the character expansion coefficient does correspond to the commutative OPE of two parallel half line defects in $\cL$.

Given the IR formula for half line defect Schur index we would like to point out a general property of half line defect index in Argyres-Douglas theories. Line defect generators in Argyres-Douglas theories can be labeled as $L_{\rho i}$ where the index $i$ is related to the underlying discrete symmetry of the theory. In particular, suppose $L_{\rho j}$ and $L_{\rho i}$ are two half line defect generators that are related by a monodromy action, namely
\begin{equation}
F(L_{\rho j})=M(q)F(L_{\rho i})M^{-1}(q).
\end{equation}
Then according to the IR formula
\begin{align*}
\cI_{L_{\rho j}}(q)& =(q)_\infty^{2r}\text{Tr}[F(L_{\rho j})S(q)\overline{S}(q)]=(q)_\infty^{2r}\text{Tr}[F(L_{\rho j})M(q)]\\
& = (q)_\infty^{2r}\text{Tr}[M(q)F(L_{\rho i})M^{-1}(q)M(q)]\\
& =\cI_{L_{\rho i}}(q).
\end{align*}
In particular this proves that Schur index with one half line defect generator insertion does not depend on the $i$-index, as first observed in some examples in \cite{Cordova:2016uwk}.

\section{Fixed points of the \texorpdfstring{$U(1)_r$}{U(1)r} action}\label{sec:U1R}

\subsection{The \texorpdfstring{$U(1)_r$}{U(1)r} action}

Because the four-dimensional theories we consider are superconformal,
they have a $U(1)_r$ symmetry in the UV. Note that the $U(1)_r$ charges
need not be integral (indeed they are not integral in Argyres-Douglas
theories), though they are rational in all examples we will consider.
Thus the action of $R_t \in U(1)_r$ is not necessarily
trivial when $t = 2\pi$, but there is some $k$ for which $R_{2 \pi k}$
is trivial.

The $U(1)_r$ symmetry of the four-dimensional superconformal theory
acts in particular on the $\half$-BPS line defects.
Recall from \cite{Gaiotto:2010be} that each $\half$-BPS line defect preserves some subalgebra of
the $\N=2$ algebra, with the different possible subalgebras parameterized
by $\zeta \in \C^\times$.
Given a line defect
$L$ preserving the subalgebra with parameter
$\zeta \in \C^\times$, a rotation $R_t \in U(1)_r$
maps $L$ to a new operator $L(t)$ preserving
the subalgebra with parameters $\e^{\I t} \zeta$.

Now suppose we consider the dimensional reduction to three dimensions
on $S^1$. The $U(1)_r$ symmetry acts on the moduli space $\cN$ of vacua
of the three-dimensional theory.
In what follows we will be particularly
interested in the $U(1)_r$-invariant vacua.

\subsection{Line defect vevs in \texorpdfstring{$U(1)_r$}{U(1)r}-invariant vacua}\label{sec:defect-vevs-invariant}

Let $\cF_L$ denote the vev of the line defect $L$ wrapped
on $S^1$. $\cF_L$ is a function on the moduli space $\cN$.
We specialize to a $U(1)_r$-invariant vacuum: after this
specialization $\cF_L$ is just a number.
Moreover, since the vacuum is invariant,
$\cF_L$ is invariant under $U(1)_r$ acting on $L$,
i.e. for any $t, t'$
\begin{equation}
\cF_{L(t)} = \cF_{L(t')}.
\end{equation}

This simple statement has surprisingly strong consequences, which put
constraints on the possible $U(1)_r$-invariant vacua, as follows.
We imagine making a small perturbation away from the invariant vacuum.
After this perturbation the UV line defect $L(t)$ can be decomposed
into IR line defects $L^{IR}_\gamma(t)$,
\begin{equation} \label{eq:ir-decomposition}
 L(t) \to \sum_\gamma \fro(L, \gamma, t) L^{IR}(t)
\end{equation}
with a corresponding decomposition of
the vev $\cF_{L(t)}$ as a sum of monomials $\cX_\gamma(t)$,
\begin{equation} \label{eq:vev-decomp}
\cF_{L(t)} = \sum_\gamma \fro(L, \gamma, t) \cX_\gamma(t).
\end{equation}
Here both sides may depend nontrivially on $t$,
since our perturbation is not $U(1)_r$ invariant.
The expansion coefficients $\fro(L, \gamma, t) \in \Z$ appearing
in \eqref{eq:vev-decomp} are the framed BPS state counts
which we discussed earlier in (\ref{framedBPScount}), evaluated
in the perturbed vacuum, and specialized to $q = 1$.

Now let us take the limit where the perturbation $\to 0$,
and optimistically assume that
the $\fro(L, \gamma, t)$ and $\cX_\gamma(t)$
remain well defined in this limit. In that case we get an
interesting equation:\footnote{We emphasize that \eqref{eq:constancy} is supposed to hold \ti{only} in
a $U(1)_r$-invariant vacuum. Indeed, when considered as functions on the whole moduli space $\cN$, $\cX_\gamma(t)$
and $\cX_\gamma(t')$ are holomorphic in different complex structures,
so they could hardly obey such a relation.}
\begin{equation} \label{eq:constancy}
\sum_\gamma \fro(L, \gamma, t) \cX_\gamma(t) = \sum_\gamma \fro(L, \gamma, t') \cX_\gamma(t').
\end{equation}

Requiring \eqref{eq:constancy} to hold for \ti{all} UV line defects $L$
gives a relation on the $\cX_\gamma(t)$.
For example, if $t'$ is sufficiently close to $t$, so that
$\fro(L, \gamma, t) = \fro(L, \gamma, t')$ for all $L$ and $\gamma$,\
then \eqref{eq:constancy} says simply that $\cX_\gamma(t) = \cX_\gamma(t')$.
More generally, though, the $\fro(L, \gamma, t)$ will jump as $t$ is varied.
Then we get a more general relation, of the form \cite{Gaiotto:2008cd,Gaiotto:2010be}
\begin{equation} \label{eq:invariance-relation}
\cX_\gamma(t') = (\cS_{t,t'} \cX)_\gamma(t).
\end{equation}
Here $\cS_{t,t'}$ denotes a birational map $(\C^\times)^n \to (\C^\times)^n$
which can be written concretely in the form
\begin{equation} \label{eq:S-factorization}
	\cS_{t,t'} = \prod_{\text{arg}(Z_\gamma)\in(t,t')}^{\curvearrowleft}T_\gamma^{\Omega(\gamma)},
\end{equation}
where $T_\gamma: (\C^\times)^n \to (\C^\times)^n$ is a transformation of the form
\cite{Kontsevich:2008fj,Gaiotto:2008cd}\footnote{$T_\gamma$ should
be thought of as the $q \to 1$ limit of the operation of conjugation by the operator $K$
appearing in \eqref{eq:ks-factor}.}
\begin{equation}
	T_\gamma: (\cX_\mu) \to (\cX_\mu (1 - \sigma(\gamma) \cX_\gamma)^{\langle\mu,\gamma\rangle})
\end{equation}
and $\sigma: \Gamma \to  \{\pm 1\}$ is a quadratic refinement of the mod $2$ intersection pairing.

The equation \eqref{eq:invariance-relation} is an interesting relation, but so far
not useful in producing a constraint:
it just relates the values of $\cX_\gamma(t)$ for different $t$.

Now let us specialize to $t' = t + \pi$.
In that case we have the key relation from \cite{Gaiotto:2009hg}
\begin{equation}
\cX_\gamma(t + \pi) = \overline{\cX_{-\gamma}(t)}
\end{equation}
so we conclude that
\begin{equation} \label{eq:fixed-half}
\cS_{t,t+\pi} \cX_\gamma(t) = \overline{\cX_{-\gamma}(t)}.
\end{equation}
This is a closed equation for the numbers $\cX_\gamma(t)$, with fixed $t$.
To make it really concrete, of course, we need some way of computing
the ``classical spectrum generator'' $\cS_{t,t+\pi}$. We could do so by
first computing the BPS spectrum (e.g. by the mutation method)
and then directly
using the definition \eqref{eq:S-factorization}, but there are also
various methods available for computing it directly. In general
theories of class $\cS$ some of these methods have appeared in
\cite{Gaiotto:2009hg,Goncharov2016,Longhi:2016wtv,Gabella:2017hpz}.
In the theories we consider, we will explain a simple method below
in \S\ref{sec:classical-monodromy-S}.

We believe that \eqref{eq:fixed-half} is likely to be a useful equation
for the study of $U(1)_r$-invariant vacua
in general $\N=2$ theories, and it would be interesting to explore
it further. For the Argyres-Douglas theories which we
consider in this paper, though, a simpler equation suffices.
Namely, instead of taking $t' = t+ \pi$ we take
$t' = t+2\pi$. Then we get the relation
\begin{equation}
\cX_\gamma(t + 2 \pi) = \cX_\gamma(t),
\end{equation}
leading to the fixed-point constraint
\begin{equation} \label{eq:fixed-monodromy}
\cS_{t,t+2\pi} \cX_\gamma(t) = \cX_\gamma(t).
\end{equation}
The constraint \eqref{eq:fixed-monodromy} has the advantage that it is purely
algebraic, not involving a complex conjugation.
\eqref{eq:fixed-half} implies \eqref{eq:fixed-monodromy},
but not the other way around: \eqref{eq:fixed-monodromy} can have additional
``spurious'' solutions not associated to actual $U(1)_r$-invariant vacua.\footnote{For
an extreme example, we could consider any superconformal theory in which the $U(1)_r$
charges are all integral, such as the $SU(2)$ gauge theory with $N_f = 4$;
in such a theory $\cS_{t,t+2\pi}$ is the identity operator, so that \eqref{eq:fixed-monodromy}
reduces to the triviality $\cX_\gamma(t) = \cX_\gamma(t)$, which of course imposes
no constraint at all on the vacuum. In contrast, even in these theories,
\eqref{eq:fixed-half} is a nontrivial constraint.}
In the Argyres-Douglas theories we consider in this paper, such spurious
solutions do not occur, as we will see directly just
by counting the number of solutions.
Thus we will use \eqref{eq:fixed-monodromy} as our criterion for
a $U(1)_r$-invariant vacuum.

There is one more point which will be important below:
we will need to keep track of some discrete information attached
to the fixed points $p \in \cN$, namely the weights of
the $U(1)_r$
action on the tangent space $T_p \cN$.
These weights are easily computable if we have a Higgs bundle
description of the fixed point as in \cite{Fredrickson-Neitzke,Fredrickson:2017yka}.
On the other hand, suppose that we only know the fixed point
as a solution of the constraint \eqref{eq:fixed-monodromy}: how then can we compute the $U(1)_r$
weights? We will use a trick, as follows. $\cS_{t,t+2\pi}$ acts as
$\exp(2 \pi \I V)$ where $V$ is a holomorphic vector field on
the twistor space of $\cN$
generating the $U(1)_r$ action. Thus we have
$\de \cS_{t,t+2\pi} = \exp(2 \pi \I V)$ acting on $T_p \cN$.
Thus, by computing $\de \cS_{t,t+2\pi}$ at the fixed point,
we can get the $U(1)_r$ weights mod $1$.

Fortunately, in
the $(A_1, A_{2N})$
cases we treat in \S\ref{sec:examples-even}, knowing the $U(1)_r$ weights mod $1$
is sufficient to determine which fixed point we are looking
at. For the $(A_1, D_{2N+1})$ cases it is not sufficient,
which will cause us some headaches in \S\ref{sec:examples-odd}.

\subsection{Classical monodromy action in Argyres-Douglas theories} \label{sec:classical-monodromy-S}

To use \eqref{eq:fixed-monodromy} in practice we need a way of computing
$\cS_{t,t+2\pi}$, which we call the \ti{classical monodromy} map.
In this section we describe a convenient way of doing so in $(A_1,A_m)$
Argyres-Douglas theories.

The starting point is to use the realization of these theories as class $\cS$ theories.
This implies that the space $\cN$ is a moduli space of flat connections --- in this case,
flat $SL(2,\C)$-connections defined on $\mathbb{CP}^1$ with an irregular singularity at $z=\infty$.
In \cite{Gaiotto:2009hg} the functions $\cX_\gamma$ appearing
in \S\ref{sec:defect-vevs-invariant} were described
from this point of view; we now review that description.

Given a point of the Coulomb branch and generic
$\zeta \in \C^\times$, \cite{Gaiotto:2009hg} gives
a construction of a triangulation of an $(m+3)$-gon,
the ``WKB triangulation.''
The vertices of this $(m+3)$-gon are asymptotic angular directions on the ``circle at infinity,''
\begin{equation}\label{WKBray}
\text{arg}(z)=\frac{2\theta+2\pi j}{m+3}, \quad j=1,\cdots,m+3,
\end{equation}
where $\theta = \arg \zeta$.
Now, given a vacuum in $\cN$ and the parameter $\zeta \in \C^\times$, there is a corresponding flat
connection $\nabla$ on $\C\PP^1$, with irregular singularity  at $z = \infty$.
For each of the $m+3$ asymptotic directions,
there is a unique $\nabla$-flat section $s_i$ whose norm is exponentially small as $z\to\infty$. Thus altogether we
get $m+3$ flat sections
\begin{equation}
	(s_1, s_2, \dots, s_{m+3}).
\end{equation}
Moreover, this tuple of flat sections is enough information
to completely determine the vacuum;
one gets coordinates on $\cN$ by computing $SL(2,\C)$-invariant
cross-ratios from the sections $s_i$.

From (\ref{WKBray}) we see that continuously varying
$\theta\to\theta+2\pi$ is equivalent to making a shift $j\to j+2$, i.e. relabeling
\begin{equation}
	(s_1, \dots, s_{m+3}) \to (s_3, s_4, \dots, s_{m+3}, s_1, s_2).
\end{equation}
This is the classical
monodromy action on $\cN$.

\begin{figure}
	\centering
	\includegraphics[scale=0.35]{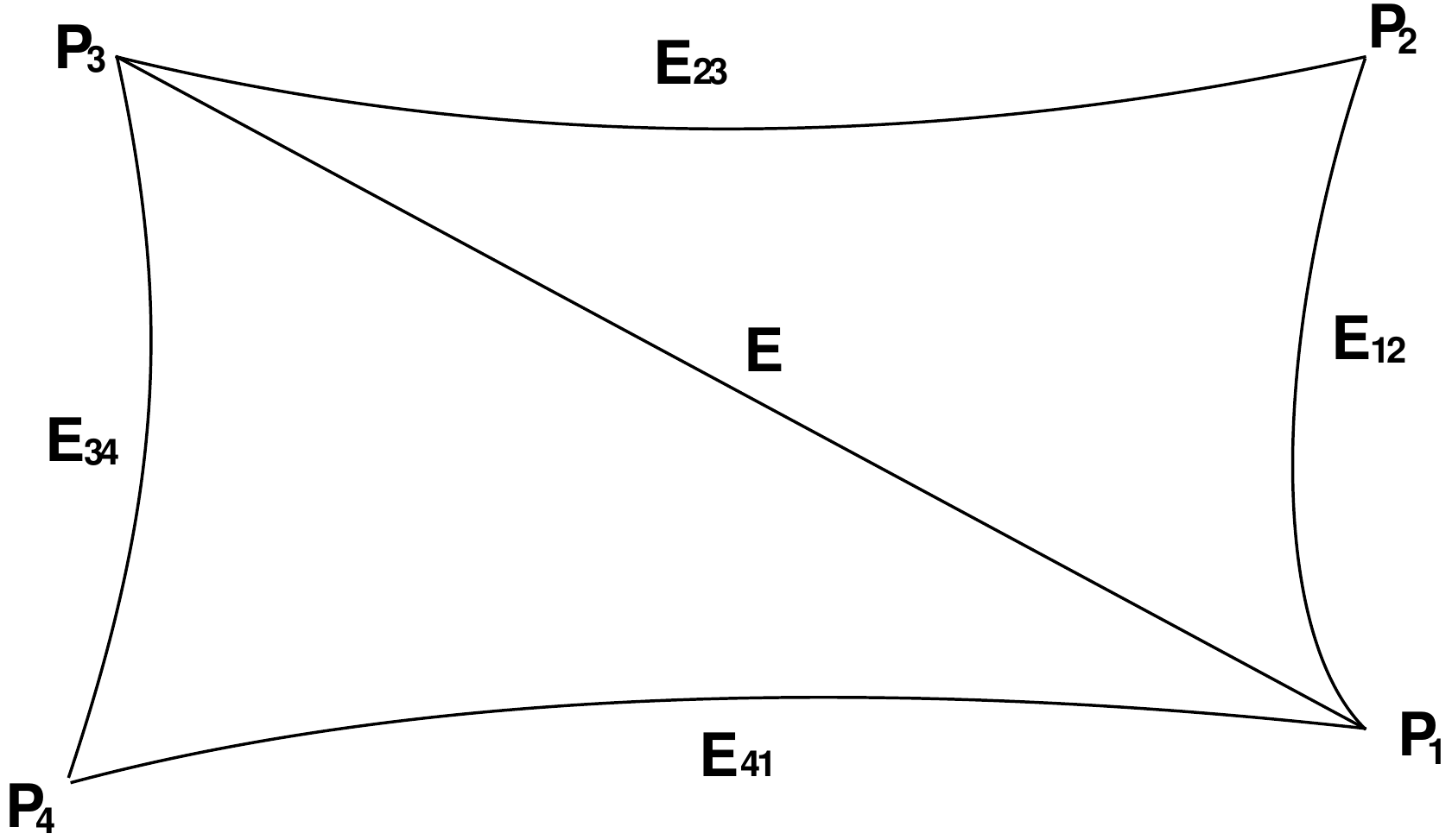}
	\caption{The quadrilateral $Q_E$ associated to edge $E$.}
	\label{Quadrilateral}
\end{figure}

Now we would like to understand concretely what this
monodromy looks like, relative to the local coordinates
$\cX_\gamma$ on $\cN$.
The first step is to explain what the $\cX_\gamma$ are.
For each internal edge $E$ of the triangulation, there is an associated coordinate function $\cX_E$. $E$ is bounded by two triangles which make up a quadrilateral $Q_E$, as
shown in Figure \ref{Quadrilateral}. Each vertex
$P_i$ is associated with a small flat section $s_i$. $\cX_E$ is then defined as:
\begin{equation}
\cX_E=-\frac{(s_1\wedge s_2)(s_3\wedge s_4)}{(s_2\wedge s_3)(s_4\wedge s_1)},
\end{equation}
where the $s_i$ are evaluated at a common point in $Q_E$.
If $E$ is a boundary edge of the $(m+3)$-gon,
by convention, we write $\cX_E=0$.
Finally to go from the $\cX_E$ to the desired $\cX_\gamma$ one uses a dictionary decribed in \cite{Gaiotto:2009hg} which maps the set of internal edges $E_i$
to a basis $(\gamma_{E_i})$ of the charge lattice $\Gamma$.

\begin{figure}
	\centering
	\includegraphics[scale=0.45]{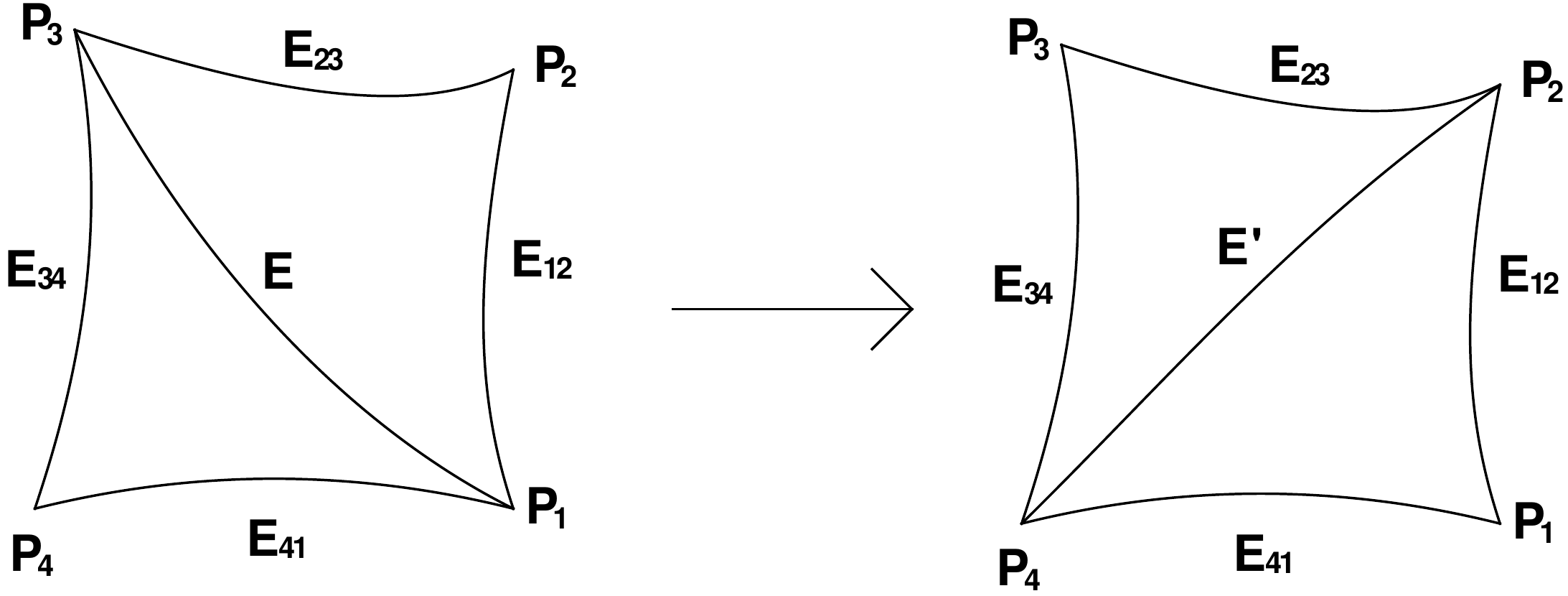}
	\caption{Action of a flip on the quadrilateral $Q_E$.}
	\label{Flip}
\end{figure}

In practice, to use this description for computing
the classical monodromy, we will need one more fact:
we need to know how the coordinates $\cX_E$ change
when we change the triangulation.
A flip of the edge $E$ is the transformation from a triangulation $T$ to another triangulation $T'$, where the edge $E=E_{13}$ in $T$ is replaced by $E'=E_{24}$ in $T'$, as in Figure
 \ref{Flip}. Using
the standard relations between cross-ratios one gets the transformation rules:
\begin{eqnarray}\label{flip}
\begin{split}
\cX_{E'}^{T'}&=\frac{1}{\cX_E^T},\quad
\cX_{E_{12}}^{T'}=\cX_{E_{12}}^T(1+\cX_E^T),\\
\cX_{E_{23}}^{T'}&=\frac{\cX_{E_{23}}^T\cX_E^T}{1+\cX_E^T},\quad
\cX_{E_{34}}^{T'}=\cX_{E_{34}}^T(1+\cX_E^T),\\
\cX_{E_{41}}^{T'}&=\frac{\cX_{E_{41}}^T\cX_E^T}{1+\cX_E^T}.
\end{split}
\end{eqnarray}
In examples below, we will compute the classical monodromy as a composition
of these flips.

For $(A_1,D_{2N+1})$ Argyres-Douglas theories the story
is very similar: the only difference is that
the Hitchin system is defined on $\mathbb{CP}^1$ with an irregular singularity at $z = \infty$ plus a regular singularity at $z = 0$.
The construction of monodromy and coordinates $\cX_\gamma$
is parallel to what we wrote above,
except that the WKB triangulations have one more ``internal'' vertex, at the location of the regular singularity.

\section{Line defects and their framed BPS states in class \texorpdfstring{$S[A_1]$}{S[A1]}}

In this paper we use two different methods for describing the algebra of
line defects in Argyres-Douglas theories of type $(A_1, \fg)$ and computing their framed BPS spectra:

\begin{itemize}
\item In \cite{Cordova:2013bza} it was proposed that generators of the ring of line defects and their framed
BPS spectra can be computed by methods of quiver quantum mechanics. The calculation of framed BPS spectra is
in parallel to the approach previously used for ordinary BPS spectra.
In simple cases this leads to an algorithm for determining the
spectrum, the ``mutation method'' as introduced
in \cite{Alim:2011kw,Gaiotto:2010be,Cecotti:2010fi,Cecotti:2011gu}.
This method is easy to implement on a computer.
We use it in \S\ref{sec:examples-even} below to compute line defect generators and their generating functions in $(A_1,A_{2N})$ Argyres-Douglas theories. However, for the $(A_1,D_{2N+1})$ Argyres-Douglas theories
which  we consider in \S\ref{sec:examples-even}, the framed BPS spectrum in general contains higher spin states, which defeat the mutation method.\footnote{In these cases the framed BPS spectra could in principle be obtained by studying the Hodge diamond of the moduli space of stable framed quiver representations \cite{Cordova:2013bza}. However, this is not as automated as the ``mutation method,'' which prompts us to use an alternative method introduced below.}

\item Alternatively, we can use the class $\cS[A_1]$ realization of the
$(A_1,A_{2N})$ or $(A_1,D_{2N+1})$ theories.
In this realization, line defect generators are in 1-to-1 correspondence with isotopy classes of simple laminations on the disc or punctured disc
\cite{Gaiotto:2010be}.
This leads to an algorithm for computing the framed BPS indices,
as described in \cite{Gaiotto:2010be}.
For our purposes in this paper, this algorithm is not quite sufficient:
we also want to know the spin content of the framed BPS spectra.
In \cite{Galakhov:2014xba,Gabella:2016zxu} a method for computing such BPS spectra in class $\cS$ theories
has been proposed, extending \cite{Gaiotto:2010be}.\footnote{The
paper \cite{Galakhov:2014xba} treated
the spin content for framed BPS spectra associated to certain \ti{interfaces} between
surface defects; \cite{Gabella:2016zxu} gave the first complete
prescription applicable directly to ordinary line defects.}
What we use in this paper is a slight extension of the method in \cite{Gabella:2016zxu}
to treat the case of an irregular singularity.

\end{itemize}

In \S \ref{sec:defects-quiver-type}-\S\ref{sec:framed-bps-from-quivers}
we review the approach via mutations; in
\S \ref{sec:defects-class-S}-\S \ref{sec:spin-class-S} we review the geometric
methods of \cite{Gaiotto:2010be,Gaiotto:2012db,Gaiotto:2012rg,Galakhov:2014xba,Gabella:2016zxu}. These two methods will be used for the
examples in \S\ref{sec:examples-even}-\S\ref{sec:examples-odd}
below.

\subsection{Line defect generators in \texorpdfstring{$\mathcal{N}=2$}{N=2} theories of quiver type}\label{sec:defects-quiver-type}

4d $\mathcal{N}=2$ theories of quiver type are $\mathcal{N}=2$ theories whose BPS spectra can be computed via a four-supercharge multi-particle quantum mechanics system encoded in a quiver \cite{Douglas:1996sw,Douglas:2000ah,Douglas:2000qw,Alim:2011kw}. In particular, Argyres-Douglas theories are examples of theories
of quiver type, as discussed e.g. in \cite{Cecotti:2010fi}.
For 4d $\mathcal{N}=2$ theories of quiver type, there is a nice way of constructing distinguished line defect generators via quiver mutation, developed in \cite{Cordova:2013bza}, which we review in this section.

Fix a point of the Coulomb branch, and fix a half-plane inside the plane of central charges:
\begin{equation}
\mathfrak{h}_\theta=\{Z\in\mathbb{C}\,|\,\theta<\text{arg}(Z)<\theta+\pi\},\quad \theta\in[0,2\pi).
\end{equation}
Then the BPS one-particle representations in the theory can be divided into ``particles'' and ``antiparticles'':
particles are those whose central charges lie in $\fh_\theta$, antiparticles are the rest.
For theories of quiver type there is a canonical positive integral basis
$\{\gamma_i\}$ for $\Gamma$, such that the cone
\begin{equation}
\mathcal{C}=\Bigg\{\Sum_{i=1}^{\text{rank}(\Gamma)}a_i\gamma_i\,|\,a_i\in\mathbb{R}_{\geq 0}\Bigg\}
\end{equation}
contains the charges of all BPS particles.
We call such a basis a \ti{seed}. The corresponding quiver has one node
for each basis charge $\gamma_i$, with the number of arrows from $\gamma_i$ to
$\gamma_j$ given by $\langle \gamma_i,\gamma_j \rangle$.

Correspondingly, in the half-plane $\mathfrak{h}_\theta$ there is a cone $Z(\mathcal{C})$ given by the central charge function $Z$.
The cone of particles is piecewise constant as one varies the parameter $\theta$ or the point of the Coulomb branch,
but jumps when one boundary ray $Z_{\gamma_i}$ of $Z(\mathcal{C})$ hits the boundary of $\mathfrak{h}_\theta$,
i.e. when the central charge of a BPS particle with charge $\gamma_i$ exits the particle half-plane.
At this point the quiver description also jumps discontinuously, by a process of ``mutation.''
Depending on whether $Z_{\gamma_i}$ exits $\mathfrak{h}_\theta$ on the right or on the left, the mutation is denoted as right mutation $\mu_{Ri}$ or left mutation $\mu_{Li}$. The explicit transformation of the basis charges
is \cite{Alim:2011kw,Cordova:2013bza}
\begin{eqnarray}
\mu_{Ri}(\gamma_j)&=-\delta_{ij}\gamma_j+(1-\delta_{ij})(\gamma_j-\text{Min}[\langle\gamma_i,\gamma_j\rangle,0]\gamma_i),\\
\mu_{Li}(\gamma_j)&=-\delta_{ij}\gamma_j+(1-\delta_{ij})(\gamma_j+\text{Max}[\langle\gamma_i,\gamma_j\rangle,0]\gamma_i).
\end{eqnarray}

Now let us see how the quiver technology is related to the spectrum of line defects
in the theory.
Recall that at low energy a UV line defect $L$
decomposes into a sum of IR line defects, as in \eqref{eq:ir-decomposition}.
Among these IR line defects, the one with the smallest
$\text{Re}(Z_\gamma/\zeta)$ corresponds to the ground state of the
UV line defect. The charge of this line defect is called the core charge
of the UV line defect. One could define a RG map which maps the UV
line defect to its core charge $\gamma_c$. As discussed in
\cite{Cordova:2013bza,Gaiotto:2010be} the RG map is a bijection in
$\mathcal{N}=2$ theories of quiver type. This nice property allows one to
identify the set of UV line defects with the IR charge lattice $\Gamma$.

The RG map is piecewise constant and jumps at the locus where
$\text{Re}(Z_{\gamma}/\zeta)=0$ for some $\gamma$, which is the same locus
where quiver mutation happens. In particular when $\gamma$ itself is the
charge of some BPS state the jump of $\gamma_c$ is given by (\cite{Cordova:2013bza}):
\begin{equation}\label{corejump}
\mu_{Ri}(\gamma_c)=\gamma_c-\text{Min}[\langle\gamma_i,\gamma_c\rangle,0]\gamma_i,\quad \mu_{Li}(\gamma_c)=\gamma_c+\text{Max}[\langle\gamma_i,\gamma_c\rangle,0]\gamma_i.
\end{equation}

For a given seed $\{\gamma_i\}$ and its associated particle cone $\mathcal{C}$, there exists a dual cone $\check{\mathcal{C}}$ defined as:
\begin{equation}
\check{\mathcal{C}}=\Bigg\{\check{\gamma}\in\Gamma_u\otimes_{\mathbb{Z}}\mathbb{R}|\langle\check{\gamma},\gamma\rangle\geq 0\quad\forall\gamma\in\mathcal{C}\Bigg\}.
\end{equation}
Using the inverse of the RG map we see that the
integral points of $\check{\mathcal{C}}$ correspond to a distinguished
set of UV line defects by the inverse of the RG map. Within this set,
the OPE relations turn out to be extremely simple. Indeed,
if $\gamma_i$ the core charge of a UV line defect $L_i$, and all
$\gamma_i\in\check{\mathcal{C}}$, then we have simply \cite{Cordova:2013bza}
\begin{equation}\label{LDOPEs}
L_1 L_2 = q^{\frac{\langle\gamma_1,\gamma_2\rangle}{2}} L_3,
\end{equation}
where $\gamma_3=\gamma_1+\gamma_2$.

Now pick a point of the Coulomb branch and a particle half-plane $\mathfrak{h}_\theta$. This fixes an
initial seed $\mathfrak{s}$. In addition to the dual cone $\check{\mathcal{C}}_\mathfrak{s}$, there are other dual cones $\check{\mathcal{C}}_{\mu(\mathfrak{s})}$, corresponding to the seeds $\mu(\mathfrak{s})$ mutated from $\mathfrak{s}$. In
these other dual cones the line defect OPE also has the nice form (\ref{LDOPEs}).
To put everything in the same footing one can trivialize $\Gamma$ using the initial seed $\mathfrak{s}$,
then mutate $\check{\mathcal{C}}_{\mu(\mathfrak{s})}$ back to $\mathfrak{s}$ using
(\ref{corejump}).
After so doing, one has a collection of dual cones meeting along codimension-one faces
in $\mathbb{Z}^{\text{rank}(\Gamma)}\otimes_{\mathbb{Z}}\mathbb{R}$.
In a general $\N=2$ theory, the dual cones obtained in this way cover only some subset of the charge lattice.
For Argyres-Douglas theories, however, there are only finitely many dual cones, and they fill up
the full charge lattice \cite{Cordova:2013bza}.
Thus the full set of UV line defects is generated by
the line defects whose core charges lie at the boundaries of the dual cones.

Concretely, in the $(A_1,A_{2N})$ Argyres-Douglas theories, although the boundaries of dual
cones are in general codimension-$1$ hyperplanes, these hyperplanes intersect at
half-lines, such that line defects with core charges along those half-lines
generate the whole space of UV line defects. In these theories we thus
obtain a unique and canonical choice of line defect generators, which is
very convenient for computational purposes. (In the $(A_1,A_2)$ theory we
have already mentioned these generators in \S\ref{sec:intro-example}.)

In contrast, in the $(A_1,D_{2N+1})$ Argyres-Douglas theories,
due to the flavor symmetry, the dual cone picture does not quite give a
unique choice of UV line defect generators.
In these theories we will use the class $\cS$ picture instead.

\subsection{Framed BPS states from framed quivers} \label{sec:framed-bps-from-quivers}

In $\cN=2$ theories of quiver type,
framed BPS spectra associated to line defects
can be computed using framed quivers \cite{Cordova:2013bza}.\footnote{As emphasized in \cite{Cordova:2013bza}, this method
does not \ti{in general} produce the correct framed BPS spectrum, but it does work
for a large class of theories including Argyres-Douglas theories.} One extends the charge lattice
$\Gamma$ by an extra direction spanned by a new ``infinitely heavy''
flavor charge $\gamma_F$,
which has zero pairing with all charges. The line defect with core charge $\gamma_c$
is then regarded as a particle carrying the charge $\gamma_c+\gamma_F$,
and framed BPS states supported by the defect
are similarly regarded as particles with charges
of the form
\begin{equation}\label{halo}
\gamma_c+\gamma_F+\gamma_h, \quad\text{where } \gamma_h=\Sum_{i=1}^{\text{rank}(\Gamma)}a_i\gamma_i, \quad a_i\in\mathbb{Z}_{\geq 0}.
\end{equation}
One then defines a new ``framed quiver,'' obtained by adding to the original quiver a new framing node representing the bare line defect and corresponding arrows. The framed BPS states are given by the unframed BPS states of the framed quiver whose charges are of the form (\ref{halo}).

BPS states in quiver quantum mechanics can be conveniently computed by the ``mutation method'' as introduced
in \cite{Alim:2011kw,Gaiotto:2010be,Cecotti:2010fi,Cecotti:2011gu}. Concretely, we first fix a point in the Coulomb branch and a choice of half-plane $\mathfrak{h}_\theta$,
then rotate $\mathfrak{h}_\theta$ counterclockwise\footnote{The choice of counterclockwise vs. clockwise is
just a convention.} until $\theta$ has increased by $\pi$.
In this process the original seed undergoes a series of right mutations $\mu_{Ri}$, and for each mutation the
node $\gamma_i$ that exits to the right of $\mathfrak{h}_\theta$ corresponds to a BPS particle.
Conversely each BPS particle will be rightmost at some stage of the rotation, so the $\gamma_i$
obtained in this way exhaust all BPS particles in this chamber. In \cite{Alim:2011kw} this method was
applied to the ordinary BPS quiver to compute the ordinary (vanilla, unframed) BPS spectrum;
here instead we apply it to the framed quiver constructed above, to get
the framed BPS spectrum.

\subsection{Line defects in class \texorpdfstring{$\cS[A_1]$}{S[A1]} theories} \label{sec:defects-class-S}

In class $\cS[A_1]$ theories there is a natural geometric picture of the
$\half$-BPS line defects:
they correspond to paths (up to homotopy) on the internal Riemann surface $C$
\cite{Alday:2009fs,Drukker:2009id,Drukker:2009tz,Gaiotto:2010be}.
For class $\cS[A_1]$ theories with irregular punctures, one has to consider not
only closed paths but also certain combinations of open paths, called
\ti{laminations} in \cite{Gaiotto:2010be} (following \cite{Fock-Goncharov}
where the same combinations of open paths were considered.)

The laminations we consider are drawn on a disc, which
we think of as the complex plane compactified by adding the ``circle at
infinity.'' The boundary circle is divided into arcs by
marked points corresponding to the Stokes directions
(see \cite{Gaiotto:2010be} for more on this.) Then
a lamination is a collection of paths on the disc,
carrying integer weights,
subject to some conditions \cite{Fock-Goncharov,Gaiotto:2010be}:
the sum of weights meeting each boundary arc must be zero, and
all paths with negative weights must be deformable into a small
neighborhood of the boundary.

\subsection{Framed BPS indices in class \texorpdfstring{$\cS[A_1]$}{S[A1]} theories, without spin}
\label{sec:framed-class-S}

In \cite{Gaiotto:2010be}, a scheme is presented for computing the
framed BPS indices associated to a given line
defect in a theory of class $\cS[A_1]$, without spin information.
In this scheme one needs two pieces of data:
\begin{itemize}
\item the lamination representing the line defect,
\item the WKB triangulation determined by the chosen point of the Coulomb branch
and phase of the line defect.
\end{itemize}
It is easiest to illustrate this rule by an example.
So, consider the triangulation of the once-punctured triangle
and the lamination shown in Figure
\ref{fig:d3-wkb-lam}. (This example arises in the $(A_1,D_3)$ theory
considered in \S\ref{sec:A1D3} below: it corresponds to the line
defect called $B_2$ there.)

 \begin{figure}
 	\centering
 	\includegraphics[scale=0.18]{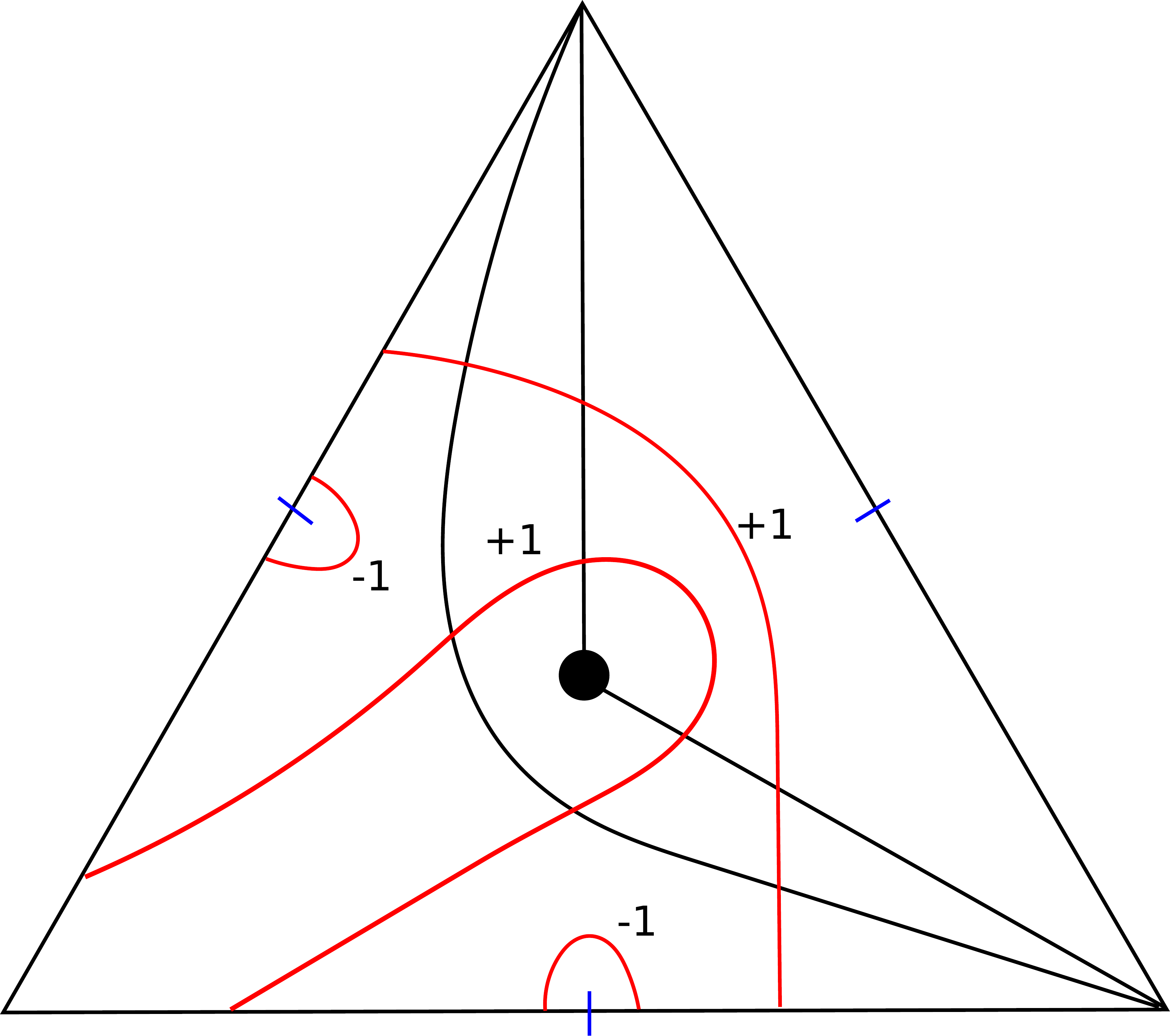}
 	\caption{An example of a WKB triangulation of the once-punctured
 	triangle and a lamination, corresponding to the line defect $B_2$
 	in the $(A_1,D_3)$ Argyres-Douglas theory.}
 	\label{fig:d3-wkb-lam}
 \end{figure}

We fix an orientation of each component of the lamination.
Then we divide each component of the lamination into arcs crossing
triangles. To each arc we assign the matrix $L$ ($R$)
if the arc turns left (right),\footnote{The matrices we present here
are the \ti{transpose} of the matrices in \cite{Gaiotto:2010be}, and correspondingly
we take the products in the reverse of the order taken in
\cite{Gaiotto:2010be}; this corresponds to the usual order
of composition of parallel transports, and makes the construction
directly compatible with \cite{Gaiotto:2012rg}, which
will be useful below.}
\begin{equation}
 L = \begin{pmatrix} 1 & 0 \\ 1 & 1 \end{pmatrix}, \qquad R = \begin{pmatrix} 1 & 1 \\ 0 & 1 \end{pmatrix}.
\end{equation}
When the lamination crosses an internal edge $E_i$ we assign the matri{}x
\begin{equation}
 M_E = \begin{pmatrix} \sqrt{\cX_E} & 0 \\ 0 & 1 / \sqrt{\cX_E} \end{pmatrix}.
\end{equation}
To the initial and final points of each component we assign the vectors
\begin{equation}
 E^R = \begin{pmatrix} 0 & 1 \end{pmatrix}, \quad E^L = \begin{pmatrix} 1 & 0 \end{pmatrix}, \quad
 B^R = \begin{pmatrix} 1 \\ 0 \end{pmatrix}, \quad B^L = \begin{pmatrix} 0 \\ 1 \end{pmatrix},
\end{equation}
choosing $L$ or $R$ according to whether the endpoint is on the left or
the right of the marked point of the boundary edge.
Then we multiply these matrices in order,
with the beginning of the path corresponding to the rightmost matrix,
to get a number for each component. If the component has weight
$k$ we raise this number to the $k$-th power. Finally we multiply the
contributions from all components to get the vev.

In the example of Figure \ref{fig:d3-wkb-lam} above, the contribution from the left long component with weight $+1$ is
\begin{multline}
	E^R L M_{E_2} L M_{E_3} R M_{E_1} L M_{E_2} L B^R = \\
	\frac{1}{\sqrt{\cX_1 \cX_3}} + \frac{1}{\sqrt{\cX_1 \cX_3} \cX_2} + 2 \frac{\sqrt{\cX_3}}{\sqrt{\cX_1}} + \sqrt{\cX_1 \cX_3} + \frac{\sqrt{\cX_3}}{\sqrt{\cX_1} \cX_2} + \frac{\cX_2 \sqrt{\cX_3}}{\sqrt{\cX_1}} + \cX_2 \sqrt{\cX_1 \cX_3}.
\end{multline}
Similarly, the contribution from the right long component with weight $+1$ is $\sqrt{\cX_3/\cX_1}$. The short components with weight $-1$ contribute 1. The total contribution from this lamination is
\begin{equation}\label{eq:component-contrib}
\frac{1}{\cX_1}+\frac{1}{\cX_1\cX_2}+\cX_3+2\frac{\cX_3}{\cX_1}+\frac{\cX_3}{\cX_1\cX_2}+\cX_2\cX_3+\frac{\cX_2\cX_3}{\cX_1}.
\end{equation}
Thus \eqref{eq:component-contrib} gives the generating
function of framed BPS states associated to this line defect,
without spin information.

\subsection{Framed BPS indices in class \texorpdfstring{$\cS[A_1]$}{S[A1]} theories, with spin} \label{sec:spin-class-S}

We continue with our example from \S\ref{sec:framed-class-S}.
Incorporating the spin information requires us to
take each term in \eqref{eq:component-contrib} and assign it the
correct power of $q$. The work of
\cite{Galakhov:2014xba,Gabella:2016zxu} provides
a rule for determining these powers.
The first step is to associate
the terms in \eqref{eq:component-contrib} to arcs on
a branched double cover $\Sigma$
of the disc\footnote{The double cover $\Sigma$ is the
Seiberg-Witten curve of the $\cN=2$ theory at a point of
its Coulomb branch,
or the corresponding spectral curve of the Hitchin system.}
following the ``path lifting'' rules of
\cite{Gaiotto:2012rg}, as follows.

\begin{figure}
 	\centering
 	\includegraphics[scale=0.17]{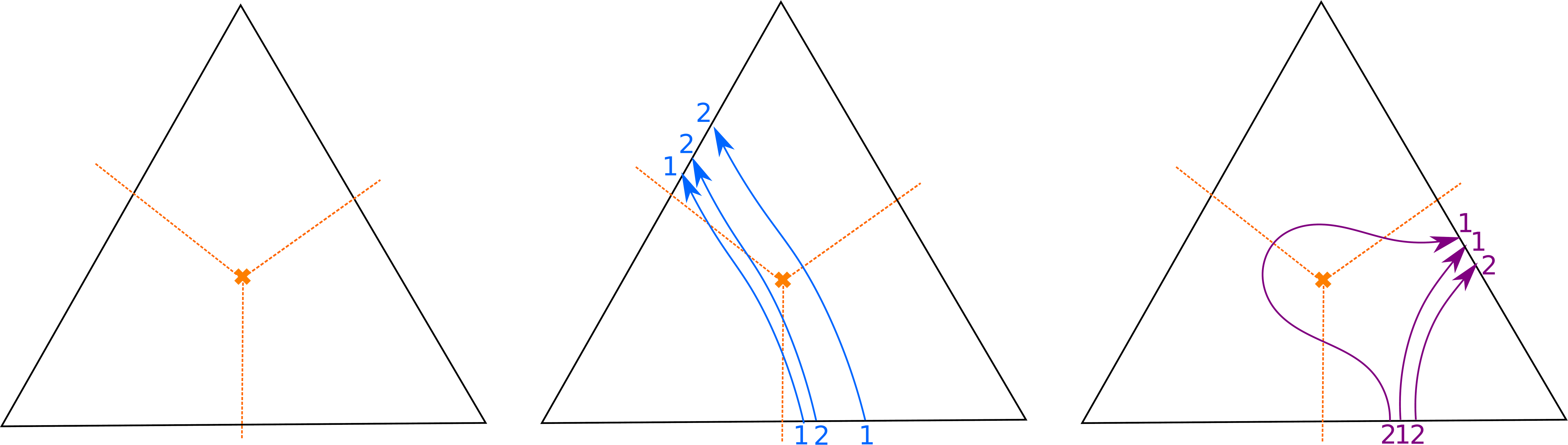}
 	\caption{Left: a triangle with branch point and branch cuts marked.
 	Middle: lifted left-turn paths. Right: lifted right-turn paths.}
 	\label{fig:cuts-and-paths}
 \end{figure}

The double cover $\Sigma$ is
presented concretely: in each triangle we fix one branch
point and three branch
cuts, as in the left side of Figure \ref{fig:cuts-and-paths};
the double cover has sheets labeled $1$ and $2$, and at
each cut sheet $1$ is glued to sheet $2$ and vice versa.
Next, note that each term in \eqref{eq:component-contrib} comes from products of two specific chains of matrix elements:
e.g. the term $\frac{1}{\cX_1}$ comes from product of two contributions. As an example, the first contribution comes from taking the $(2,2)$ entries of the matrices
from the beginning to the second-to-last $L$, then taking the $(2,1)$ entry
of that $L$, then the $(1,1)$ entries of all the rest.
Each of these matrix elements corresponds to an arc on the double cover,
which we regard as a ``lift'' of the corresponding arc of the
lamination.
In Figure \ref{fig:cuts-and-paths} we show three arcs
corresponding to the three nonzero matrix elements
of each of $L$ and $R$;
the arc for the $(i,j)$ matrix element begins on sheet
$j$ and ends on sheet $i$.

\begin{figure}
	\centering
	\includegraphics[scale=0.2]{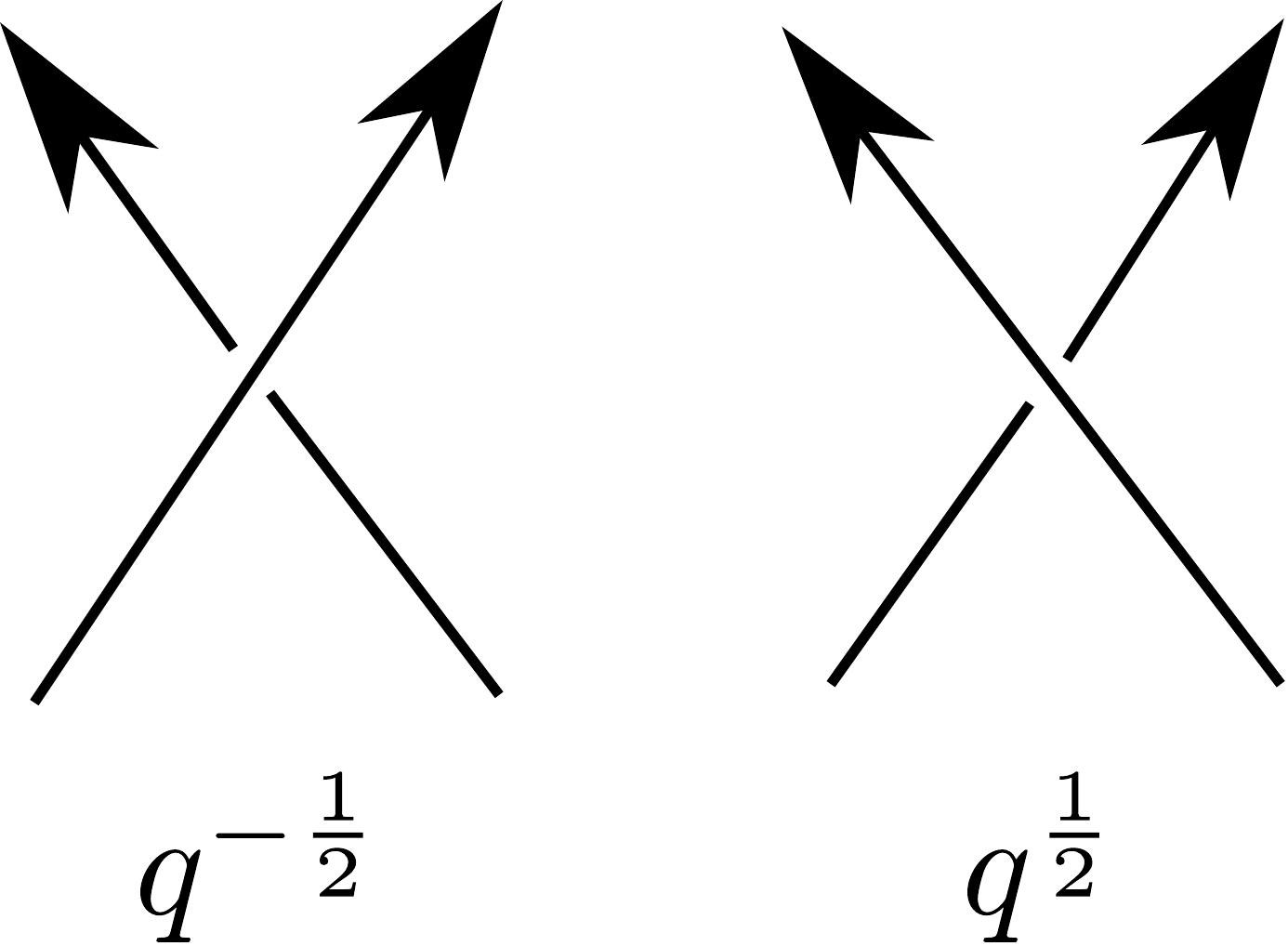}
	\caption{Rules for assigning powers of $q$ to self-crossings
	of the lifted path.}
	\label{fig:q-rules}
\end{figure}

Concatenating these arcs gives a long path $P$ on $\Sigma$,
associated to the term in \eqref{eq:component-contrib}
which we are studying.
If $P$ has no self-intersections then we assign this term the factor $q^0$.
If there are self-intersections then each contributes a factor
$q^\half$ or $q^{-\half}$, according to Figure \ref{fig:q-rules}, where
the arc which appears later in the path is drawn higher.

\begin{figure}
 	\centering
 	\includegraphics[scale=0.22]{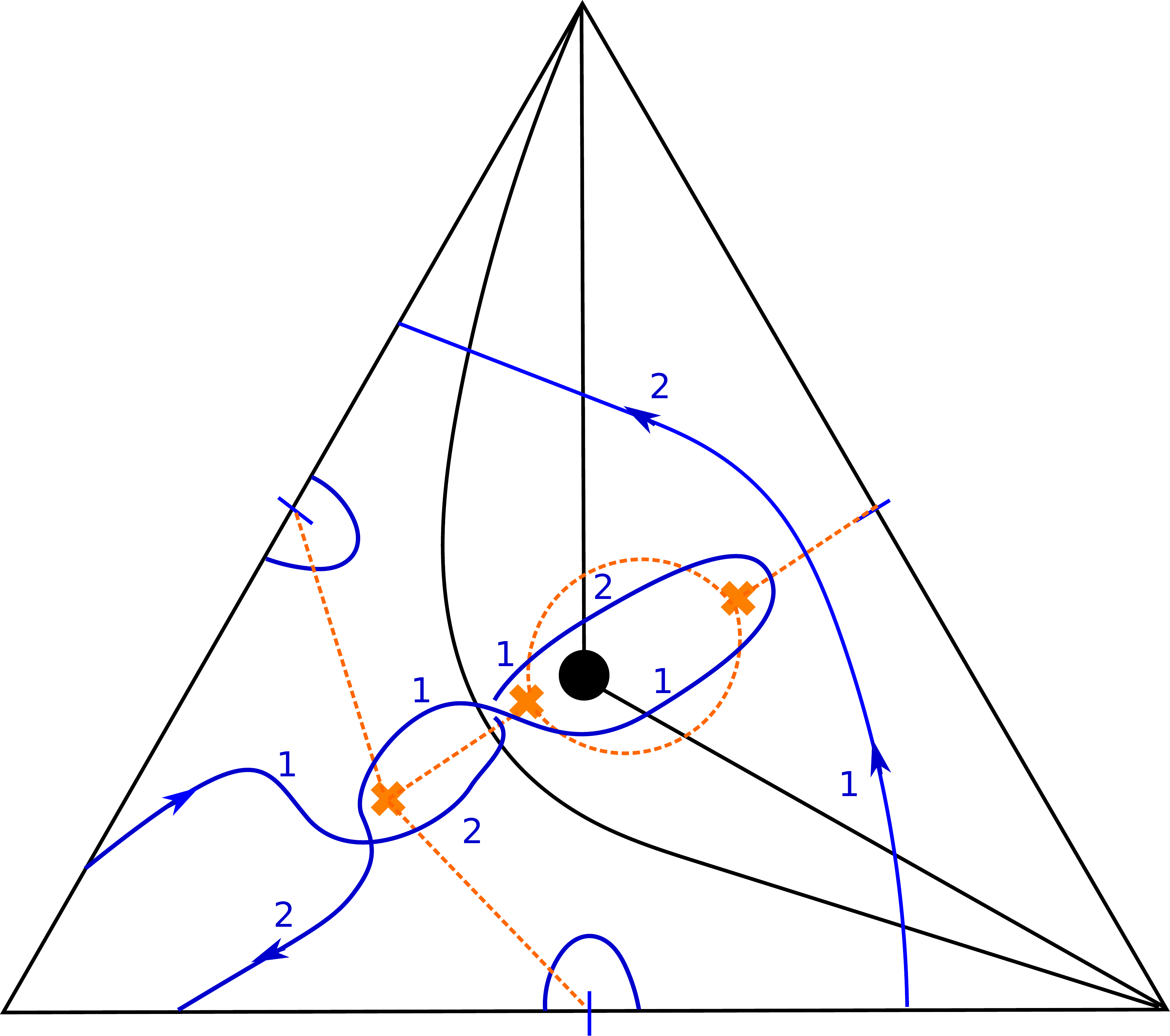}
 	\caption{One of the lifted paths contributing to the term \eqref{eq:term-to-study}.}
 	\label{fig:example-lifts}
 \end{figure}

To illustrate how this works, we consider the term
\begin{equation} \label{eq:term-to-study}
	2 \frac{\cX_3}{\cX_1}
\end{equation}
in \eqref{eq:component-contrib}.
The factor $2$ here means \eqref{eq:term-to-study}
is a sum of two contributions, associated to two different
lifted paths: we show one of them in Figure \ref{fig:example-lifts}.
There is one crossing in Figure \ref{fig:example-lifts}, where
both strands are lifted to sheet $1$.\footnote{The
projection of the path to the base has two crossings, but at one
of these crossings the two strands are lifted to different sheets,
so it is not a crossing for the lifted path.}
Comparing this crossing to Figure
\ref{fig:q-rules}, we see that this term should be weighted by
$q^{\half}$. Drawing a similar picture for the other contribution to \eqref{eq:term-to-study} we see that it gets weighted by
$q^{-\half}$. Thus altogether \eqref{eq:term-to-study} is replaced
by
\begin{equation}
	(q^\half + q^{-\half}) \frac{\cX_3}{\cX_1},
\end{equation}
which tells us that the $2$
framed BPS states with charge $\gamma_3 - \gamma_1$
come in a $2$-dimensional multiplet of the rotation group $SO(3)$.
Carrying out similar computations for the other terms one finds
(as expected)
that all of them come with the factor $q^0$, i.e. they are in
the trivial representation of $SO(3)$.
Thus altogether the $q$-deformed version of the generating function \eqref{eq:component-contrib}
turns out to be
\begin{equation} \label{eq:q-framed-bps}
	\frac{1}{\cX_1}+\frac{1}{\cX_1\cX_2}+\cX_3+(q^\half+q^{-\half})\frac{\cX_3}{\cX_1}+\frac{\cX_3}{\cX_1\cX_2}+\cX_2\cX_3+\frac{\cX_2\cX_3}{\cX_1}.
\end{equation}
This is exactly the generating function for the line defect generator $B_2$ in \S\ref{sec:A1D3} below.

\section{\texorpdfstring{$(A_1,A_{2N})$}{(A1,A2N)} Argyres-Douglas theories}\label{sec:examples-even}

In this section we present the results of explicit computations verifying the commutativity
\eqref{eq:commutativity} in the Argyres-Douglas theories of type $(A_1,A_2)$,
$(A_1,A_4)$ and $(A_1,A_6)$.

\subsection{\texorpdfstring{$(A_1,A_2)$}{(A1,A2)} Argyres-Douglas theory}\label{sec:A1A2}
\begin{figure}
	\centering
	\includegraphics[scale=0.15]{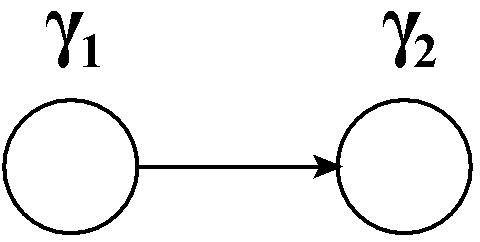}
	\caption{A BPS quiver for ($A_1,A_2$) Argyres-Douglas theory.}
	\label{A2quiver}
\end{figure}

We consider $(A_1,A_2)$ Argyres-Douglas theory and choose the chamber\footnote{In all the examples considered in this paper, to simplify computation, we always work in a chamber for which the number of number of BPS particles is the minimum possible --- with one exception in the case of
$(A_1,A_6)$ as noted below.}
represented by the BPS quiver in Figure \ref{A2quiver} containing two BPS particles: (in increasing central charge phase order)
\begin{equation}
\gamma_1,\gamma_2.
\end{equation}
There are five non-identity line defect generators. Assuming the line defect phase is smaller than the phases of all BPS particles, the generating functions are \cite{Cordova:2013bza,Gaiotto:2010be}:
\begin{eqnarray} \label{eq:FLi}
\begin{split}
F(L_1)&=X_{\gamma_1},\\
F(L_2)&=X_{\gamma_2}+X_{\gamma_1+\gamma_2},\\
F(L_3)&=X_{-\gamma_1}+X_{-\gamma_1+\gamma_2}+X_{\gamma_2},\\
F(L_4)&=X_{-\gamma_1-\gamma_2}+X_{-\gamma_1},\\
F(L_5)&=X_{-\gamma_2}.
\end{split}
\end{eqnarray}
In the geometric picture these generators $L_i$ correspond to five laminations which are rotated into each other under the monodromy action. As a result their generating functions are related to each other by the action of powers
of the monodromy operator.

The Schur index with $L_i$ inserted is computed via \cite{Cordova:2016uwk}:
\begin{equation}
\cI_{L_i}(q)=(q)_\infty^2\text{Tr}[F(L_i)S(q)\overline{S}(q)], \quad S(q)=E_q(X_{\gamma_1})E_q(X_{\gamma_2}).
\end{equation}
The corresponding $2d$ chiral algebra is the $(2,5)$ minimal model \cite{Cordova:2015nma,Beem:2013sza,Beem:2014zpa}, which has two primaries: the vacuum $\Phi_{1,1}$ and $\Phi_{1,2}$ with weight $-1/5$. In general, characters of $\Phi_{s,r}$ in the $(p,p')$ minimal model ($1\leq s\leq p-1$, $1\leq r\leq p'-1$) are given by \cite{D.Francesco}:
\begin{eqnarray}
\begin{split}
\chi_{s,r}(q)&=q^{-\frac{(rp-sp')^2-(p-p')^2}{4pp'}+\frac{1}{24}(1-\frac{6(p-p')^2}{pp'})}\bigg(K_{s,r}^{p,p'}(q)-K_{-s,r}^{p,p'}(q)\bigg),\\
K_{s,r}^{p,p'}(q)&=\frac{1}{q^{\frac{1}{24}}(q)_\infty}\Sum_{n\in\mathbb{Z}}q^{\frac{(2pp'n+pr-p's)^2}{4pp'}}.
\end{split}
\end{eqnarray}
\par
The line defect Schur index $\cI_{L_i}(q)$ does not depend on the index $i$ and admits the following character expansion \cite{Cordova:2016uwk}:
\begin{equation}
\cI_{L}(q)=q^{-\frac{1}{2}}\big(\chi_{1,1}(q)-\chi_{1,2}(q)\big).
\end{equation}
Similarly, the Schur index with two $L_i$ inserted is given by \cite{Cordova:2016uwk}:
\begin{equation}
\cI_{L_iL_j}(q)=(q)_\infty^2\text{Tr}[F(L_i)F(L_j)S(q)\overline{S}(q)].
\end{equation}
Unlike $\cI_{L_i}(q)$, $\cI_{L_iL_j}(q)$ does depend on $i$ and $j$, though this dependence disappears
in the limit $q\to 1$. Expansions of $\cI_{L_iL_j}(q)$ in terms of characters are given as follows:
\begin{eqnarray}
\begin{split}
\cI_{L_iL_i}(q)=\cI_{L_iL_{i-1}}(q) & =  (q^{-1}+q^{-2})\chi_{1,1}(q)-q^{-2}\chi_{1,2}(q),\\
\cI_{L_iL_{i+1}}(q)=\cI_{L_iL_{i-2}}(q) & =  (1+q^{-1})\chi_{1,1}(q)-q^{-1}\chi_{1,2}(q),\\
\cI_{L_iL_{i+2}}(q) & =  2\chi_{1,1}(q)-\chi_{1,2}(q).
\end{split}
\end{eqnarray}
The map $f$ is given by
\begin{equation}\label{A2f}
I\xrightarrow{f} [\Phi_{1,1}],\quad
L_i\xrightarrow{f}[L]:=[\Phi_{1,1}]-[\Phi_{1,2}].
\end{equation}
Moreover,
\begin{equation}\label{A22LD}
L_iL_j\xrightarrow{f}[LL]:=2[\Phi_{1,1}]-[\Phi_{1,2}].
\end{equation}
Recall that the non-trivial fusion rule in $(2,5)$ minimal model is given by
\begin{equation}
[\Phi_{1,2}]\times[\Phi_{1,2}]=[\Phi_{1,1}]+[\Phi_{1,2}].
\end{equation}
Combining with (\ref{A2f}) and (\ref{A22LD}) we have
\begin{equation}\label{A2prod}
[LL]=[L]\times[L],
\end{equation}
as first observed in \cite{Cordova:2016uwk}.

\begin{figure}
	\centering
	\includegraphics[scale=0.16]{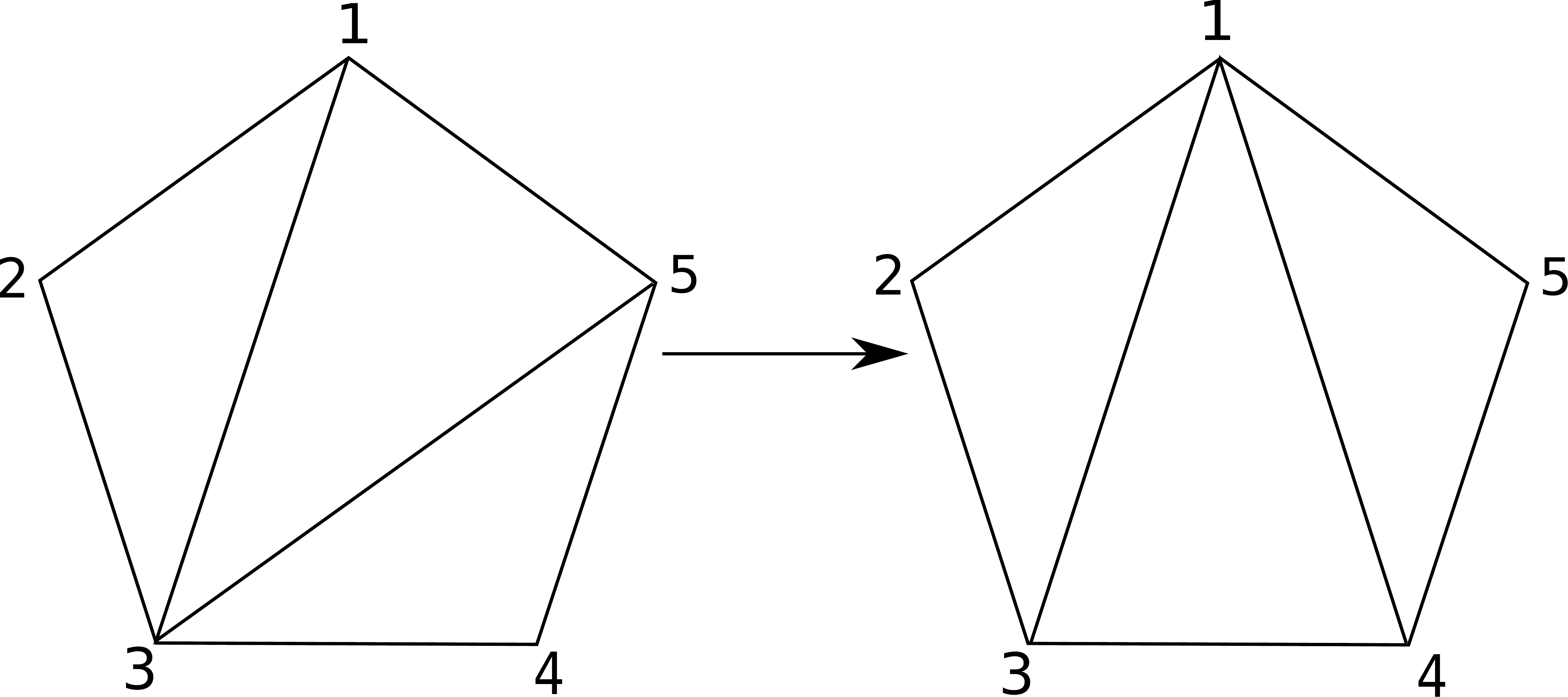}
	\caption{The classical monodromy action in the $(A_1,A_2)$ theory, which rotates the triangulation of
	the pentagon clockwise by $2$ units, is equivalent to a single flip which replaces
	the $35$ edge by a $14$ edge.}
	\label{5gon}
\end{figure}

Next we consider the fixed points of $U(1)_r$.
For this purpose we found it convenient to use the geometric picture
as described in \S\ref{sec:classical-monodromy-S}.
The classical monodromy action $M$ is directly given by a single flip: see Figure \ref{5gon}. According to (\ref{flip}) the concrete transformation is given by
\begin{equation} \label{eq:classical-moonodromy-A1A2}
\cX_{\gamma_1}\to\frac{1}{\cX_{\gamma_2}},\quad \cX_{\gamma_2}\to\frac{\cX_{\gamma_1}\cX_{\gamma_2}}{1+\cX_{\gamma_2}}.
\end{equation}
Thus the fixed locus is
\begin{equation}
\cX_{\gamma_1}^2-\cX_{\gamma_1}-1=0,\quad \cX_{\gamma_2}=\frac{1}{\cX_{\gamma_1}}.
\end{equation}
This locus consists of two points, which we label I, II.
At these points the $X_\gamma$ evaluate to:
\begin{equation}
\text{I}: (\cX_{\gamma_1}, \cX_{\gamma_2}) = \left( \frac{1-\sqrt{5}}{2}, -\frac{1+\sqrt{5}}{2} \right),\quad
\text{II}: (\cX_{\gamma_1}, \cX_{\gamma_2}) = \left( \frac{1+\sqrt{5}}{2}, -\frac{1-\sqrt{5}}{2} \right).
\end{equation}
To construct the map $g: \cL \to \cO(F)$, for any line defect generator $L_i$
we evaluate $F({L_i})$ at these two fixed points,
using \eqref{eq:FLi}.
As expected, the dependence on $L_i$ disappears in the process:
\begin{equation}\label{A2g}
L_i\xrightarrow{g}(F_{L_i}^{\text{I}},F_{L_i}^{\text{II}}) = \left(\frac{1-\sqrt{5}}{2},\frac{1+\sqrt{5}}{2}\right).
\end{equation}
Of course we also have the trivial line defect, whose vev
is $1$ at every fixed point:
\begin{equation}
	1 \xrightarrow{g} (1, 1).
\end{equation}

Finally, we follow the recipe described in \S\ref{sec:ver-diag}, \S\ref{sec:verlinde-fixed} to construct the isomorphism $h: \cV \to \cO(F)$.
We need the fusion matrices, which are given by\footnote{Our convention is to order the primaries as $(\Phi_{1,1},\Phi_{1,2})$.}
\begin{equation}
N_{\Phi_{1,1}}=\begin{pmatrix}
1 & 0\\
0 & 1
\end{pmatrix}, \qquad
N_{\Phi_{1,2}}=\begin{pmatrix}
0 & 1\\
1 & 1
\end{pmatrix}.
\end{equation}
The modular $S$-matrix is \cite{D.Francesco}:
\begin{equation}
S = \frac{2}{\sqrt{5}}\begin{pmatrix}
-\sin\frac{2\pi}{5} & \sin\frac{4\pi}{5}\\
\sin\frac{4\pi}{5} & \sin\frac{2\pi}{5}
\end{pmatrix}.
\end{equation}
Thus the fusion matrices are diagonalized by the $S$ matrix,
\begin{equation} \label{eq:fusion-diag-25}
\hat{N}_{\Phi_{1,1}} = SN_{\Phi_{1,1}} S^{-1}=\begin{pmatrix}
1 & 0\\
0 & 1
\end{pmatrix}, \qquad
\hat{N}_{\Phi_{1,2}} = SN_{\Phi_{1,2}} S^{-1}=\begin{pmatrix}
\frac{1-\sqrt{5}}{2} & 0\\
0 & \frac{1+\sqrt{5}}{2}
\end{pmatrix}.
\end{equation}
As we explained in \S\ref{sec:ver-diag}-\S\ref{sec:verlinde-fixed},
the map $h$ takes each of $\Phi_{1,1}$ and $\Phi_{1,2}$
to its eigenvalues. So, it takes
$h(\Phi_{1,1}) = (1,1)$ and \ti{either}
$h(\Phi_{1,2}) = (\frac{1 - \sqrt{5}}{2}, \frac{1 + \sqrt{5}}{2})$
\ti{or} $h(\Phi_{1,2}) = (\frac{1 + \sqrt{5}}{2}, \frac{1 - \sqrt{5}}{2})$.
To decide which is the right ordering, we need to
know the dictionary between $U(1)_r$ fixed points and
eigenspaces of the fusion operators. These eigenspaces
themselves correspond to primary fields, so equivalently,
we need the dictionary between
the fixed points I, II and the primary
fields $\Phi_{1,1}$, $\Phi_{1,2}$.
This dictionary is determined by the table below:
\begin{center}
	\begin{tabular}{ |c|c|c|c| }
		\hline
		fixed point & weights of $M$ & weights of $U(1)_r$ & primary field\\ \hline
		I & $\e^{2\pi \I(3/5)}, \e^{2\pi \I(2/5)}$ & $\frac{3}{5}, \frac{2}{5}$ & $\Phi_{1,2}$ \\ \hline
		II & $\e^{2\pi \I(6/5)}, \e^{-2\pi \I(1/5)}$ & $\frac{6}{5}, -\frac{1}{5}$ & $\Phi_{1,1}$ \\ \hline
	\end{tabular}
\end{center}
In this table, to determine the weights of $M$ at each fixed
point,
we computed directly the linearization of the classical
 monodromy \eqref{eq:classical-moonodromy-A1A2}.
On the other side, the dictionary between primary fields
and $U(1)$ weights is taken from \cite{Fredrickson:2017yka}.
At any rate, we can now read off that $\Phi_{1,1}$ corresponds to fixed point II and $\Phi_{1,2}$ corresponds to fixed point I.
Combining this with \eqref{eq:fusion-diag-25}, $h$ is given by:
\begin{equation}
[\Phi_{1,1}]\xrightarrow{h}(1,1), \quad [\Phi_{1,2}]\xrightarrow{h}\Big(\frac{1+\sqrt{5}}{2},\frac{1-\sqrt{5}}{2}\Big).
\end{equation}
Composing this with $f$ from \eqref{A2f} we have
\begin{equation}
L_i\xrightarrow{h\circ f}\Big(\frac{1-\sqrt{5}}{2},\frac{1+\sqrt{5}}{2}\Big).
\end{equation}
Comparing this with (\ref{A2g}) we see that the diagram
indeed commutes.

\subsection{An intermission on the homomorphism property}\label{sec:homomophism-property}

To make sure $f$ is a homomorphism, (\ref{A2prod}) needs to hold not only for the generators $L_i$ but also for arbitrary line defects. This would involve checking e.g.
\begin{equation}
[LLL]\stackrel{?}{=}[L]\times[L]\times[L]
\end{equation}
and similar relations for higher number of line defect generators\footnote{We would like to comment that the product of $F(L)$ is associative (due to associativity of the quantum torus algebra of $X_\gamma$) and so is the fusion product.}. As an example let us consider the case of three line defect generators. The line defect Schur index is given by
\begin{equation}
\cI_{L_iL_jL_k}(q)=(q)^2_\infty\text{Tr}[F(L_i)F(L_j)F(L_k)S(q)\overline{S}(q)].
\end{equation}
There are many relations between $\cI_{L_iL_jL_k}$,
\begin{align*}
& \cI_{L_{i-1}L_iL_{i+2}}=\cI_{L_{i-1}L_iL_{i+1}}=\cI_{L_{i-2}L_iL_{i+1}},\\
& \cI_{L_iL_iL_{i+2}}=\cI_{L_{i-2}L_iL_{i+2}}=\cI_{L_{i-2}L_iL_i},\\
& \cI_{L_{i+2}L_iL_{i+1}}=\cI_{L_iL_iL_{i+1}}=\cI_{L_{i+2}L_iL_i}=\cI_{L_{i-1}L_iL_i}=\cI_{L_iL_iL_{i-2}}=\cI_{L_{i-1}L_iL_{i-2}},\\
& \cI_{L_{i+1}L_iL_i}=\cI_{L_iL_iL_i}=\cI_{L_iL_iL_{i-1}}=q^{-2} \cI_{L_{i-1}L_iL_{i-2}},\\
& \cI_{L_{i+1}L_iL_{i+1}}=\cI_{L_{i-1}L_iL_{i-1}}=\cI_{L_{i+1}L_iL_{i-1}}=q^{-1} \cI_{L_{i-1}L_iL_{i-2}},\\
& \cI_{L_{i+2}L_iL_{i+2}}=\cI_{L_{i+1}L_iL_{i+2}}=\cI_{L_{i+2}L_iL_{i-1}}=\cI_{L_{i-2}L_iL_{i-1}}=\cI_{L_{i+1}L_iL_{i-2}}=\cI_{L_{i-2}L_iL_{i-2}}.
\end{align*}
The independent indices admit the following character expansions,
\begin{align*}
\cI_{L_{i-2}L_iL_{i+1}} & =q^{-\frac{1}{2}}\big((1+2q)\chi_{1,1}(q)-(1+q)\chi_{1,2}(q)\big),\\
\cI_{L_{i-2}L_iL_i} & =q^{-\frac{1}{2}}\big((2+q)\chi_{1,1}(q)-2\chi_{1,2}(q)\big),\\
\cI_{L_{i-1}L_iL_{i-2}} & =q^{-\frac{1}{2}}\big((1+q^{-1}+q^{-2})\chi_{1,1}(q)-(1+q^{-2})\chi_{1,2}(q)\big),\\
\cI_{L_{i+2}L_iL_{i-2}} & =q^{-\frac{1}{2}}\big(3\chi_{1,1}(q)-2\chi_{1,2}(q)\big),\\
\cI_{L_{i-2}L_iL_{i-2}} & =q^{-\frac{1}{2}}\big((2+q^{-1})\chi_{1,1}(q)-(1+q^{-1})\chi_{1,2}(q)\big).
\end{align*}
We immediately see that
\begin{equation}
L_iL_jL_k\xrightarrow{f}[LLL]:=3[\Phi_{1,1}]-2[\Phi_{1,2}]=[L]\times[L]\times[L].
\end{equation}
\par
In principal, to prove that $f$ is a homomorphism we need to repeat the above calculation for arbitrary number of line defect generator insertions. We are not able to prove it in this paper. Instead we offer some arguments about why we believe $f$ is indeed a homomorphism. We have seen explicitly that the images of $L_iL_j$ and $L_iL_jL_k$ under $f$ does not depend on the index $i$. In other examples that we consider in this paper we also checked the image of $L_{\rho i}L_{\mu j}$\footnote{Here $\rho,\mu$ label different types of line defect generators, see \S \ref{sec:A1A4}, \S \ref{sec:A1A6}, \S\ref{sec:A1D3}, \S\ref{sec:A1D5}} does not depend on $i$. Although we don't have a proof for now, we conjecture this phenomenon is general, i.e. the image of $L_{\rho_1 i_1}L_{\rho_2 i_2}\dots L_{\rho_n i_n }$ under $f$ does not depend on $i_1,\dots,i_n$. Combining this conjecture with relations between line defect generating functions one could see that $f$ is indeed a homomorphism.\par
We revisit the situation of three line defect generators. To compute the image of $L_iL_jL_k$ under $f$ we could pick any three line defect generators. Let's recall the following relation between $F(L_i)$ \cite{Gaiotto:2010be,Cordova:2013bza}:
\begin{equation}\label{A2rel}
F(L_i)F(L_{i+2})=1+q^{\frac{1}{2}}F(L_{i+1}),
\end{equation}
from which follows $[L]\times[L]=[\Phi_{1,1}]+[L]$.\footnote{As discussed in \S\ref{sec:comments}, in $(A_1,A_{2N})$ theories the line defect generators themselves correspond to a basis which also realizes fusion rules.} Schur index with insertion of $L_i,L_{i+2},L_k$ is then given by
\begin{equation}
\cI_{L_iL_{i+2}L_k}(q)=\cI_{L_k}(q)+q^{\frac{1}{2}}\cI_{L_{i+1}L_k}(q),
\end{equation}
from which it follows that
\begin{equation}
[LLL]=[L]+[LL]=[L]\times[L]\times[L].
\end{equation}
Similarly one could consider insertion of more line defect generators. By the conjecture, to compute the image of $L_{i_1}\dots L_{i_n}$ under $f$, it doesn't matter what $i_1,\dots,i_n$ are. Then we could again use (\ref{A2rel}) to reduce the number of line defect generators. Moreover this process is consistent with the fusion rules such that
\begin{equation}
[L\dots L]=[L] \times \dots \times [L].
\end{equation}

For other Argyres-Douglas theories that we are considering in this paper, there are always enough relations between $F(L_{\alpha i})$ such that the same argument goes through provided our conjecture would hold. \par

\subsection{\texorpdfstring{$(A_1,A_4)$}{(A1,A4)} Argyres-Douglas theory} \label{sec:A1A4}

\begin{figure}
	\centering
	\includegraphics[scale=0.15]{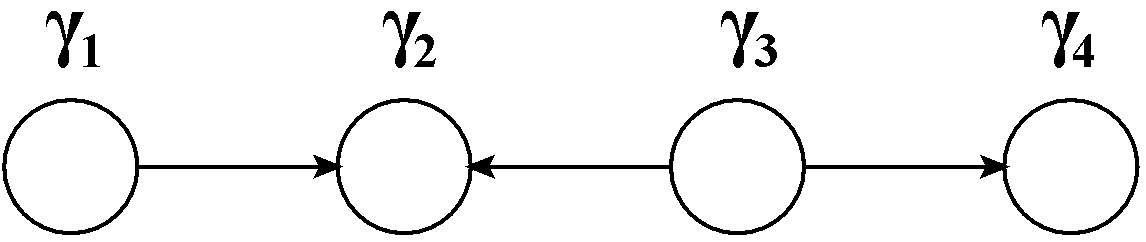}
	\caption{A BPS quiver for ($A_1,A_4$) Argyres-Douglas theory.}
	\label{A4quiver}
\end{figure}

We consider the $(A_1,A_4)$ Argyres-Douglas theory. We choose a chamber represented by the BPS quiver shown in Figure \ref{A4quiver}. Moreover our choice is made such that there are four BPS particles in this chamber. Their charges are (in increasing central charge phase order):
\begin{equation}
\gamma_1,\gamma_3,\gamma_2,\gamma_4
\end{equation}

Line defect generators in $(A_1,A_4)$ Argyres-Douglas theory and their generating functions were computed in \cite{Cordova:2016uwk}. For completeness we reproduce their results here.
Starting from the initial seed, we apply all possible left
mutations to generate other seeds. There are in total $42$ seeds.
Correspondingly there are $42$ dual cones. Each dual cone is bounded
by four half-hyperplanes. Moreover, every three out of the four
half-hyperplanes intersect at a half line. In total there are four
such half-lines for each dual cone and they form edges of the dual
cone. Each edge corresponds to the core charge of one line defect generator.
For example, the dual cone for the initial seed is given by:
\begin{equation}
\check{\mathcal{C}}_{\{\gamma_1,\gamma_2,\gamma_3,\gamma_4\}}=\bigg\{\Sum_{i=1}^4a_i\gamma_i\mid a_2\leq 0,a_1+a_3\geq 0,a_2+a_4\leq 0,a_3\geq 0\bigg\}.
\end{equation}
Then we get four line defect generators whose core charges are given by
\begin{equation}
\gamma_1,-\gamma_1+\gamma_3,-\gamma_2+\gamma_4,-\gamma_4.
\end{equation}
Repeating this procedure for all $42$ dual cones we get $14$ edges. Thus the line defects in $(A_1,A_4)$ Argyres-Douglas theory are generated by the identity operator and $14$ nontrivial generators. Recall that the $(2,7)$ minimal model has two non-vacuum modules; therefore we have an expected multiplicity of $7$.
In the class $\cS$ realization of the theory this would correspond
to the $\mathbb{Z}_7$ symmetry of the $7$-gon.

We assume that the line defect phase is smaller than the phases of all vanilla BPS particles,
and calculate the generating function using consecutive right mutations on the framed quiver. For example, the line defect generator with core charge $\gamma_c=\gamma_1-\gamma_3$ goes through the following mutation sequence:
\begin{eqnarray}
\begin{split}
&\{\gamma_1,\gamma_2,\gamma_3,\gamma_4,\gamma_c\}\xrightarrow{\mu^R_{\gamma_c}}\{\gamma_1,\gamma_2,\gamma_3,\gamma_4+\gamma_c,-\gamma_c\}\xrightarrow{\mu^R_{\gamma_4+\gamma_c}}\\
&\{\gamma_1,\gamma_2,\gamma_3+\gamma_4+\gamma_c,-\gamma_4-\gamma_c,\gamma_4\}\xrightarrow{\mu^R_{\gamma_3+\gamma_4+\gamma_c}}\{\gamma_1,\gamma_2,-\gamma_3-\gamma_4-\gamma_c,\gamma_3,\gamma_4\},
\end{split}
\end{eqnarray}
which implies that its generating function is
\begin{equation*}
F(L)=X_{\gamma_1-\gamma_3}+X_{\gamma_1-\gamma_3+\gamma_4}+X_{\gamma_1+\gamma_4}.
\end{equation*}
The generating functions for all 14 line defect generators are (as given also in \cite{Cordova:2016uwk}):
\begin{align*}
F(A_1)&=X_{-\gamma_2+\gamma_4},\\
F(A_2)&=X_{-\gamma_1+\gamma_3},\\
F(A_3)&=X_{\gamma_2-\gamma_4}+X_{\gamma_1+\gamma_2-\gamma_4},\\
F(A_4)&=X_{\gamma_1-\gamma_3-\gamma_4}+X_{\gamma_1-\gamma_3},\\
F(A_5)&=X_{-\gamma_1-\gamma_4}+X_{-\gamma_1+\gamma_2-\gamma_4}+X_{\gamma_2-\gamma_4},\\
F(A_6)&=X_{-\gamma_1-\gamma_2+\gamma_4}+X_{-\gamma_1+\gamma_4}+X_{-\gamma_1+\gamma_3+\gamma_4},\\
F(A_7)&=X_{\gamma_1-\gamma_3}+X_{\gamma_1-\gamma_3+\gamma_4}+X_{\gamma_1+\gamma_4},\\
F(B_1)&=X_{\gamma_1},\\
F(B_2)&=X_{-\gamma_4},\\
F(B_3)&=X_{-\gamma_1-\gamma_2}+X_{-\gamma_1},\\
F(B_4)&=X_{\gamma_4}+X_{\gamma_3+\gamma_4},\\
F(B_5)&=X_{-\gamma_1}+X_{-\gamma_1+\gamma_2}+X_{\gamma_2}+X_{-\gamma_1+\gamma_2+\gamma_3}+X_{\gamma_2+\gamma_3},\\
F(B_6)&=X_{-\gamma_2-\gamma_3}+X_{-\gamma_3}+X_{-\gamma_2-\gamma_3+\gamma_4}+X_{-\gamma_3+\gamma_4}+X_{\gamma_4},\\	F(B_7)&=X_{-\gamma_3-\gamma_4}+X_{\gamma_2-\gamma_3-\gamma_4}+X_{\gamma_1+\gamma_2-\gamma_3-\gamma_4}+X_{-\gamma_3}+X_{\gamma_2-\gamma_3}+X_{\gamma_1+\gamma_2-\gamma_3}\\
&\ \ +X_{\gamma_2}+X_{\gamma_1+\gamma_2}.
\end{align*}
The generating functions for $A_i$ ($B_i$) are related to each other by the action of powers of the
monodromy operator. The Schur index with line defect $A_i$ ($B_i$) inserted is computed using \cite{Cordova:2016uwk}
\begin{equation}
\cI_{A_i}(q)=(q)_\infty^4\text{Tr}[F(A_i)S(q)\overline{S}(q)],\quad \cI_{B_i}(q)=(q)_\infty^4\text{Tr}[F(B_i)S(q)\overline{S}(q)]
\end{equation}
where in this particular chamber $S(q)$ is given by
\begin{equation}
S(q)=E_q(X_{\gamma_1})E_q(X_{\gamma_3})E_q(X_{\gamma_2})E_q(X_{\gamma_4}).
\end{equation}
As described in \cite{Cordova:2016uwk}, the Schur index with one line defect inserted does not depend on $i \in \{1,\dots,7\}$:
\begin{eqnarray}
\begin{split}
\cI_{A}(q)&=q+q^4+q^5+q^6+2q^7+2q^8+3q^9+3q^{10}+\cdots,\\
\cI_{B}(q)&=-q^{\frac{1}{2}}-q^{\frac{5}{2}}-q^{\frac{7}{2}}-q^{\frac{9}{2}}-2q^{\frac{11}{2}}-3q^{\frac{13}{2}}-3q^{\frac{15}{2}}-4q^{\frac{17}{2}}-5q^{\frac{19}{2}}+\cdots.
\end{split}
\end{eqnarray}

The chiral algebra in this case is the $(2,7)$ Virasoro minimal model \cite{Cordova:2015nma,Beem:2013sza,Beem:2014zpa}. There are three primary fields: the vacuum $\Phi_{1,1}$, $\Phi_{1,2}$ with weight $-2/7$ and $\Phi_{1,3}$ with weight $-3/7$.
Line defect Schur indices admit the following expansions in terms of characters:
\begin{eqnarray}
\begin{split}
\cI_{A}(q)&=q^{-1}\big(\chi_{1,3}(q)-\chi_{1,2}(q)\big),\\
\cI_{B}(q)&=q^{-\frac{1}{2}}\big(\chi_{1,1}(q)-\chi_{1,2}(q)\big).
\end{split}
\end{eqnarray}
The map $f$ between the line defect algebra $\cL$ and the Verlinde algebra $\cV$ is then given by:
\begin{eqnarray}
\begin{split}
I&\xrightarrow{f} [\Phi_{1,1}], \\
A_{i}&\xrightarrow{f} [A]=[\Phi_{1,3}]-[\Phi_{1,2}],\\
B_{i}& \xrightarrow{f} [B]=[\Phi_{1,1}]-[\Phi_{1,2}].
\end{split}
\end{eqnarray}
The non-trivial fusion rules in the $(2,7)$ Virasoro minimal model are:
\begin{eqnarray}
\begin{split}
[\Phi_{1,2}]\times[\Phi_{1,2}]&=[\Phi_{1,1}]+[\Phi_{1,3}],\\
[\Phi_{1,3}]\times[\Phi_{1,3}]&=[\Phi_{1,1}]+[\Phi_{1,2}]+[\Phi_{1,3}],\\
[\Phi_{1,2}]\times[\Phi_{1,3}]&=[\Phi_{1,2}]+[\Phi_{1,3}].
\end{split}
\end{eqnarray}
As first checked in \cite{Cordova:2016uwk},
\begin{eqnarray}
\begin{split}
[AA]&=[A]\times[A],\\
[BB]&=[B]\times[B],\\
[AB]&=[A]\times[B],
\end{split}
\end{eqnarray}
which gives evidence $f$ is indeed a homomorphism $\cL \to \cV$.

\begin{figure}
	\centering
	\includegraphics[scale=0.15]{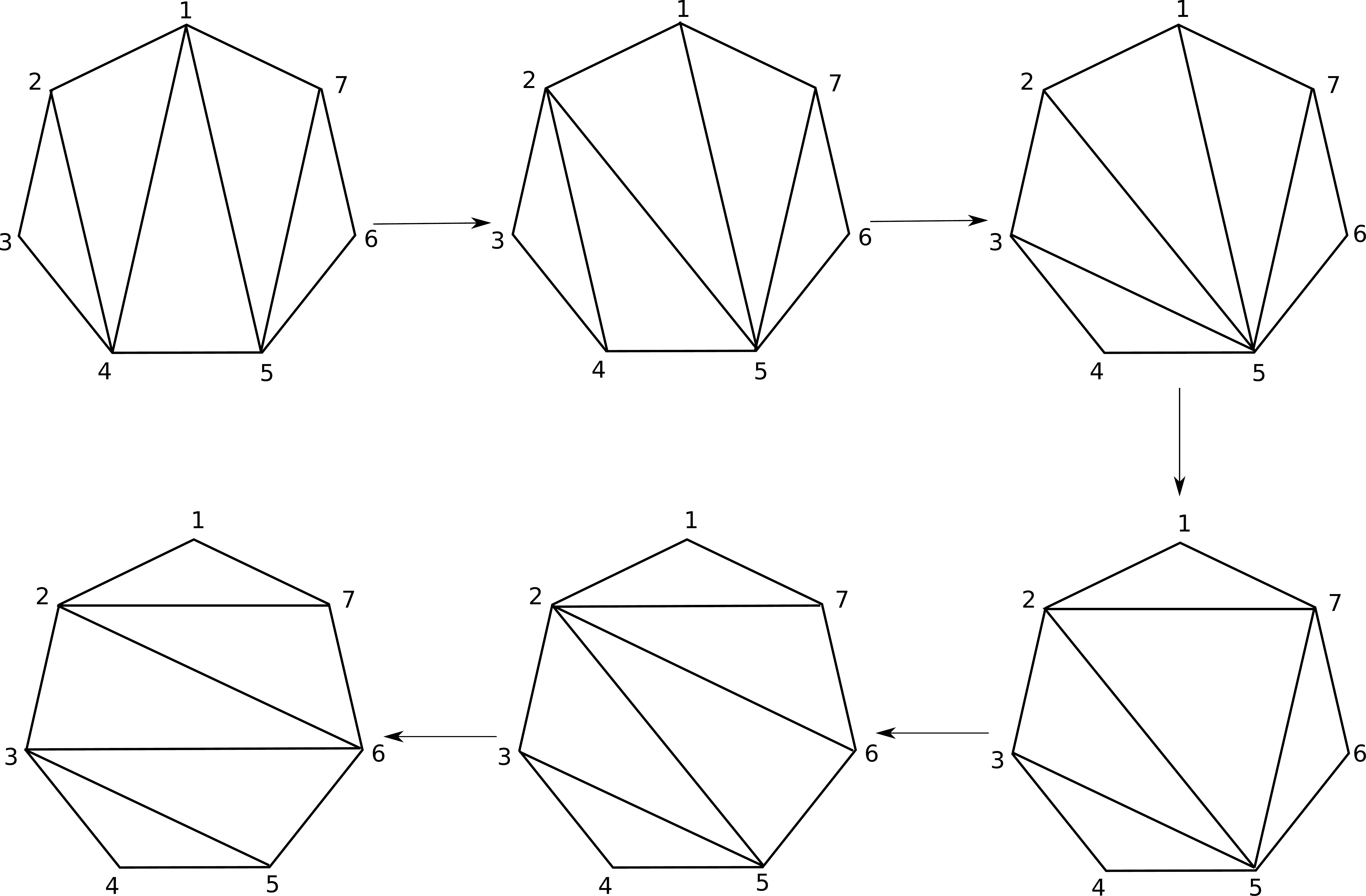}
	\caption{The classical monodromy action in the $(A_1,A_4)$ theory is realized by
	a sequence of flips of triangulations of the $7$-gon. The initial triangulation
	differs from the final one by a clockwise rotation by $2$ units.}
	\label{7gon}
\end{figure}

Now we turn to study the fixed points under the classical monodromy action $M$. By doing a series of flips (see Figure \ref{7gon}, the initial zigzag triangulation corresponds to the BPS quiver in Figure \ref{A4quiver} using the dictionary in \cite{Gaiotto:2009hg}. The monodromy action is given as follows:
\begin{eqnarray}
\begin{split}
\cX_{\gamma_1}& \rightarrow\frac{1+\cX_{\gamma_2}+\cX_{\gamma_4}+\cX_{\gamma_2}\cX_{\gamma_4}+\cX_{\gamma_2}\cX_{\gamma_3}\cX_{\gamma_4}}{\cX_{\gamma_2}\cX_{\gamma_3}},\\
\cX_{\gamma_2}& \rightarrow\frac{\cX_{\gamma_1}\cX_{\gamma_2}\cX_{\gamma_3}}{(1+\cX_{\gamma_2}+\cX_{\gamma_2}\cX_{\gamma_3})[1+\cX_{\gamma_4}+\cX_{\gamma_2}(1+\cX_{\gamma_1})(1+\cX_{\gamma_4}+\cX_{\gamma_3}\cX_{\gamma_4})]},\\
\cX_{\gamma_3}&\rightarrow\frac{(1+\cX_{\gamma_2}+\cX_{\gamma_1}\cX_{\gamma_2})[1+\cX_{\gamma_4}+\cX_{\gamma_2}(1+\cX_{\gamma_4}+\cX_{\gamma_3}\cX_{\gamma_4})]}{\cX_{\gamma_1}\cX_{\gamma_2}\cX_{\gamma_3}\cX_{\gamma_4}},\\
\cX_{\gamma_4}&\rightarrow\frac{\cX_{\gamma_3}\cX_{\gamma_4}}{1+\cX_{\gamma_4}+\cX_{\gamma_2}(1+\cX_{\gamma_1})(1+\cX_{\gamma_4}+\cX_{\gamma_3}\cX_{\gamma_4})}.
\end{split}
\end{eqnarray}
There are exactly three fixed points, which we label I, II, III.
On the fixed points $\cX_\gamma$ evaluate to
\begin{eqnarray}
\begin{split}
\cX_{\gamma_4}&:(\alpha_1,\alpha_2,\alpha_3),\\
\cX_{\gamma_3}&:(4+\alpha_1-2\alpha_1^2,4+\alpha_2-2\alpha_2^2,4+\alpha_3-2\alpha_3^2 ),\\
\cX_{\gamma_2}&:(\alpha_1-\alpha_1^2,\alpha_2-\alpha_2^2,\alpha_3-\alpha_3^2),\\
\cX_{\gamma_1}&:(2+\alpha_1-\alpha_1^2,2+\alpha_2-\alpha_2^2,2+\alpha_3-\alpha_3^2 ),
\end{split}
\end{eqnarray}
where $\alpha_i$ are the three roots of the cubic equation
\begin{equation}
\alpha^3-\alpha^2-2\alpha+1=0.
\end{equation}
Concretely,
\begin{align*}
\alpha_1&=\frac{1}{3}\big(1-\frac{7}{a}(-1)^{1/3}+a(-1)^{2/3}\big),\quad
\alpha_2=\frac{1}{3}\big(1+\frac{7}{a}(-1)^{2/3}-a(-1)^{1/3}\big),\\
\alpha_3&= \frac{1}{3}\big(1+\frac{7}{a}+a\big), \quad\text{with}\quad
a=\bigg(\frac{7}{2}\bigg)^{\frac{1}{3}}\big(-1+\I 3\sqrt{3}\big)^{\frac{1}{3}}.
\end{align*}
Evaluating the $F({A_i})$ at the fixed points we find that the values are independent
of $i=1, \dots, 7$, and similarly for $F({B_i})$, as expected. Concretely, we get
\begin{eqnarray}\label{A4g}
	\begin{split}
		A_i&\xrightarrow{g}\bigg(\frac{1}{1-\alpha_1},\frac{1}{1-\alpha_2},\frac{1}{1-\alpha_3}\bigg),\\
		B_i&\xrightarrow{g}\bigg(\frac{1}{\alpha_1},\frac{1}{\alpha_2},\frac{1}{\alpha_3}\bigg).
	\end{split}
\end{eqnarray}
Finally we want to construct $h$.
We have the following Verlinde matrices for $[\Phi_{1,2}]$ and $[\Phi_{1,3}]$:
\begin{equation}
N_{\Phi_{1,2}}=\begin{pmatrix}
0 & 1 & 0\\
1 & 0 & 1\\
0 & 1 & 1
\end{pmatrix},\quad
N_{\Phi_{1,3}}=\begin{pmatrix}
0 & 0 & 1\\
0 & 1 & 1\\
1 & 1 & 1
\end{pmatrix}.
\end{equation}
As before, we obtain $h$ by simultaneously diagonalizing $N_{\Phi_{1,2}}$ and $N_{\Phi_{1,3}}$ using $S$-matrix and then comparing with the correspondence between $U(1)$ fixed points and primaries of $(2,7)$ Virasoro minimal model. The $S$-matrix for the (2,7) minimal models is \cite{D.Francesco}:
\begin{equation}
S=\frac{2}{\sqrt{7}}\begin{pmatrix}
\text{cos}\frac{3\pi}{14} & -\text{cos}\frac{\pi}{14} & \text{sin}\frac{\pi}{7}\\
-\text{cos}\frac{\pi}{14} & -\text{sin}\frac{\pi}{7} & \text{cos}\frac{3\pi}{14}\\
\text{sin}\frac{\pi}{7} & \text{cos}\frac{3\pi}{14} & \text{cos}\frac{\pi}{14}\\
\end{pmatrix}.
\end{equation}
$N_{\Phi_{1,2}}$ and $N_{\Phi_{1,3}}$ are simultaneously diagonalized by $S$:
\begin{equation}\label{A4eigen}
SN_{\Phi_{1,2}}S^{-1}=\begin{pmatrix}
\alpha_1 & 0 & 0\\
0 & \alpha_2 & 0\\
0 & 0 & \alpha_3
\end{pmatrix},\quad
SN_{\Phi_{1,3}}S^{-1}=\begin{pmatrix}
\beta_1 & 0 & 0\\
0 & \beta_2 & 0\\
0 & 0 & \beta_3
\end{pmatrix},
\end{equation}
where
\begin{align*}
\beta_1 & =\frac{1}{3}\big(2+\frac{7}{b}(-1)^{2/3}-b(-1)^{1/3}\big),\quad
\beta_2=\frac{1}{3}\big(2-\frac{7}{b}(-1)^{1/3}+b(-1)^{2/3}\big)\\
\beta_3 & = \frac{1}{3}\big(2+\frac{7}{b}+b\big), \quad\text{with}\quad
b=\bigg(\frac{7}{2}\bigg)^{\frac{1}{3}}\big(1+\I 3\sqrt{3}\big)^{\frac{1}{3}}.
\end{align*}

According to \cite{Fredrickson-Neitzke,Fredrickson:2017yka}, the corresponding wild Hitchin moduli space has exactly three $U(1)_r$-fixed points, each of which corresponds to a primary field in the $(2,7)$ minimal model:
\begin{center}
	\begin{tabular}{ |c|c|c|c| }
		\hline
		fixed point & weights of $M$ & $U(1)_r$ weights & primary field\\ \hline
		I & $\e^{2\pi \I(3/7)}, \e^{2\pi \I(4/7)}, \e^{2\pi \I(5/7)},\e^{2\pi \I(2/7)}$ & $\frac{3}{7}, \frac{4}{7},\frac{5}{7},\frac{2}{7}$ & $\Phi_{1,3}$ \\ \hline
		II & $\e^{2\pi \I(8/7)}, \e^{-2\pi \I(1/7)}, \e^{2\pi \I(10/7)},\e^{-2\pi \I(3/7)}$ & $\frac{8}{7}, -\frac{1}{7},\frac{10}{7},-\frac{3}{7}$ & $\Phi_{1,1}$ \\ \hline
		III & $\e^{2\pi \I(8/7)}, \e^{-2\pi \I(1/7)}, \e^{2\pi \I(5/7)},\e^{2\pi \I(2/7)}$ & $\frac{8}{7}, -\frac{1}{7},\frac{5}{7},\frac{2}{7}$ & $\Phi_{1,2}$ \\ \hline
	\end{tabular}
\end{center}
Using this table and (\ref{A4eigen}), the isomorphism $h$ between $\cV$ and $\cO(F)$ is:
\begin{eqnarray}
\begin{split}
&[\Phi_{1,1}]\xrightarrow{h}(1,1,1),\\
&[\Phi_{1,2}]\xrightarrow{h}(\alpha_3,\alpha_1,\alpha_2),\\
&[\Phi_{1,3}]\xrightarrow{h}(\beta_3,\beta_1,\beta_2).
\end{split}
\end{eqnarray}
The image of $A_i$ and $B_i$ under $h\circ f$ is then:
\begin{eqnarray}\label{4.16}
\begin{split}
& A_i\xrightarrow{h\circ f}(\beta_3-\alpha_3,\beta_1-\alpha_1,\beta_2-\alpha_2),\\
& B_i\xrightarrow{h\circ f}(1-\alpha_3,1-\alpha_1,1-\alpha_2).
\end{split}
\end{eqnarray}
Although it is not obvious, one can check that this indeed agrees with (\ref{A4g}), so the diagram commutes,
as desired.

\subsection{\texorpdfstring{$(A_1,A_6)$}{(A1,A6)} Argyres-Douglas theory} \label{sec:A1A6}

\begin{figure}
	\centering
	\includegraphics[scale=0.08]{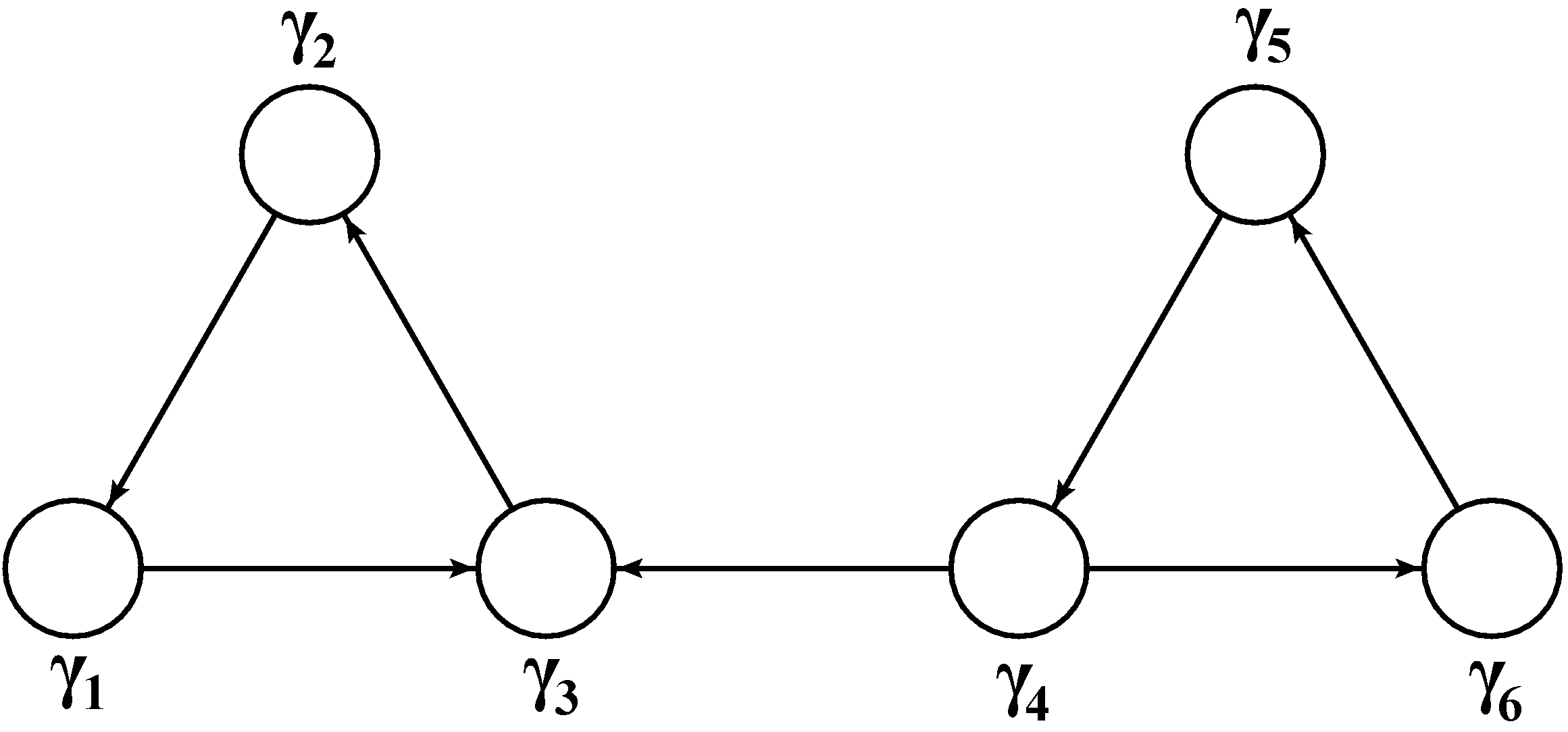}
	\caption{A BPS quiver for ($A_1,A_6$) Argyres-Douglas theory.}
	\label{A6quiver}
\end{figure}

Here we consider the $(A_1,A_6)$ Argyres-Douglas theory.
This theory has a new feature:
at one of the fixed points (fixed point I below), some of the cluster
coordinates $\cX_\gamma$ associated to the canonical chamber blow up. This being so, computing the fixed points of the classical monodromy in that chamber actually misses
one fixed point. Thus, with the benefit of hindsight,
we choose a different chamber, whose BPS quiver is shown in Figure \ref{A6quiver}.

There are eight BPS particles in this chamber, with the following charges (in increasing central charge phase order):
\begin{equation}
\gamma_4, \gamma_6, \gamma_4+\gamma_5, \gamma_5, \gamma_3, \gamma_1+\gamma_3, \gamma_2, \gamma_1.
\end{equation}

Quiver mutation starting from this chamber
generates in total 429 seeds. After mutating back to the original seed the 429 dual cones
span the whole charge lattice. Each dual cone is bounded by six half-hyperplanes. Every five of the six half-hyperplanes intersect at a half line which forms an edge of the dual cone and there are six edges for each dual cone. For example, the six edges of the dual cone for the initial seed $\check{C}_{\{\gamma_1,\gamma_2,\gamma_3,\gamma_4,\gamma_5,\gamma_6\}}$ are spanned by:
\begin{eqnarray}
\begin{split}
&\gamma_2+\gamma_4+\gamma_5+\gamma_6, -\gamma_1+\gamma_4+\gamma_5+\gamma_6,\gamma_4+\gamma_5+\gamma_6,\\
&-\gamma_1-\gamma_2-\gamma_3,-\gamma_1-\gamma_2-\gamma_3+\gamma_6,-\gamma_1-\gamma_2-\gamma_3-\gamma_5.
\end{split}
\end{eqnarray}
Repeating this for all $429$ dual cones we get in total $27$ edges. Correspondingly there are $27$ nontrivial line defect generators in the $(A_1,A_6)$ theory. The $(2,9)$ minimal model has three non-vacuum modules, so there is a multiplicity of $9$, corresponding to the $\mathbb{Z}_9$ symmetry of the $9$-gon.
Assuming that the line defect phase is smaller than central charge phases of all vanilla BPS particles, their generating functions are:
\begin{align*}
F(A_1) & = X_{\gamma_1+\gamma_2+\gamma_3-\gamma_6}+X_{\gamma_1+\gamma_2+\gamma_3+\gamma_5-\gamma_6}+X_{\gamma_1+\gamma_2+\gamma_3+\gamma_5},\\
F(A_2) & = X_{-\gamma_2-\gamma_4-\gamma_5-\gamma_6}+X_{\gamma_1-\gamma_2-\gamma_4-\gamma_5-\gamma_6}+X_{\gamma_1-\gamma_4-\gamma_5-\gamma_6},\\
F(A_3) & = X_{-\gamma_1-\gamma_2-\gamma_3-\gamma_5},\\
F(A_4) & = X_{\gamma_2+\gamma_4+\gamma_5+\gamma_6},\\
F(A_5) & = X_{-\gamma_1-\gamma_2-\gamma_3+\gamma_6},\\
F(A_6) & = X_{\gamma_1+\gamma_2+\gamma_3+\gamma_5}+X_{\gamma_1+\gamma_2+\gamma_3+\gamma_5+\gamma_6}+X_{\gamma_1+\gamma_2+\gamma_3+\gamma_4+\gamma_5+\gamma_6},\\
F(A_7) & = X_{\gamma_1-\gamma_4-\gamma_5-\gamma_6}+X_{\gamma_1+\gamma_2-\gamma_4-\gamma_5-\gamma_6}+X_{\gamma_1+\gamma_2+\gamma_3-\gamma_4-\gamma_5-\gamma_6}+X_{\gamma_1+\gamma_2+\gamma_3-\gamma_5-\gamma_6}\\
& \ \ +X_{\gamma_1+\gamma_2+\gamma_3-\gamma_6},\\
F(A_8) & = X_{-\gamma_1-\gamma_2-\gamma_3-\gamma_4-\gamma_5-\gamma_6}+X_{-\gamma_1-\gamma_2-\gamma_4-\gamma_5-\gamma_6}+X_{-\gamma_2-\gamma_4-\gamma_5-\gamma_6},\\
F(A_9) & = X_{-\gamma_1+\gamma_4+\gamma_5+\gamma_6},\\
F(B_1) & = X_{-\gamma_5-\gamma_6}+X_{-\gamma_6},\\
F(B_2) & = X_{\gamma_1}+X_{\gamma_1+\gamma_2},\\
F(B_3) & = X_{\gamma_5}+X_{\gamma_5+\gamma_6},\\
F(B_4) & = X_{-\gamma_1-\gamma_2}+X_{-\gamma_2},\\
F(B_5) & = X_{\gamma_6}+X_{\gamma_4+\gamma_6},\\
F(B_6) & = X_{-\gamma_1-\gamma_3}+X_{-\gamma_1},\\
F(B_7) & = X_{-\gamma_2-\gamma_3-\gamma_4-\gamma_5}+X_{-\gamma_3-\gamma_4-\gamma_5}+X_{-\gamma_4-\gamma_5}+X_{-\gamma_5},\\
F(B_8) & = X_{-\gamma_4-\gamma_6}+X_{-\gamma_4}+X_{\gamma_3-\gamma_4-\gamma_6}+X_{\gamma_3-\gamma_4}+X_{\gamma_3}+X_{\gamma_1+\gamma_3-\gamma_4-\gamma_6}+X_{\gamma_1+\gamma_3-\gamma_4}+X_{\gamma_1+\gamma_3},\\
F(B_9) & = X_{\gamma_2}+X_{\gamma_2+\gamma_3}+X_{\gamma_2+\gamma_3+\gamma_4}+X_{\gamma_2+\gamma_3+\gamma_4+\gamma_5},\\
F(C_1) & = X_{-\gamma_1-\gamma_2-\gamma_3},\\
F(C_2) & = X_{\gamma_4+\gamma_5+\gamma_6},\\
F(C_3) & = X_{\gamma_1+\gamma_2+\gamma_3}+X_{\gamma_1+\gamma_2+\gamma_3+\gamma_4}+X_{\gamma_1+\gamma_2+\gamma_3+\gamma_4+\gamma_5},\\
F(C_4) & = X_{-\gamma_2-\gamma_3-\gamma_4-\gamma_5-\gamma_6}+X_{-\gamma_3-\gamma_4-\gamma_5-\gamma_6}+X_{-\gamma_4-\gamma_5-\gamma_6},\\
F(C_5) & = X_{\gamma_1-\gamma_4-\gamma_6}+X_{\gamma_1-\gamma_4}+X_{\gamma_1+\gamma_2-\gamma_4-\gamma_6}+X_{\gamma_1+\gamma_2-\gamma_4}+X_{\gamma_1+\gamma_2+\gamma_3-\gamma_4-\gamma_6}\\
& \ \ +X_{\gamma_1+\gamma_2+\gamma_3-\gamma_4}+X_{\gamma_1+\gamma_2+\gamma_3},\\
F(C_6) & = X_{-\gamma_4-\gamma_5-\gamma_6}+X_{\gamma_3-\gamma_4-\gamma_5-\gamma_6}+X_{\gamma_3-\gamma_5-\gamma_6}+X_{\gamma_3-\gamma_6}+X_{\gamma_1+\gamma_3-\gamma_4-\gamma_5-\gamma_6}\\
& \ \ +X_{\gamma_1+\gamma_3-\gamma_5-\gamma_6}+X_{\gamma_1+\gamma_3-\gamma_6},\\
F(C_7) & = X_{-\gamma_1-\gamma_2-\gamma_3-\gamma_4-\gamma_5}+X_{-\gamma_1-\gamma_2-\gamma_4-\gamma_5}+X_{-\gamma_1-\gamma_2-\gamma_5}+X_{-\gamma_2-\gamma_4-\gamma_5}+X_{-\gamma_2-\gamma_5},\\
F(C_8) & = X_{\gamma_2+\gamma_5}+X_{\gamma_2+\gamma_5+\gamma_6}+X_{\gamma_2+\gamma_3+\gamma_5}+X_{\gamma_2+\gamma_3+\gamma_5+\gamma_6}+X_{\gamma_2+\gamma_3+\gamma_4+\gamma_5+\gamma_6},\\
F(C_9) & = X_{-\gamma_1-\gamma_3+\gamma_6}+X_{-\gamma_1+\gamma_6}+X_{-\gamma_1+\gamma_4+\gamma_6}.
\end{align*}

In this chosen chamber the spectrum generator $S(q)$ is given by
\begin{align*}
S(q)& =E_q(X_{\gamma_4})E_q(X_{\gamma_6})E_q(X_{\gamma_4+\gamma_5})E_q(X_{\gamma_5})E_q(X_{\gamma_3})E_q(X_{\gamma_1+\gamma_3})E_q(X_{\gamma_2})E_q(X_{\gamma_1})\\
 & =\Sum_{l_1,\cdots,l_8=0}
^\infty\frac{(-1)^{\sum_{i=1}^8l_i}q^{\frac{A}{2}}}{(q)_{l_1}\dots (q)_{l_8}}X_{(l_1+l_7)\gamma_1+l_2\gamma_2+(l_3+l_7)\gamma_3+(l_4+l_8)\gamma_4+(l_5+l_8)\gamma_5+l_6\gamma_6},
\end{align*}
where
\begin{equation}
A=\Sum_{i=1}^8l_i-l_1(l_7-l_2+l_3)+l_3(l_2+l_4+l_8-l_7)-l_4(l_8+l_5-l_6-l_7)+l_8(l_7-l_5)+l_5l_6.
\end{equation}
For sufficiently large enough $N$ the truncated $S_N(q)$ stabilizes to
\begin{align*}
S_N(q)& =1-\Sum_{i=1}^6X_{\gamma_i}q^{\frac{1}{2}}+(X_{2\gamma_1}+X_{2\gamma_2}+X_{2\gamma_3}+X_{\gamma_1+\gamma_2+\gamma_3}+X_{2\gamma_4}+X_{\gamma_1+\gamma_4}\\
& \ \ +X_{\gamma_2+\gamma_4}+X_{2\gamma_5}+X_{\gamma_1+\gamma_5}+X_{\gamma_2+\gamma_5}+X_{\gamma_3+\gamma_5}+X_{2\gamma_6}+X_{\gamma_1+\gamma_6}+X_{\gamma_2+\gamma_6}\\
& \ \ +X_{\gamma_3+\gamma_6}+X_{\gamma_4+\gamma_5+\gamma_6})q+\dots
\end{align*}
The Schur index with line defect $L$ ($L=A_i,B_i,C_i$) inserted is given by
\begin{equation}
\cI_L(q)=(q)_\infty^6\text{Tr}[F(L)S(q)\overline{S}(q)].
\end{equation}
In particular the line defect Schur index forgets the $i$ index as expected:
\begin{eqnarray}
\begin{split}
\cI_{A}(q)&=-q^{\frac{3}{2}}(1+q^3+q^4+q^5+2q^6+2q^7+3q^8+\cdots),\\
\cI_{B}(q)&=-q^{\frac{1}{2}}(1+q^2+q^3+2q^4+2q^5+3q^6+4q^7+6q^8+\cdots),\\
\cI_C(q)&= q(1+q^2+q^3+q^4+2q^5+3q^6+3q^7+5q^8+\cdots).
\end{split}
\end{eqnarray}

The chiral algebra in this case is conjectured to be the $(2,9)$ Virasoro minimal model \cite{Cordova:2015nma,Beem:2013sza,Beem:2014zpa}. There are four primary fields: $\Phi_{1,1}$ which is the vacuum, $\Phi_{1,2}$ with weight $-1/3$, $\Phi_{1,3}$ with weight $-5/9$, and $\Phi_{1,4}$ with weight $-2/3$. The line defect Schur indices have the following expansions in terms of the characters:
\begin{eqnarray}
\begin{split}
\cI_{A}(q)&=q^{-\frac{3}{2}}\big(\chi_{1,3}(q)-\chi_{1,4}(q)\big),\\
\cI_{B}(q)&=q^{-\frac{1}{2}}\big(\chi_{1,1}(q)-\chi_{1,2}(q)\big),\\
\cI_C(q)&=q^{-1}\big(-\chi_{1,2}(q)+\chi_{1,3}(q)\big).
\end{split}
\end{eqnarray}
Thus the map $f$ between the line defect OPE algebra $\cL$ and the Verlinde algebra $\cV$ of the $(2,9)$ minimal model is:
\begin{eqnarray}\label{f9gon}
\begin{split}
I&\xrightarrow{f}[\Phi_{1,1}],\\
A_i&\xrightarrow{f}[A]=[\Phi_{1,3}]-[\Phi_{1,4}],\\
B_i&\xrightarrow{f}[B]=[\Phi_{1,1}]-[\Phi_{1,2}],\\
C_i&\xrightarrow{f}[C]=-[\Phi_{1,2}]+[\Phi_{1,3}].
\end{split}
\end{eqnarray}
Non-trivial fusion rules in the $(2,9)$ minimal model are given by:
\begin{eqnarray}
\begin{split}
[\Phi_{1,2}]\times[\Phi_{1,2}]&=[\Phi_{1,1}]+[\Phi_{1,3}],\\
[\Phi_{1,2}]\times[\Phi_{1,3}]&=[\Phi_{1,2}]+[\Phi_{1,4}],\\
[\Phi_{1,2}]\times[\Phi_{1,4}]&=[\Phi_{1,3}]+[\Phi_{1,4}],\\
[\Phi_{1,3}]\times[\Phi_{1,3}]&=[\Phi_{1,1}]+[\Phi_{1,3}]+[\Phi_{1,4}],\\
[\Phi_{1,3}]\times[\Phi_{1,4}]&=[\Phi_{1,2}]+[\Phi_{1,3}]+[\Phi_{1,4}],\\
[\Phi_{1,4}]\times[\Phi_{1,4}]&=[\Phi_{1,1}]+[\Phi_{1,2}]+[\Phi_{1,3}]+[\Phi_{1,4}].
\end{split}
\end{eqnarray}
Using these fusion rules one can check that $[AA]=[A]\times[A]$, $[AB]=[A]\times[B]$, and $[BB]=[B]\times[B]$.
\begin{figure}
	\centering
	\includegraphics[scale=0.25]{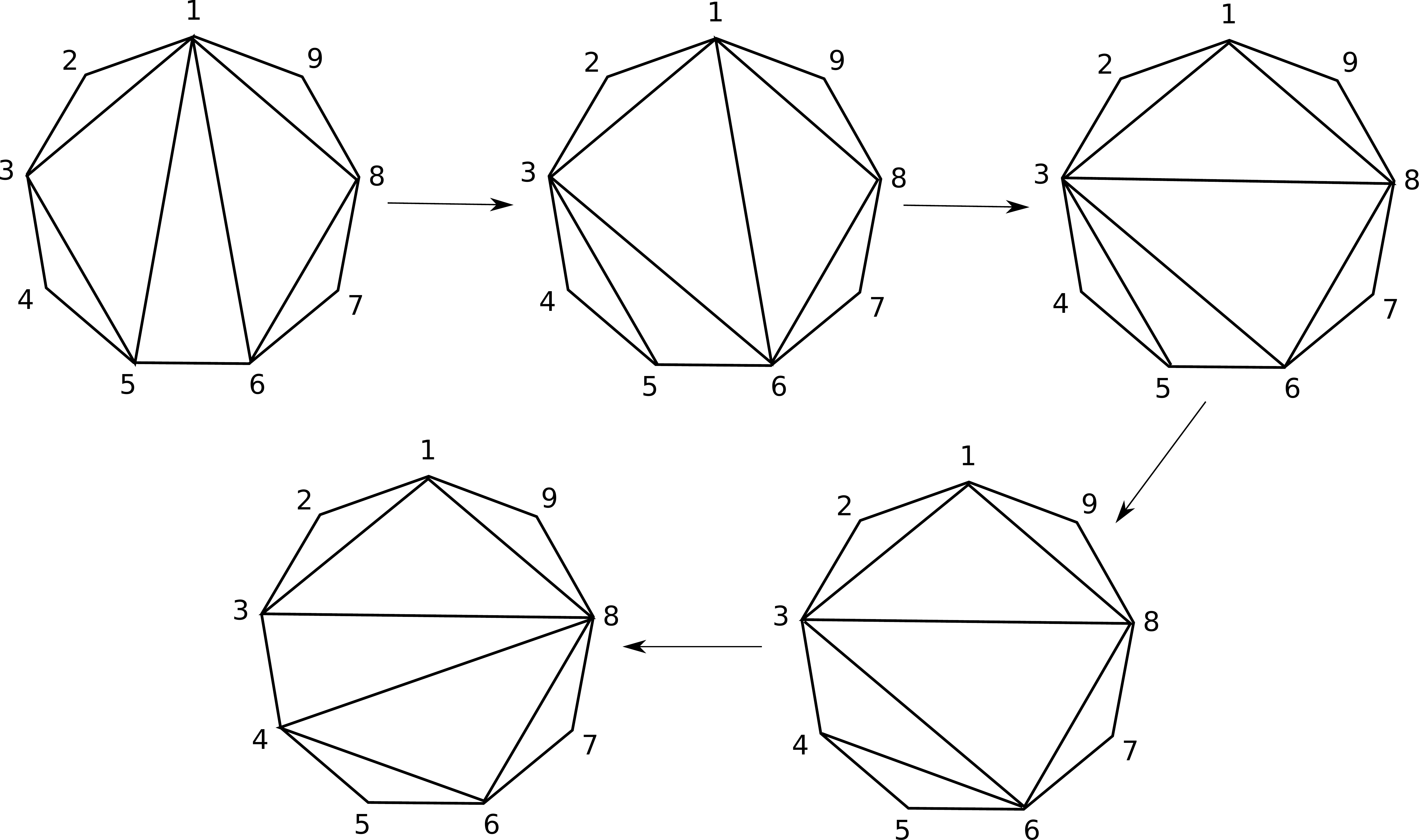}
	\caption{Monodromy action via a sequence of flips of triangulations of the $9$-gon.}
	\label{9gon}
\end{figure}

Now we study the fixed points under the classical monodromy action. By considering the sequence of flips shown in Figure \ref{9gon}
we compute that the classical monodromy is:
\begin{eqnarray}
\begin{split}
\cX_{\gamma_1}&\rightarrow \cX_{\gamma_2}(1+\cX_{\gamma_3}+\cX_{\gamma_3}\cX_{\gamma_4}),\quad\quad\quad\quad\quad
\cX_{\gamma_2}\rightarrow\frac{\cX_{\gamma_3}\cX_{\gamma_4}\cX_{\gamma_5}}{1+\cX_{\gamma_3}+\cX_{\gamma_3}\cX_{\gamma_4}},\\
\cX_{\gamma_3}&\rightarrow\frac{\cX_{\gamma_1}}{1+\cX_{\gamma_3}(1+\cX_{\gamma_4})(1+\cX_{\gamma_1})},\\
\cX_{\gamma_4}&\rightarrow\frac{(1+\cX_{\gamma_3}+\cX_{\gamma_3}\cX_{\gamma_4})(1+\cX_{\gamma_3}+\cX_{\gamma_3}\cX_{\gamma_1})}{\cX_{\gamma_3}\cX_{\gamma_4}\cX_{\gamma_1}},\\
\cX_{\gamma_5}&\rightarrow\frac{\cX_{\gamma_6}[1+\cX_{\gamma_3}(1+\cX_{\gamma_4})(1+\cX_{\gamma_1})]}{1+\cX_{\gamma_3}+\cX_{\gamma_3}\cX_{\gamma_1}},\quad
\cX_{\gamma_6}\rightarrow\frac{\cX_{\gamma_4}}{1+\cX_{\gamma_3}(1+\cX_{\gamma_4})(1+\cX_{\gamma_1})}.
\end{split}
\end{eqnarray}
There are exactly four fixed points which we label I, II, III, IV. At the fixed points $X_\gamma$ evaluate to:
\begin{align*}
\cX_{\gamma_1} & : (-1,\alpha_1,\alpha_2,\alpha_3),\quad
\cX_{\gamma_2}  : (-1,1-\alpha_2,1-\alpha_3,1-\alpha_1),\\
\cX_{\gamma_3} & : (-1,\alpha_2,\alpha_3,\alpha_1),\quad
\cX_{\gamma_4} : (-1,1-\alpha_3,1-\alpha_1,1-\alpha_2),\\
\cX_{\gamma_5} & : (-1,\alpha_1,\alpha_2,\alpha_3),\quad
\cX_{\gamma_6}: (-1,1-\alpha_2,1-\alpha_3,1-\alpha_1),
\end{align*}
where
\begin{equation*}
\alpha_1=(-1)^{\frac{4}{9}}-(-1)^{\frac{5}{9}},\quad
\alpha_2=(-1)^{\frac{8}{9}}-(-1)^{\frac{1}{9}},\quad
\alpha_3=(-1)^{\frac{2}{9}}-(-1)^{\frac{7}{9}}.
\end{equation*}
The line defect vevs evaluated at the fixed points satisfy:
\begin{equation}
F({A_i})=F({A_j}),\quad F({B_i})=F({B_j}),\quad F({C_i})=F({C_j}).
\end{equation}
Explicitly, the evaluation map is:
\begin{eqnarray}\label{g9gon}
\begin{split}
A_i&\xrightarrow{g}\big(1,-\alpha_3,-\alpha_1,-\alpha_2),\\
B_i&\xrightarrow{g}\big(0,1+\alpha_1,1+\alpha_2,1+\alpha_3\big),\\
C_i&\xrightarrow{g}\big(- 1,1-\alpha_3,1-\alpha_1,1-\alpha_2\big).
\end{split}
\end{eqnarray}
The fusion matrices for $[\Phi_{1,2}]$, $[\Phi_{1,3}]$ and $[\Phi_{1,4}]$ are:
\begin{equation}
N_{\Phi_{1,2}}=\begin{pmatrix}
0 & 1 & 0& 0\\
1 & 0 & 1 & 0\\
0 & 1 & 0 & 1\\
0 & 0 & 1 & 1
\end{pmatrix},\quad
N_{\Phi_{1,3}}=\begin{pmatrix}
0 & 0 & 1 & 0\\
0 & 1 & 0 & 1\\
1 & 0 & 1 & 1\\
0 & 1 & 1 & 1
\end{pmatrix},\quad
N_{\Phi_{1,4}}=\begin{pmatrix}
0 & 0 & 0 & 1\\
0 & 0 & 1 & 1\\
0 & 1 & 1 & 1\\
1 & 1 & 1 & 1
\end{pmatrix}.
\end{equation}
The $S$-matrix for (2,9) minimal model is given by \cite{D.Francesco}:
\begin{equation}
S=\frac{2}{3}\begin{pmatrix}
-\text{sin}\frac{2\pi}{9} & \text{cos}\frac{\pi}{18} & -\text{sin}\frac{\pi}{3} & \text{sin}\frac{\pi}{9}\\
\text{cos}\frac{\pi}{18} & -\text{sin}\frac{\pi}{9} & -\text{sin}\frac{\pi}{3} & \text{sin}\frac{2\pi}{9}\\
-\text{sin}\frac{\pi}{3} & -\text{sin}\frac{\pi}{3} & 0 & \text{sin}\frac{\pi}{3}\\
\text{sin}\frac{\pi}{9} & \text{sin}\frac{2\pi}{9} & \text{sin}\frac{\pi}{3} & \text{cos}\frac{\pi}{18}
\end{pmatrix}.
\end{equation}
The fusion matrices are simultaneously diagonalized by $S$:
\begin{eqnarray}\label{A6eigen}
\begin{split}
SN_{\Phi_{1,2}}S^{-1}&=\begin{pmatrix}
-\alpha_3 & 0 & 0 & 0\\
0 & -\alpha_1 & 0 & 0\\
0 & 0 & 1 & 0\\
0 & 0 & 0 & -\alpha_2
\end{pmatrix},\quad
SN_{\Phi_{1,3}}S^{-1}=\begin{pmatrix}
1+\alpha_1 & 0 & 0 & 0\\
0 & 1+\alpha_2 & 0 & 0\\
0 & 0 & 0 & 0\\
0 & 0 & 0 & 1+\alpha_3
\end{pmatrix},\\
SN_{\Phi_{1,4}}S^{-1}&=\begin{pmatrix}
1-\alpha_3 & 0 & 0 & 0\\
0 & 1-\alpha_1 & 0 & 0\\
0 & 0 & -1 & 0\\
0 & 0 & 0 & 1-\alpha_2
\end{pmatrix}.
\end{split}
\end{eqnarray}

According to \cite{Fredrickson-Neitzke,Fredrickson:2017yka}, the correspondence between $U(1)_r$-fixed points in $\cN$ and the
primaries of the $(2,9)$ Virasoro minimal model is:
\begin{center}
	\begin{tabular}{ |c|c|c| }
		\hline
		fixed point & $U(1)$ weights & primary field\\ \hline \hline
		I  & $\frac{4}{9},\frac{5}{9},\frac{7}{9},\frac{2}{9},\frac{10}{9},-\frac{1}{9}$ & $\Phi_{1,3}$ \\ \hline
		II & $\frac{7}{9},\frac{2}{9},\frac{10}{9}, -\frac{1}{9},\frac{4}{3},-\frac{1}{3}$ & $\Phi_{1,2}$ \\ \hline
		III & $\frac{1}{3}, \frac{2}{3},\frac{4}{9},\frac{5}{9},\frac{7}{9},\frac{2}{9}$  & $\Phi_{1,4}$ \\ \hline
		IV  & $\frac{4}{3}, -\frac{1}{3},\frac{10}{9},-\frac{1}{9},\frac{14}{9},-\frac{5}{9}$  & $\Phi_{1,1}$ \\ \hline
	\end{tabular}
\end{center}
Based on this table and (\ref{A6eigen}), the isomorphism $h: \cV \to \cO(F)$ is:
\begin{eqnarray}\label{h9gon}
\begin{split}
[\Phi_{1,1}]&\xrightarrow{h}\big(1,1,1,1\big),\\
[\Phi_{1,2}]&\xrightarrow{h}\big(1,-\alpha_1,-\alpha_2,-\alpha_3\big),\\
[\Phi_{1,3}]&\xrightarrow{h}\big(0,1+\alpha_2,1+\alpha_3,1+\alpha_1\big),\\
[\Phi_{1,4}]&\xrightarrow{h}\big(-1,1-\alpha_1,1-\alpha_2,1-\alpha_3\big).
\end{split}
\end{eqnarray}
Combining (\ref{f9gon}), (\ref{g9gon}) and (\ref{h9gon}) confirms that $h\circ f=g$ in the $(A_1,A_6)$ Argyres-Douglas theory.

\section{\texorpdfstring{$(A_1,D_{2N+1})$}{(A1,D(2N+1))} Argyres-Douglas theories}\label{sec:examples-odd}

In this section we present the results of explicit computations verifying the commutativity
\eqref{eq:commutativity} in the Argyres-Douglas theories of type $(A_1,D_3)$ and
$(A_1,D_5)$, with the appropriate modifications to take care of the flavor symmetry
in these theories.

\subsection{\texorpdfstring{$(A_1,D_3)$}{(A1,D3)} Argyres-Douglas theory} \label{sec:A1D3}

\begin{figure}
	\centering
	\includegraphics[scale=0.12]{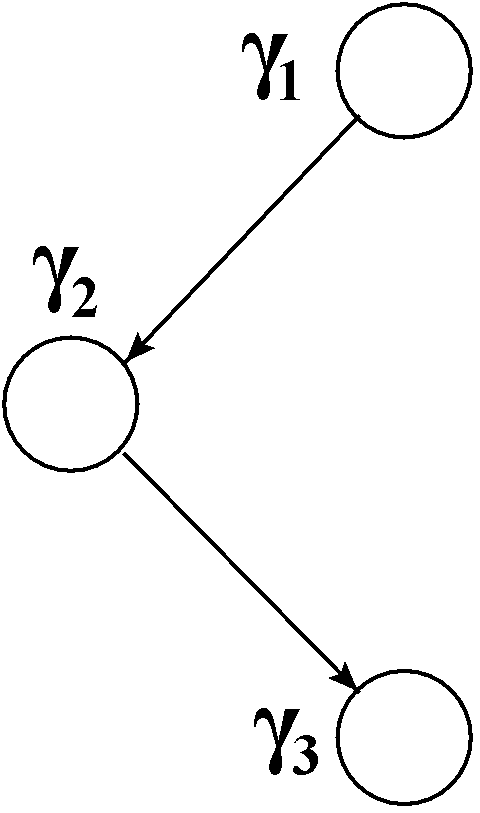}
	\caption{A BPS quiver for the ($A_1,D_3$) Argyres-Douglas theory.}
	\label{D3quiver}
\end{figure}
We consider $(A_1,D_3)$ Argyres-Douglas theory. This is equivalently the $(A_1,A_3)$ Argyres-Douglas theory. Line defect generators and their generating functions in this description were studied in \cite{Cordova:2016uwk,Gaiotto:2010be}. Line defect Schur indices and the relation to the Verlinde algebra were studied in \cite{Cordova:2016uwk}. Here we use the $(A_1,D_3)$ description instead.

We choose a chamber where the BPS quiver is as in Figure \ref{D3quiver}, containing BPS particles with charges (in increasing phase order):
\begin{equation*}
\gamma_1,\gamma_2,\gamma_3.
\end{equation*}
Note that $\gamma_1+\gamma_3$ has zero Dirac pairing with any charge,
and thus is a pure flavor charge.

\begin{figure}
	\begin{subfigure}{.45\textwidth}
	\centering
	\includegraphics[width=.8\linewidth]{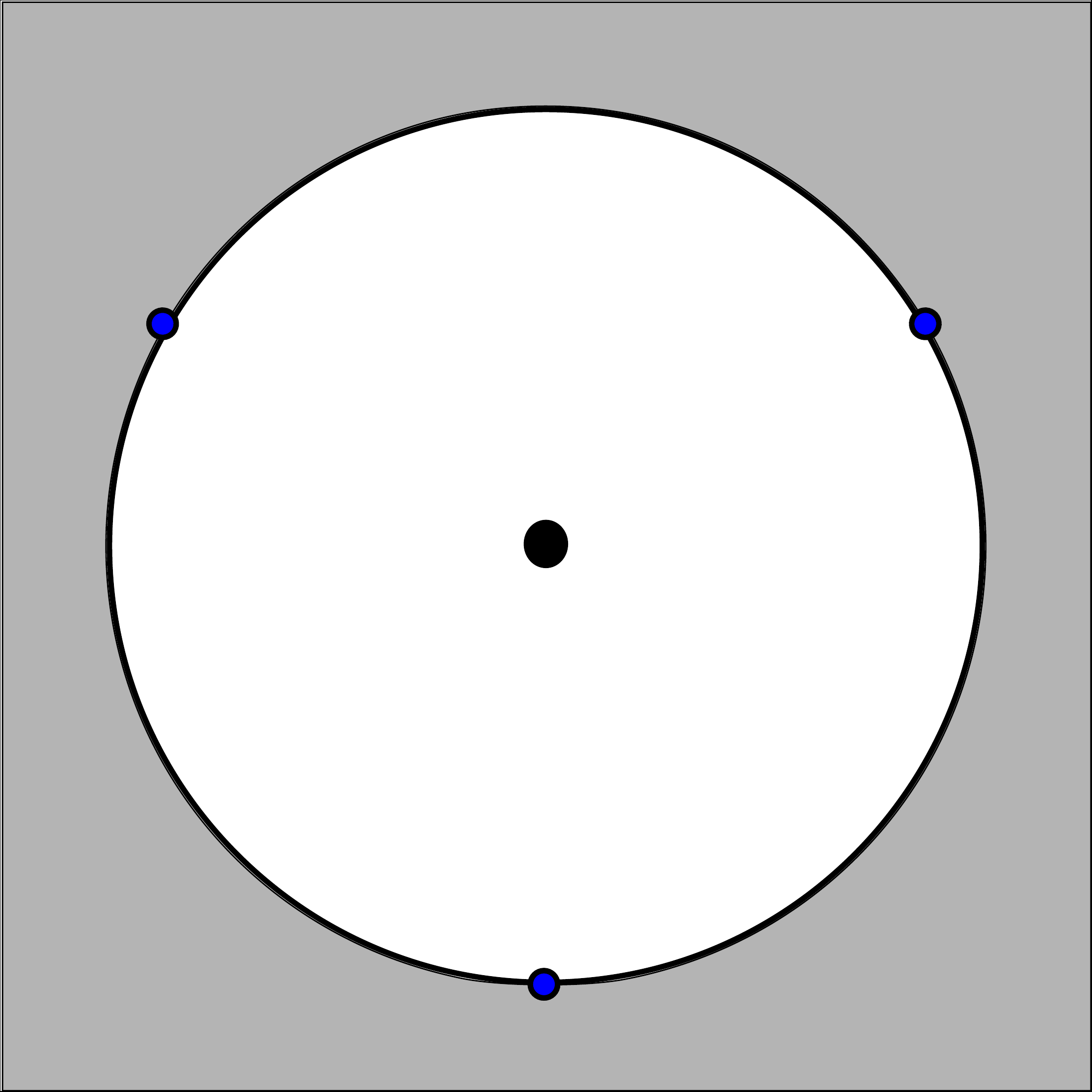}
	\caption{}
	\label{D3FIG1}
     \end{subfigure}
     \begin{subfigure}{.45\textwidth}
	\centering
	\includegraphics[width=.8\linewidth]{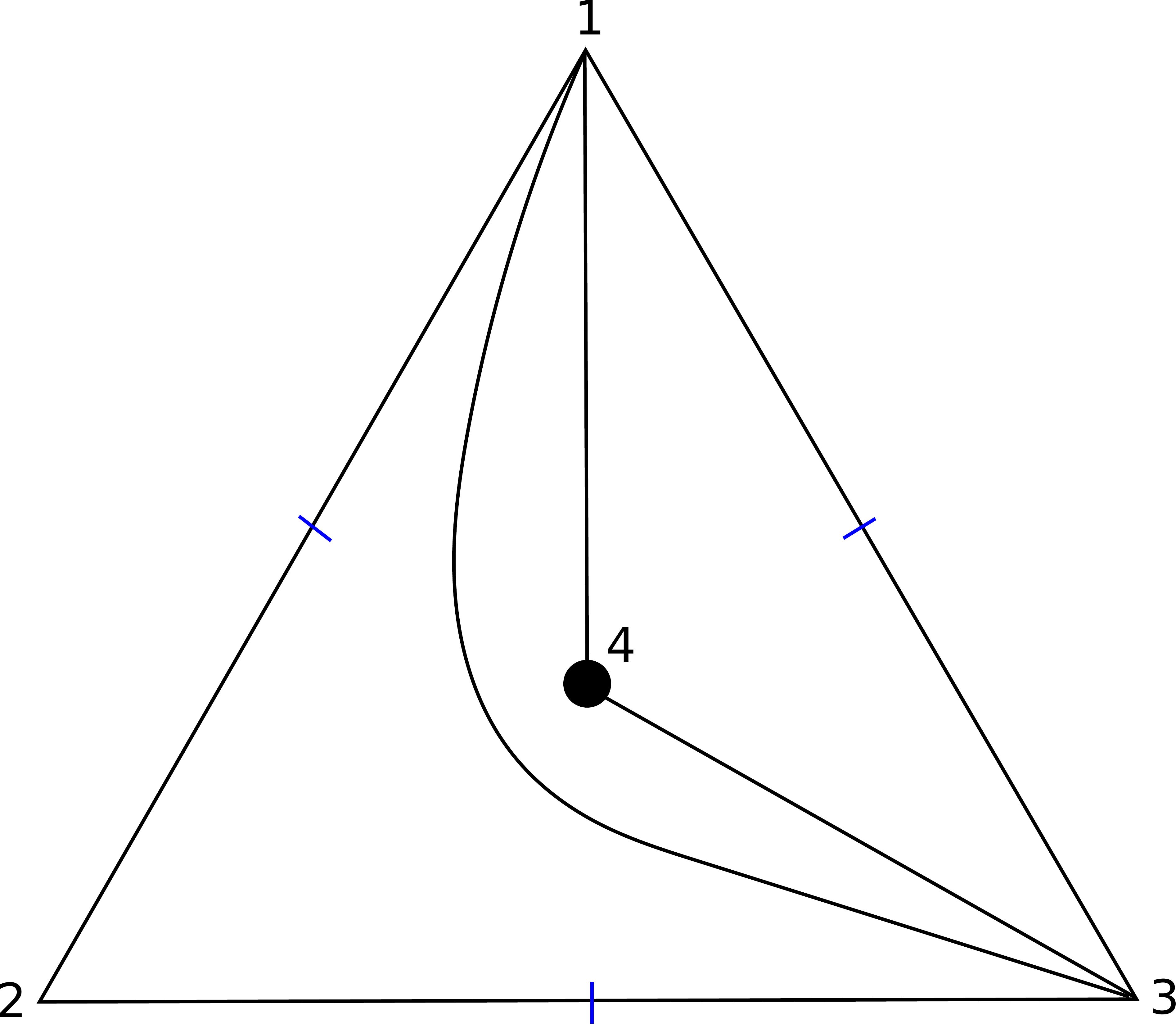}
	\caption{}
	\label{D3WKB}
\end{subfigure}
\caption{(a): $\mathbb{CP}^1\setminus D_\infty$ where $D_\infty$ is a disk around $z=\infty$ bounded by $S^1$ with three marked points colored in blue. The regular singularity at $z=0$ is colored in black.
(b): A triangulation in the $(A_1,D_3)$ Argyres-Douglas theory. There are three boundary edges. The blue marks correspond to the positions of three Stokes rays. }
\end{figure}

The corresponding Hitchin system is defined on $\mathbb{CP}^1$,
with one irregular singularity at $z=\infty$ and one regular singularity at $z=0$.
There are three Stokes rays emerging from the irregular singularity.
Correspondingly there are three marked points on the $S^1$ bounding the
cut-out disc around $z=\infty$, as in Figure \ref{D3FIG1}.
The WKB triangulation for the chosen chamber is shown in Figure
\ref{D3WKB}. Here $\cX_{\gamma_1}$ corresponds to edge 14,
$\cX_{\gamma_2}$ corresponds to edge 13, and $\cX_{\gamma_3}$ corresponds
to edge 34.

\begin{figure}
	\begin{subfigure}{.45\textwidth}
		\centering
		\includegraphics[width=.8\linewidth]{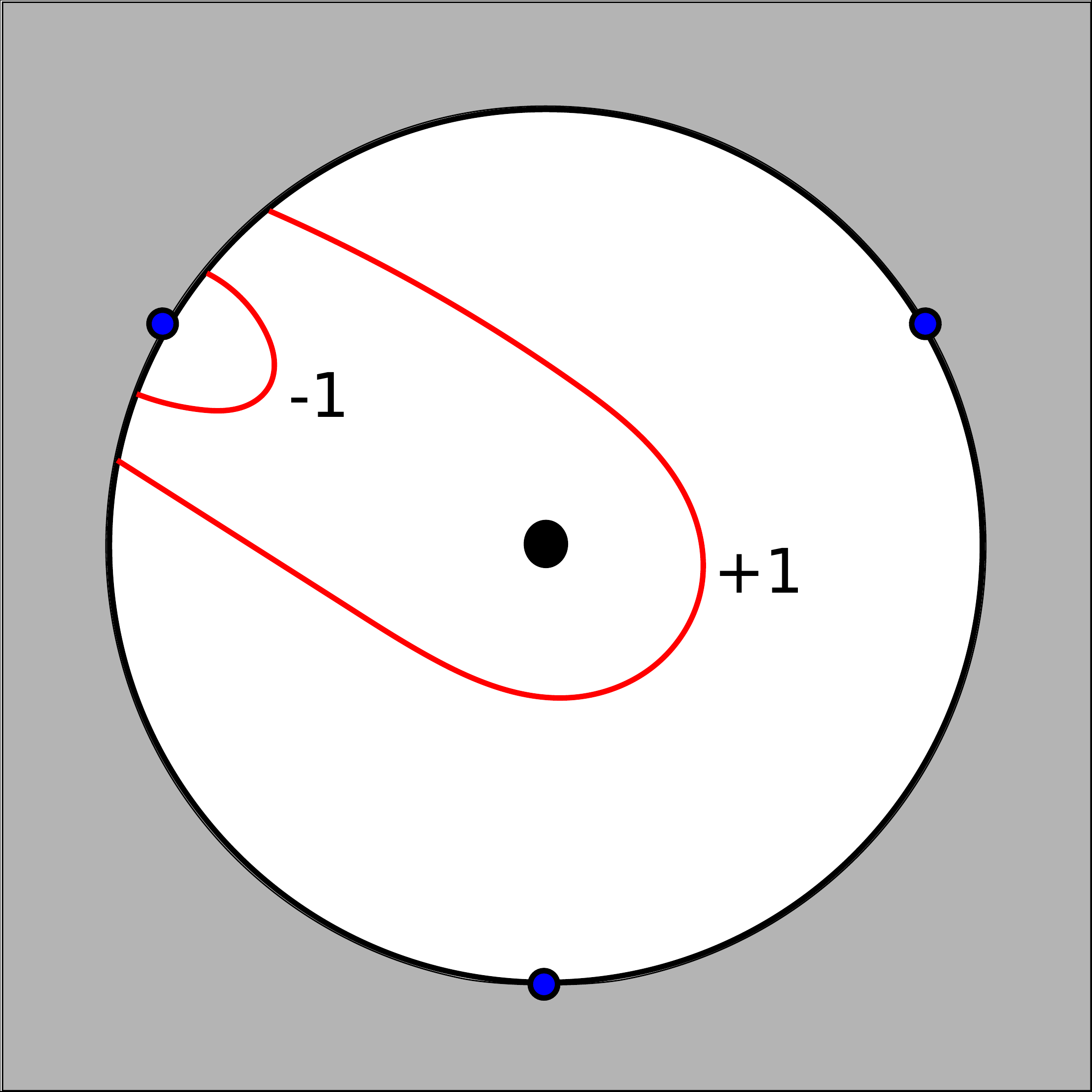}
		\caption{Type $A$}
		\label{D3lamA}
	\end{subfigure}
	\begin{subfigure}{.45\textwidth}
		\centering
		\includegraphics[width=.8\linewidth]{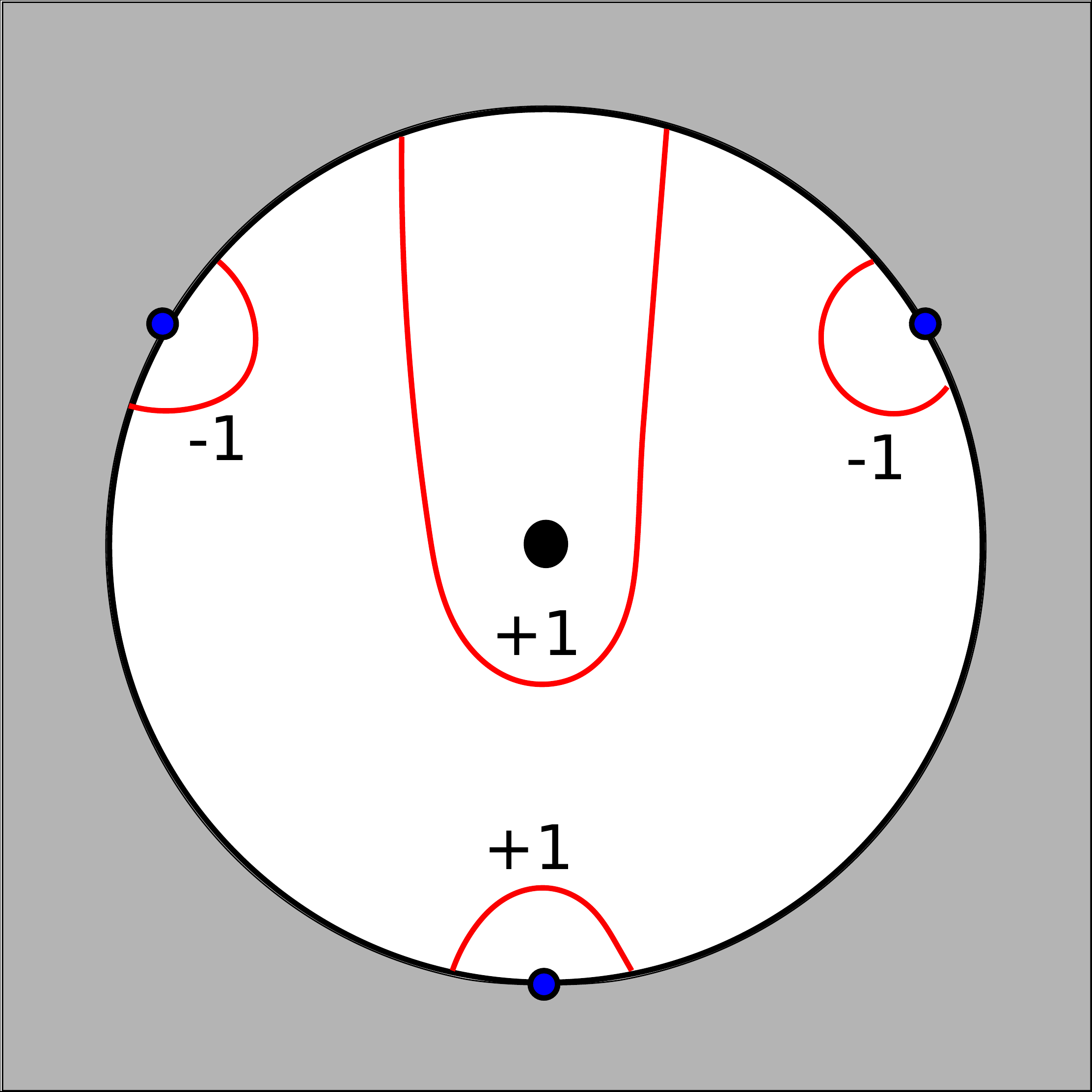}
		\caption{Type $B$}
		\label{D3lamB}
	\end{subfigure}\\
	\begin{center}
	\begin{subfigure}{.45\textwidth}
		\centering
		\includegraphics[width=.8\linewidth]{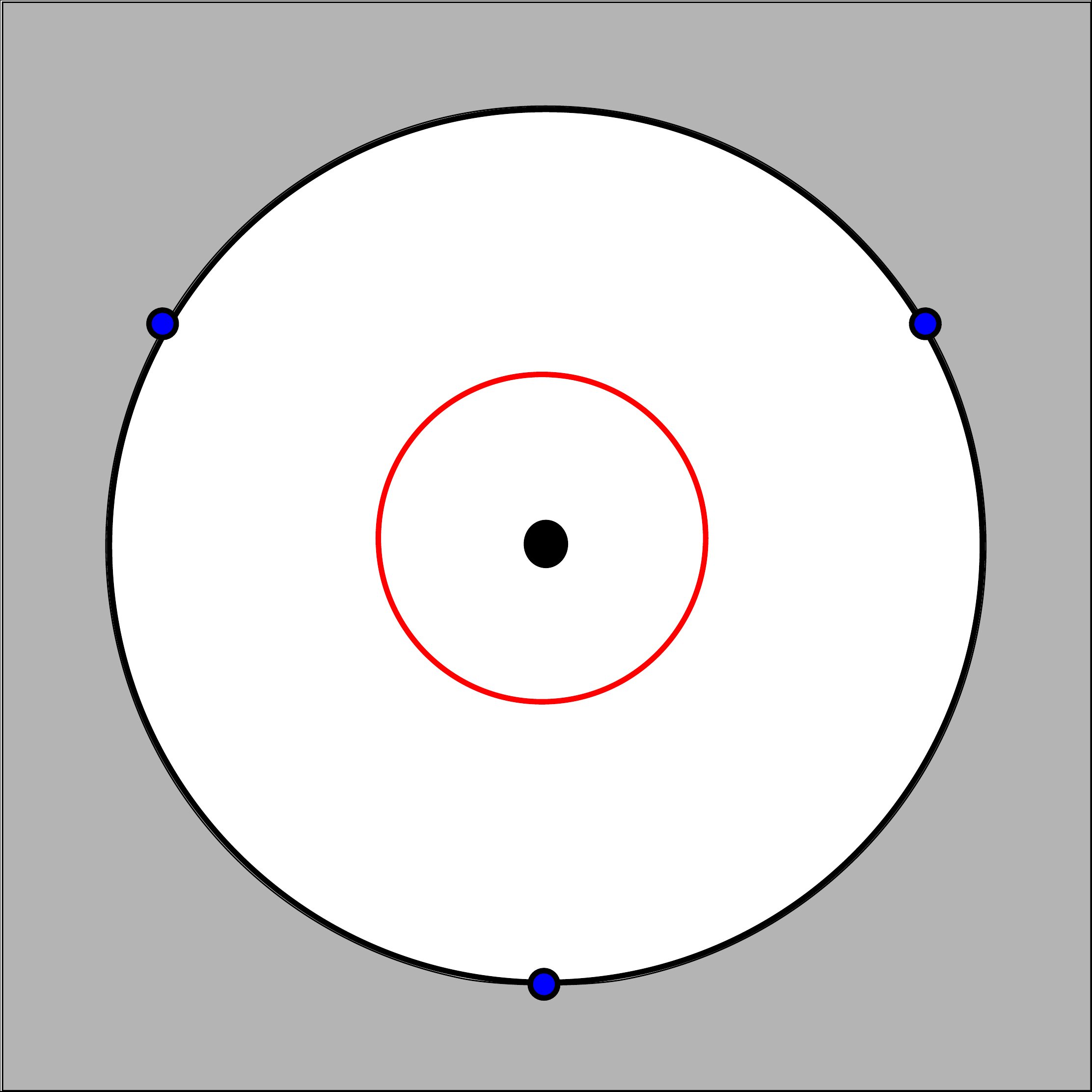}
		\caption{Type $C$}
		\label{D3lamC}
	\end{subfigure}
	\end{center}
	\caption{Three types of laminations in $(A_1,D_3)$ Argyres-Douglas theory.}
    \label{D3lam}
\end{figure}

Now we use the method reviewed in \S\ref{sec:defects-class-S} to describe a generating
set of line defects.
There are seven generators, including a pure flavor line defect $C$
whose corresponding lamination is a loop around the regular singularity.
The other six generators come in two types, $A$ and $B$, corresponding to two
different kinds of laminations: see Figure \ref{D3lam}.
We denote the six generators as $A_i, B_i \  (i=1, 2, 3)$,
where $A_1$ and $B_1$ correspond to the laminations shown in Figure \ref{D3lam}.
The lamination for $A_{i+1}$ ($B_{i+1}$) is given by rotating the lamination
for $A_i$ ($B_i$) counterclockwise by $2\pi/3$.
The flavor charge is normalized to be $(\gamma_1+\gamma_3)/2$,
and the corresponding $X_\gamma$ is equal to the $SU(2)$ flavor fugacity $z$:
\begin{equation}
z = X_{\frac{\gamma_1+\gamma_3}{2}}.
\end{equation}
Moreover we define
\begin{equation}
X_{\gamma'}:=X_{\frac{\gamma_1-\gamma_3}{2}}.
\end{equation}
We computed generating functions of line defect generators using the method reviewed in \S\ref{sec:defects-class-S}. They are listed below (these differ slightly from the analogous formulas in \cite{Cordova:2016uwk} because we are computing
in a different chamber):
\begin{align*}
F({A_1}) & = z^{-1}X_{-{\gamma_2}}+X_{-\gamma'}+X_{-\gamma'-\gamma_2},\\
F({A_2}) & = X_{-\gamma'}+X_{-\gamma'+\gamma_2}+z X_{\gamma_2},\\
F({A_3}) & = X_{\gamma'},\\
F({B_1}) & = X_{-\gamma_2}+z^{-1}X_{-\gamma_2+\gamma'},\\
F({B_2}) & = X_{-2\gamma'+\gamma_2}+X_{-2\gamma'-\gamma_2}+z X_{-\gamma'+\gamma_2}+(q^{\frac{1}{2}}+q^{-\frac{1}{2}})X_{-2\gamma'}+(z+z^{-1})X_{-\gamma'}+z^{-1}X_{-\gamma'-\gamma_2},\\
F({B_3}) & = X_{\gamma_2}+z X_{\gamma_2+\gamma'},\\
F(C) & =z+z^{-1}.
\end{align*}
The pure flavor line defect $C$ is a Wilson line in the fundamental representation of the $SU(2)$ flavor symmetry.

The Schur index with one line defect $L$ inserted is computed as
\begin{equation}
\cI_{L}(q,z)=(q)_\infty^2\text{Tr}[F(L)S(q)\overline{S}(q)], \quad \text{with}\quad
S(q)=E_q(X_{\gamma_1})E_q(X_{\gamma_2})E_q(X_{\gamma_3}).
\end{equation}
As usual the Schur indices with defects $A_i$ and $B_i$ inserted do not depend on the index $i$; concretely (these do match
\cite{Cordova:2016uwk}, as they should since they are chamber-independent):
\begin{align*}
\cI_{A}(q,z) & = -q^{\frac{1}{2}}(\chi_{\textbf{2}}+\chi_{\textbf{4}} q+\chi_{\bf{2\oplus 4\oplus 6}}q^2+\chi_{\bf{2}^{\oplus \text{2}}\oplus 4^{\oplus \text{2}}\oplus \bf{6}\oplus \bf{8}}q^3+\chi_{\bf{2^{\oplus \text{3}}\oplus 4^{\oplus \text{3}}\oplus 6^{\oplus \text{3}}\oplus 8\oplus 10}}q^4\\
& \ \ +\chi_{\bf{2^{\oplus \text{4}}\oplus 4^{\oplus \text{6}}\oplus 6^{\oplus \text{4}}\oplus 8^{\oplus \text{3}}\oplus 10\oplus 12}}q^5+\cdots),\\
\cI_{B}(q,z) & = -q^{\frac{1}{2}}(1+\chi_{\bf{3}} q^2+\chi_{\bf{1\oplus 3}}q^3+\chi_{\bf{1\oplus 3\oplus 5}}q^4+\chi_{\bf{1\oplus 3^{\oplus \text{2}}\oplus 5}}q^5+\chi_{\bf{1^{\oplus \text{2}}\oplus 3^{\oplus \text{3}}\oplus 5^{\oplus \text{2}}\oplus 7}}q^6+\cdots),
\end{align*}
where framed BPS states organize themselves into representations of $SU(2)$\footnote{We label irreducible $SU(2)$ representations by their dimensions.}.

The associated chiral algebra is $\widehat{\mathfrak{sl}(2)}_{-\frac{4}{3}}$\cite{Beem:2013sza,Beem:2014zpa,Cordova:2015nma,Buican:2015ina,Xie:2016evu}. There are three admissible representations \cite{D.Francesco,V.G.Kac} with highest weights:
\begin{equation}
\Phi_0=\left[-\frac{4}{3},0\right],\quad \Phi_1=\left[-\frac{2}{3},-\frac{2}{3}\right], \quad \Phi_2=\left[0,-\frac{4}{3}\right]
\end{equation}
where $\Phi_0$ is the highest weight for the vacuum module. Their characters were computed using the Kazhdan-Lusztig formula in \cite{D.Francesco,Cordova:2016uwk}. In particular the line defect Schur indices could be written as:
\begin{eqnarray}
\begin{split}
\cI_{A}(q,z)&=q^{-\frac{1}{2}}z^{-1}\big(-\chi_1(q,z)+\chi_2(q,z)\big),\\
\cI_{B}(q,z)&=q^{-\frac{1}{2}}\big(\chi_0(q,z)-\chi_1(q,z)+z^{-2}\chi_2(q,z)\big).
\end{split}
\end{eqnarray}
The expansions of $\cI_{A_iA_j}$, $\cI_{B_iB_j}$ and $\cI_{A_iB_j}$ in terms of characters are:
\begin{align*}
\cI_{A_iA_i}(q,z)=\cI_{A_iA_{i+1}}(q,z)&= (1+q^{-1})\chi_0(q,z)-q^{-1}\chi_1(q,z)+q^{-1}z^{-2}\chi_2(q,z),\\
\cI_{A_iA_{i-1}}(q,z) & = 2\chi_0(q,z)-\chi_1(q,z)+z^{-2}\chi_2(q,z),\\
\cI_{B_iB_i}(q,z)=\cI_{B_iB_{i+1}}(q,z) & = (1+q^{-1}+q^{-2})\chi_0(q,z)-[q^{-1}(1+z^{-2})+q^{-2}]\chi_1(q,z)\\
&\ \ +[q^{-1}(1+z^{-2})+q^{-2}z^{-2}]\chi_2(q,z),\\
\cI_{B_iB_{i-1}}(q,z) & = (2+q)\chi_0(q,z)-(2+z^{-2})\chi_1(q,z)+(1+2z^{-2})\chi_2(q,z),\\
\cI_{A_iB_i}(q,z) & = q^{-1}(z+z^{-1})\chi_0(q,z)-(q^{-1}+q^{-2})z^{-1}\big(\chi_1(q,z)-\chi_2(q,z)\big),\\
\cI_{A_iB_{i+1}}(q,z)=\cI_{A_iB_{i-1}}(q,z) & = (z+z^{-1})\chi_0(q,z)-(1+q^{-1})z^{-1}\big(\chi_1(q,z)-\chi_2(q,z)\big).
\end{align*}
In \cite{Cordova:2016uwk} the authors take the limit $q\to 1, z\to 1$ and relate the line defect algebra to the Verlinde-like algebra of $\widehat{\mathfrak{sl}(2)}_{-\frac{4}{3}}$. Here we keep $z$ general while taking $q\to 1$. In this limit the expansion coefficients do not depend on the $i$ index anymore, just as in the $(A_1,A_{2N})$ case. We introduce a $z$-deformed Verlinde-like algebra $\cV_z$
with the $z$-deformed modular fusion rules:
\begin{eqnarray}\label{D3fus}
\begin{split}
[\Phi_1]\times[\Phi_1]&=[\Phi_2],\\
[\Phi_1]\times[\Phi_2]&=-z^2[\Phi_0],\\
[\Phi_2]\times[\Phi_2]&=-z^2[\Phi_1].
\end{split}
\end{eqnarray}
If we take $z=1$, this reduces to the naive modular fusion rules of $\widehat{\mathfrak{sl}(2)}_{-\frac{4}{3}}$ \cite{D.Francesco,Cordova:2016uwk}.
The homomorphism $f: \cL \to \cV_z$ is given by:
\begin{eqnarray}\label{5.9}
\begin{split}
I&\xrightarrow{f}[\Phi_0],\\
A_i&\xrightarrow{f}[A]=z^{-1}\big([\Phi_2]-[\Phi_1]\big),\\
B_i&\xrightarrow{f}[B]=[\Phi_0]-[\Phi_1]+z^{-2}[\Phi_2].
\end{split}
\end{eqnarray}
$f$ is believed to be a homomorphism since
\begin{eqnarray}
\begin{split}
[AA]&=2[\Phi_0]-[\Phi_1]+z^{-2}[\Phi_2]=[A]\times[A],\\
[BB]&=3[\Phi_0]-(2+z^{-2})[\Phi_1]+(1+2z^{-2})[\Phi_2]=[B]\times[B],\\
[AB]&=(z+z^{-1})[\Phi_0]-2z^{-1}\big([\Phi_1]-[\Phi_2]\big)=[A]\times[B].
\end{split}
\end{eqnarray}
We emphasize that this holds if and only if the $z$-deformed modular fusion rules are as given in (\ref{D3fus}).

The fusion matrices for $[\Phi_1]$ and $[\Phi_2]$ are:
\begin{equation}
N_{\Phi_{1}}=\begin{pmatrix}
0 & 1 & 0 \\
0 & 0 & 1 \\
-z^2 & 0 & 0
\end{pmatrix},\quad
N_{\Phi_{2}}=\begin{pmatrix}
0 & 0 & 1 \\
-z^2 & 0 & 0 \\
0 & -z^2 & 0
\end{pmatrix}.
\end{equation}
These two matrices are simultaneously diagonalizable for $z\neq 0$, with eigenvalues:
\begin{center}
	\begin{tabular}{ |c|c|c| }
		\hline
		$\text{eigenvector}$& $\lambda_{\Phi_1}$ & $\lambda_{\Phi_2}$  \\ \hline
		$(1,-z^{2/3},z^{4/3})$ & $-z^{2/3}$ & $z^{4/3}$  \\ \hline
		$(1,(-1)^{1/3}z^{2/3},(-1)^{2/3}z^{4/3})$ & $(-1)^{1/3}z^{2/3}$ & $(-1)^{2/3}z^{4/3}$\\ \hline
		$(1,-(-1)^{2/3}z^{2/3},-(-1)^{1/3}z^{4/3})$ & $-(-1)^{2/3}z^{2/3}$ & $-(-1)^{1/3}z^{4/3}$   \\ \hline
		\end{tabular}
\end{center}

\begin{figure}
	\centering
	\includegraphics[scale=0.12]{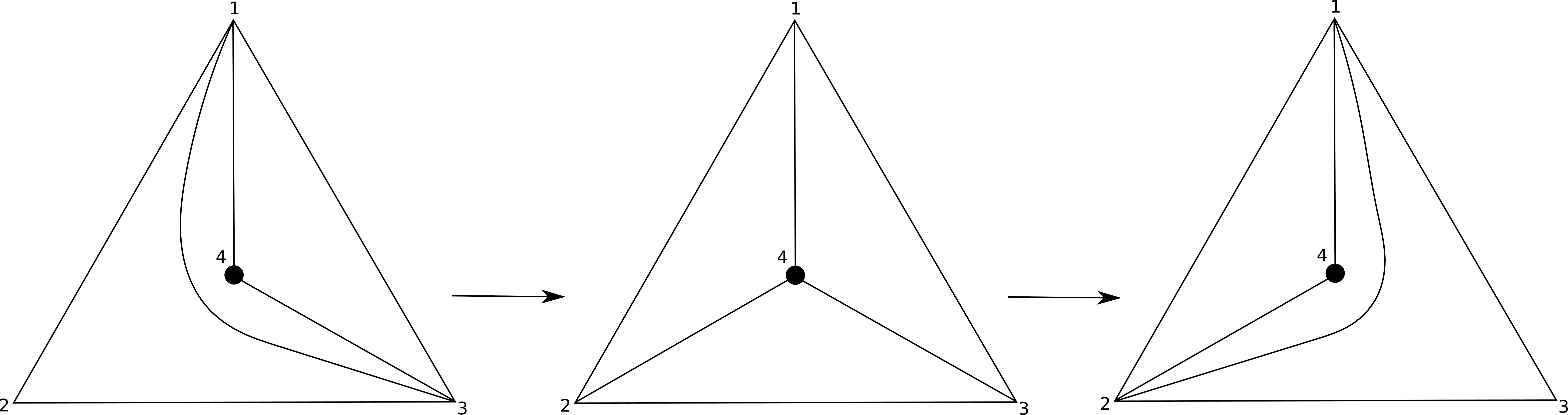}
	\caption{Classical monodromy action via two flips in $(A_1,D_3)$ Argyres-Douglas theory.}
	\label{D3flip}
\end{figure}

Now we turn to study fixed loci of the classical monodromy in this chamber.
Through a composition of two flips (see Figure \ref{D3flip}) the monodromy action is:
\begin{eqnarray}
\begin{split}
& \cX_{\gamma_1}\to\frac{1+\cX_{\gamma_3}+\cX_{\gamma_2}\cX_{\gamma_3}}{\cX_{\gamma_2}},\\
& \cX_{\gamma_2}\to \frac{1}{\cX_{\gamma_3}+\cX_{\gamma_2}\cX_{\gamma_3}},\\
& \cX_{\gamma_3}\to\frac{\cX_{\gamma_1}\cX_{\gamma_2}\cX_{\gamma_3}}{1+\cX_{\gamma_3}+\cX_{\gamma_2}\cX_{\gamma_3}}.
\end{split}
\end{eqnarray}
The fixed locus is determined by the equations
\begin{equation}
\cX_{\gamma_2}(1+\cX_{\gamma_2})\cX_{\gamma_3}=1,\quad
\cX_{\gamma_1}=\cX_{\gamma_3}(2+\cX_{\gamma_2}+\cX_{\gamma_3}+\cX_{\gamma_2}\cX_{\gamma_3}).
\end{equation}
To make connection with the flavor fugacity, we rewrite these equations in terms of $\cX_{\gamma_2}$, $z$ and $x:=\cX_{\gamma'}$;
this gives
\begin{equation}\label{D3fixed}
\cX_{\gamma_2}^3 z^2=1,\quad x=\cX_{\gamma_2}(1+\cX_{\gamma_2})z.
\end{equation}
One can check that this is exactly the same locus where $F({A_i})=F({A_j})$ and $F({B_i})=F({B_j})$. In particular,
this implies the evaluation map $g$ forgets the $i$ index as expected.

Now recall that the value of $z$ corresponds to the $SU(2)$ flavor holonomy that could be turned on when compactifying the
4d theory on
$S^1$. With this in mind we first fix $z$ and then look for the $U(1)_r$-fixed points.
For each value of $z\neq 0$, there are three $U(1)_r$-fixed points, which matches
the number of admissible representations of $\widehat{\mathfrak{sl}(2)}_{-\frac{4}{3}}$.
The evaluation map $g$ is concretely given by:
\begin{eqnarray}\label{D3g}
\begin{split}
1&\xrightarrow{g}\big(1,1,1\big),\\
A_i&\xrightarrow{g}\big(z^{1/3}+z^{-1/3},-(-1)^{1/3}z^{1/3}+(-1)^{2/3}z^{-1/3},-(-1)^{1/3}z^{-1/3}+(-1)^{2/3}z^{1/3}\big),\\
B_i&\xrightarrow{g}\big(1+z^{2/3}+z^{-2/3},1+(-1)^{2/3}z^{2/3}-(-1)^{1/3}z^{-2/3},\\
& \ \ \ \ \  1+(-1)^{2/3}z^{-2/3}-(-1)^{1/3}z^{2/3}\big).
\end{split}
\end{eqnarray}

Now, in contrast to the cases we studied in \S\ref{sec:examples-even}, in this
case the
weights of the classical monodromy action are not sufficient to distinguish
the three $U(1)_r$-fixed points, as we see from
the following table ($U(1)_r$ weights and correspondence between fixed points and primary fields taken from results of \cite{Fredrickson-Neitzke,Fredrickson:2017yka}):
\begin{center}
	\begin{tabular}{ |c|c|c|c| }
		\hline
		fixed point & weights of $M$ & weights of $U(1)_r$ & primary field\\ \hline
		I & $ -\frac{1\pm \I \sqrt{3}}{2}$ & $\frac{1}{3}, \frac{2}{3}$ & $\Phi_{1}$ \\ \hline
		II &$-\frac{1\pm \I \sqrt{3}}{2}$ & $-\frac{1}{3}, \frac{4}{3}$ & $\Phi_{0}$ \\ \hline
		III &$ -\frac{1\pm \I \sqrt{3}}{2}$ & $-\frac{1}{3}, \frac{4}{3}$ & $\Phi_{2}$ \\ \hline
	\end{tabular}
\end{center}

Thus we cannot determine \ti{a priori} which $U(1)_r$-fixed point should correspond
to which eigenspace of the fusion matrices. This gives an
$S_3$ ambiguity in constructing the map $h$.
Still, we can just try all of the $6$ possible mappings and see if one of them
works. Indeed, suppose we take:
\begin{eqnarray}
\begin{split}
[\Phi_1]&\xrightarrow{h}\big(-z^{2/3},-(-1)^{2/3}z^{2/3},(-1)^{1/3}z^{2/3}\big),\\
[\Phi_2]&\xrightarrow{h}\big(z^{4/3},-(-1)^{1/3}z^{4/3},(-1)^{2/3}z^{4/3}\big).
\end{split}
\end{eqnarray}
Combining this with (\ref{5.9}) and (\ref{D3g}), we find that
indeed $h\circ f=g$ for every $z\neq 0$.

\subsection{\texorpdfstring{$(A_1,D_5)$}{(A1,D5)} Argyres-Douglas theory}\label{sec:A1D5}

\begin{figure}
	\centering
	\includegraphics[scale=0.12]{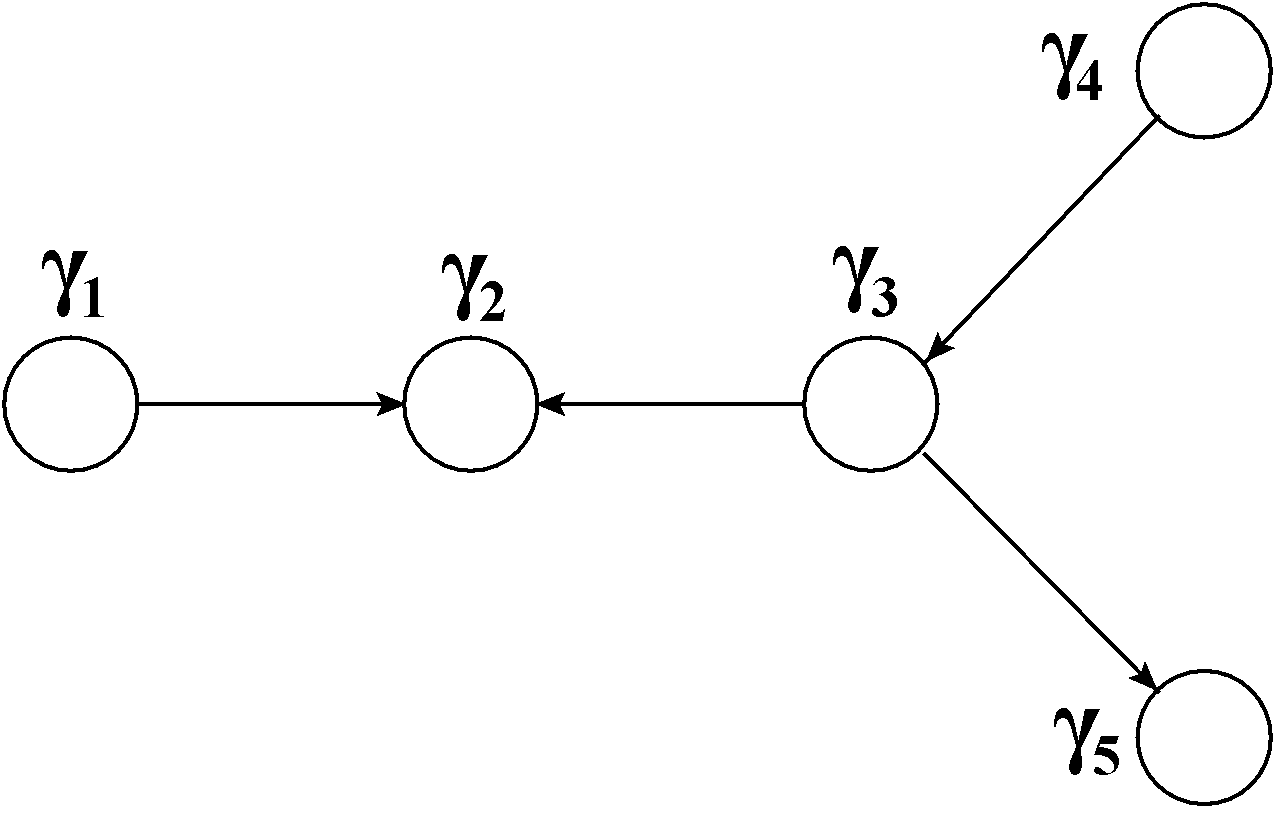}
	\caption{A BPS quiver for the ($A_1,D_5$) Argyres-Douglas theory.}
	\label{D5quiver}
\end{figure}
We choose the canonical chamber represented by the BPS quiver given in Figure \ref{D5quiver}, with five BPS particles (in increasing central charge phase order):
\begin{equation*}
\gamma_1,\gamma_4,\gamma_3,\gamma_2,\gamma_5.
\end{equation*}

\begin{figure}
	\centering
	\includegraphics[scale=0.18]{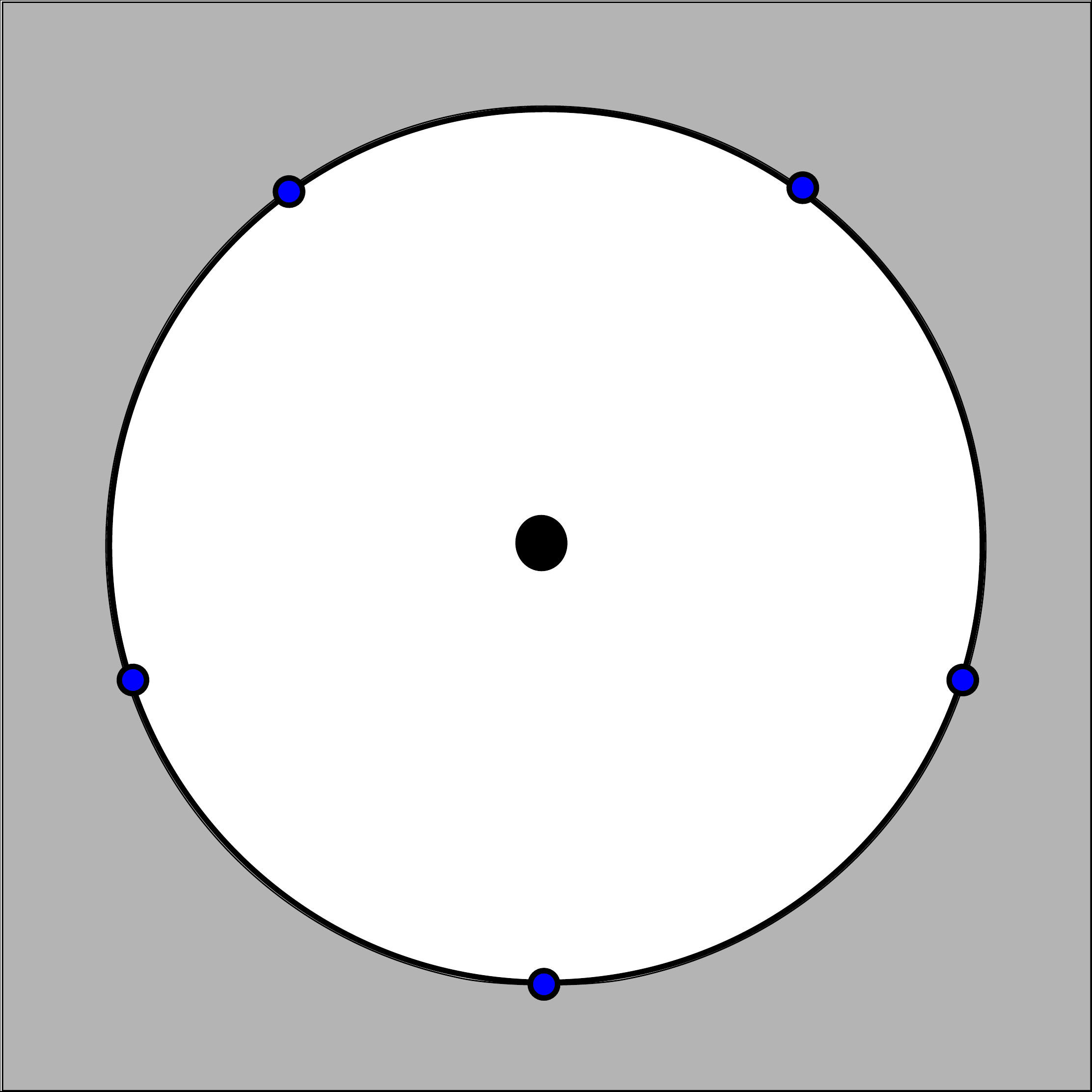}
	\caption{$\mathbb{CP}^1\setminus D_\infty$ where $D_\infty$ is a disk around $z=\infty$ bounded by $S^1$ with five marked points colored in blue. The regular singularity at $z=0$ is colored in black.}
	\label{D5FIG1}
\end{figure}

\begin{figure}
	\centering
	\includegraphics[scale=0.3]{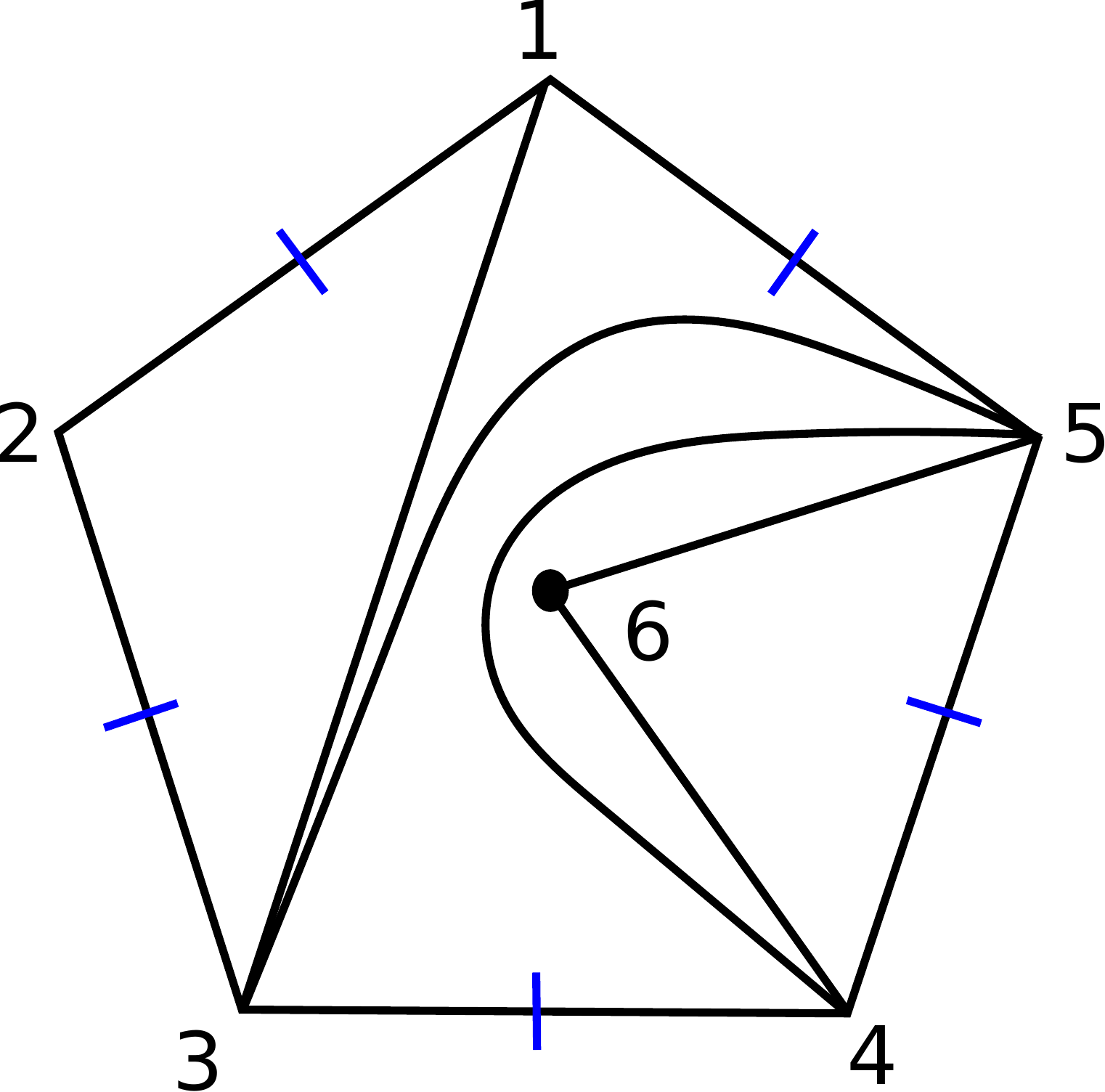}
	\caption{A triangulation in the $(A_1,D_5)$ Argyres-Douglas theory. There are five boundary edges. The blue marks correspond to positions of five Stokes rays. }
	\label{D5WKB}
\end{figure}

The corresponding Hitchin system is defined on $\mathbb{CP}^1$ with one regular singularity at $z=0$ and one irregular singularity at $z=\infty$ with five stokes rays emerging from it, i.e. there are five marked points on the $S^1$ which bounds $D_\infty$, the disk around $z=\infty$ that's cut out from $\mathbb{CP}^1$. The situation is depicted in Figure \ref{D5FIG1}. The corresponding WKB triangulation for this chamber is given in Figure \ref{D5WKB}, where $\cX_{\gamma_1}$ corresponds to edge 13, $\cX_{\gamma_2}$ corresponds to edge 35, $\cX_{\gamma_3}$ corresponds to edge 45, $\cX_{\gamma_4}$ corresponds to edge 56 and $\cX_{\gamma_5}$ corresponds to edge 46.

\begin{figure}
	\begin{subfigure}{.45\textwidth}
		\centering
		\includegraphics[width=.8\linewidth]{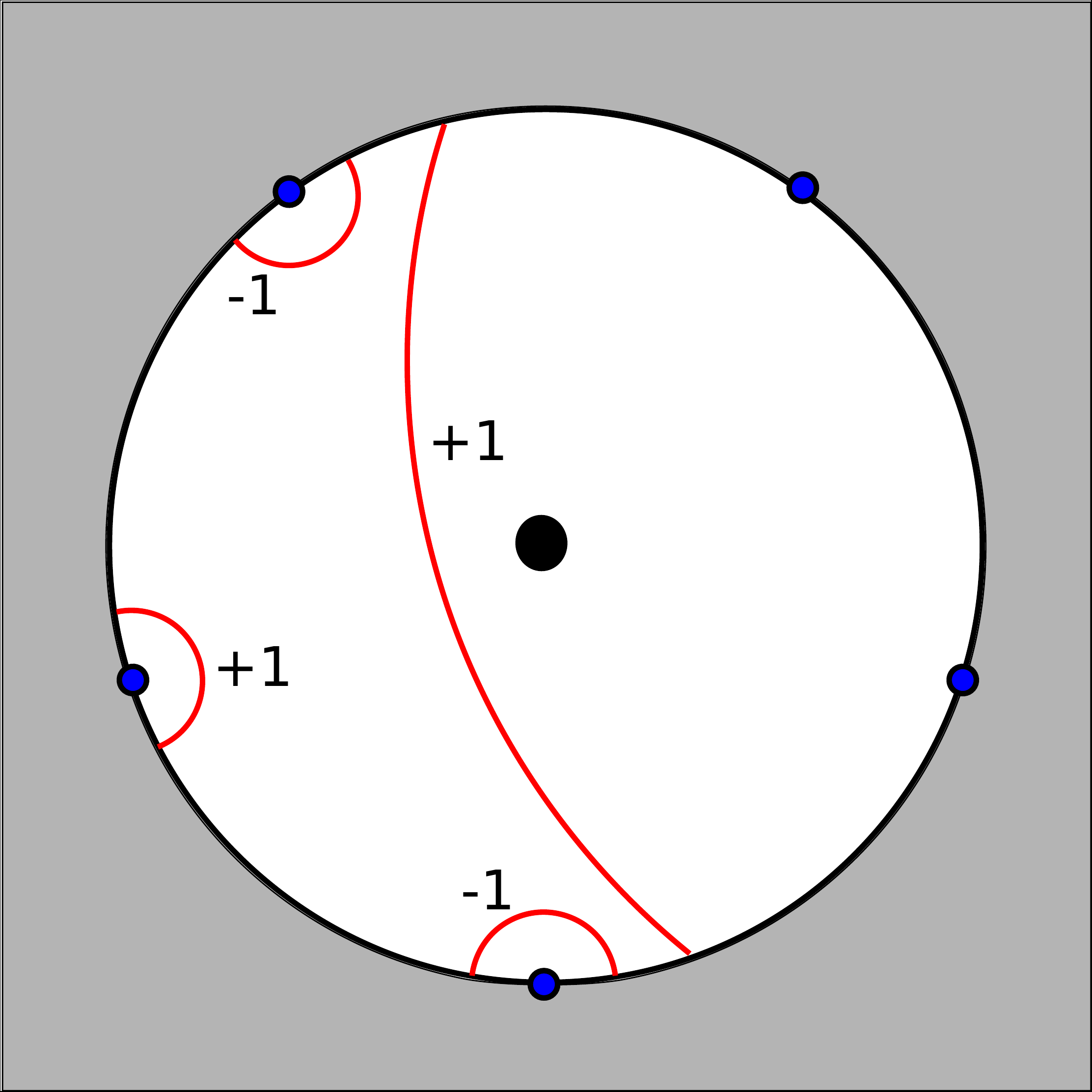}
		\caption{Type A}
		\label{D5lamA}
	\end{subfigure}
	\begin{subfigure}{.45\textwidth}
			\centering
			\includegraphics[width=.8\linewidth]{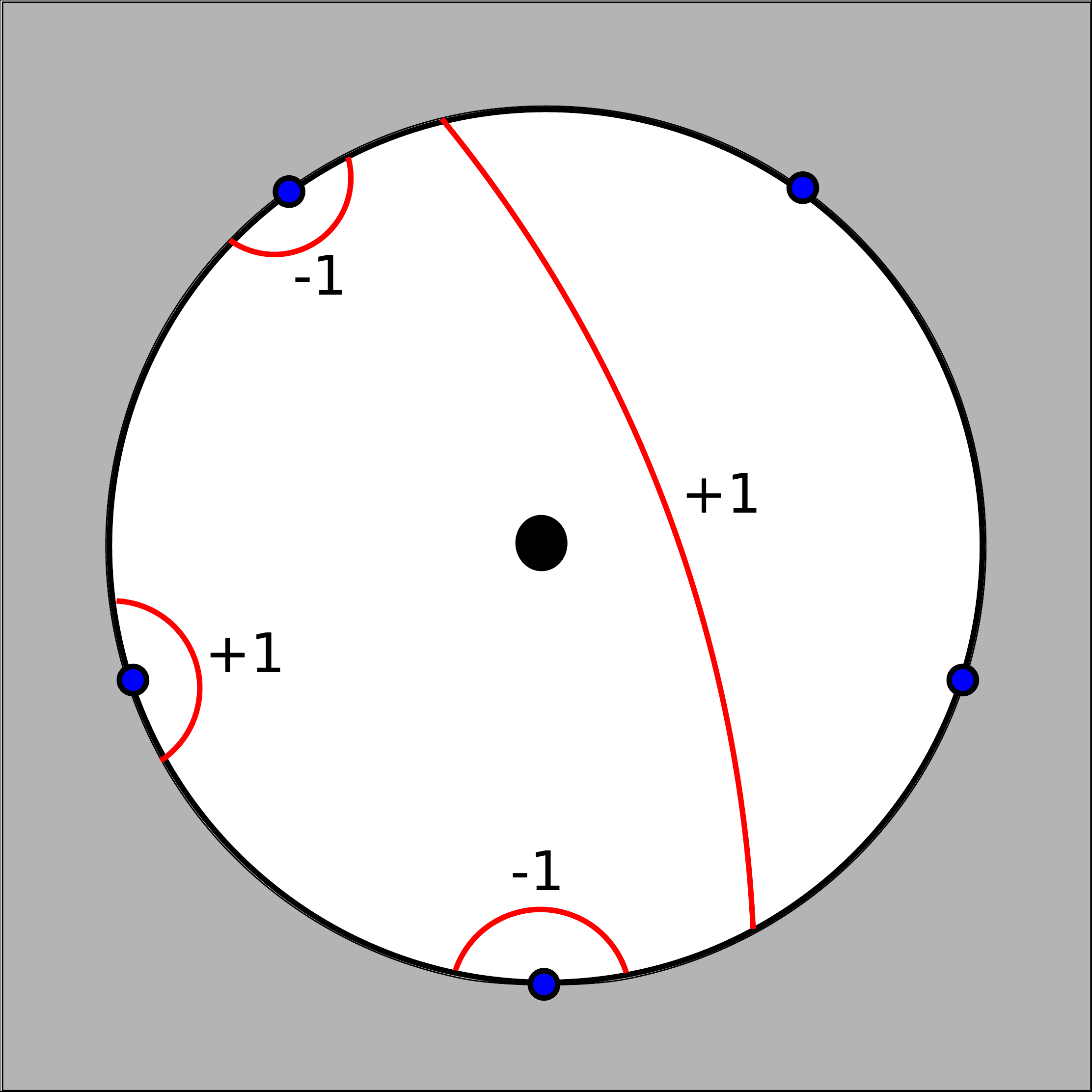}
			\caption{Type B}
			\label{D5lamB}
	\end{subfigure}
	\begin{subfigure}{.45\textwidth}
				\centering
				\includegraphics[width=.8\linewidth]{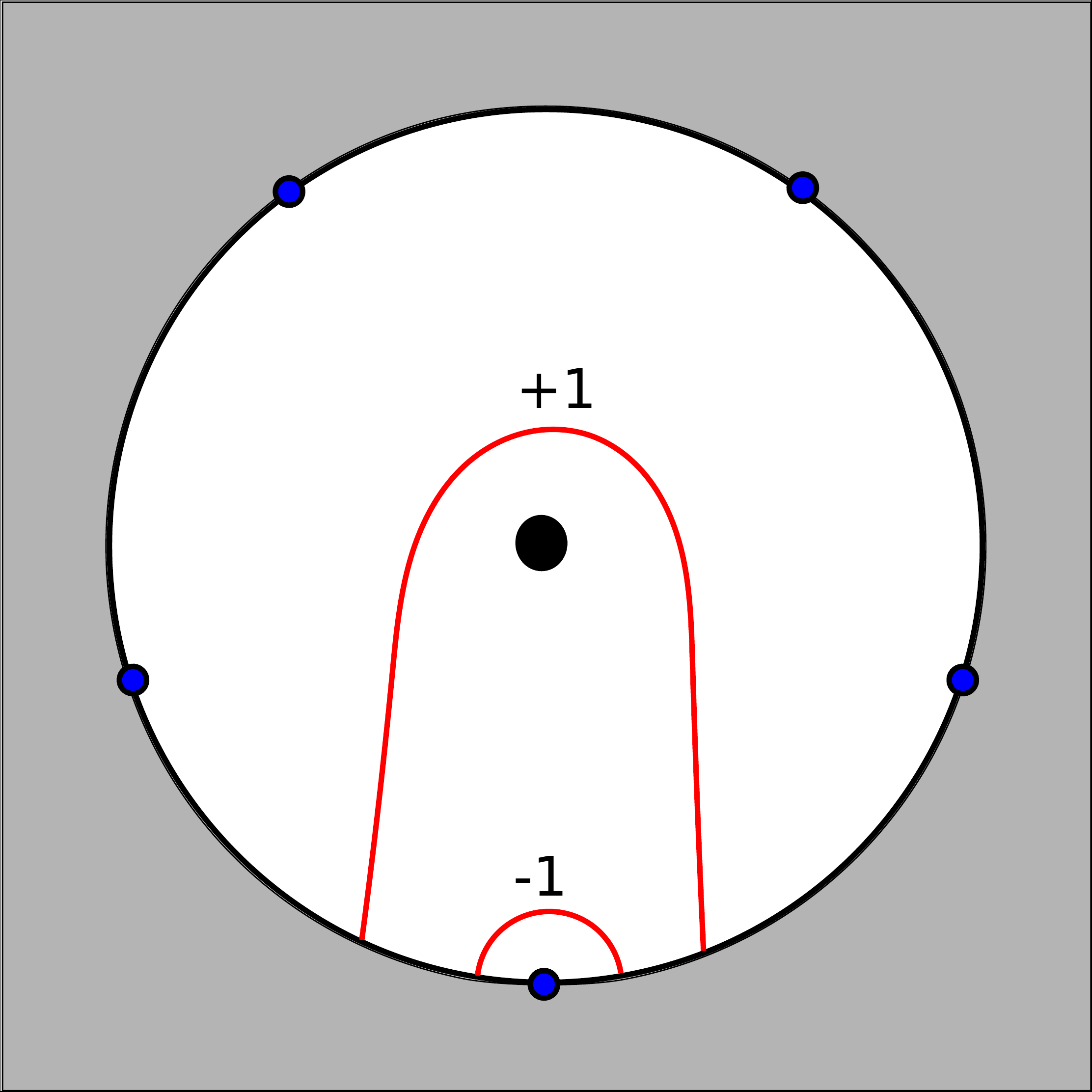}
				\caption{Type C}
				\label{D5lamC}
	\end{subfigure}
	\begin{subfigure}{.45\textwidth}
			\centering
			\includegraphics[width=.8\linewidth]{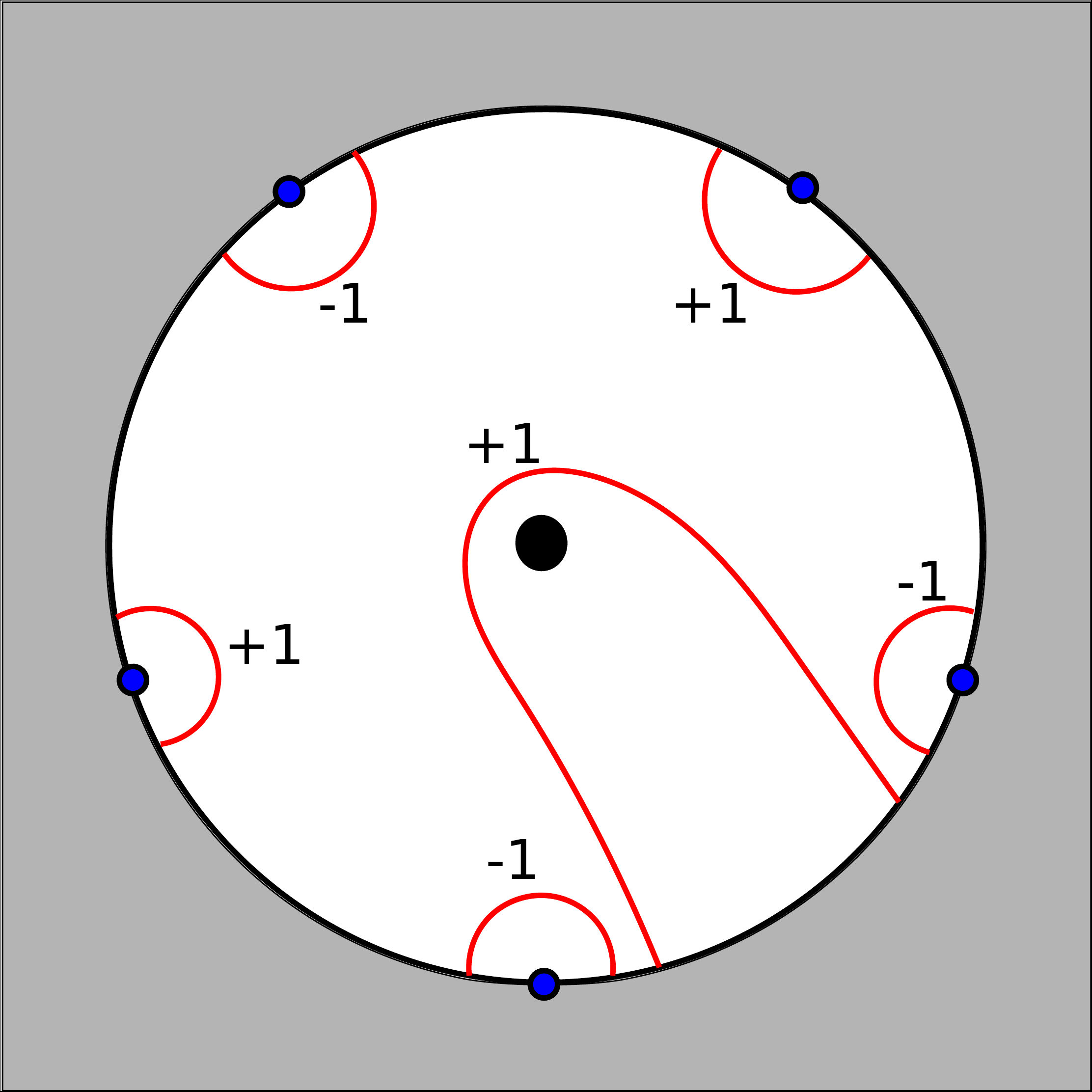}
			\caption{Type D}
			\label{D5lamD}
	\end{subfigure}
	\begin{subfigure}{.45\textwidth}
			\centering
			\includegraphics[width=.8\linewidth]{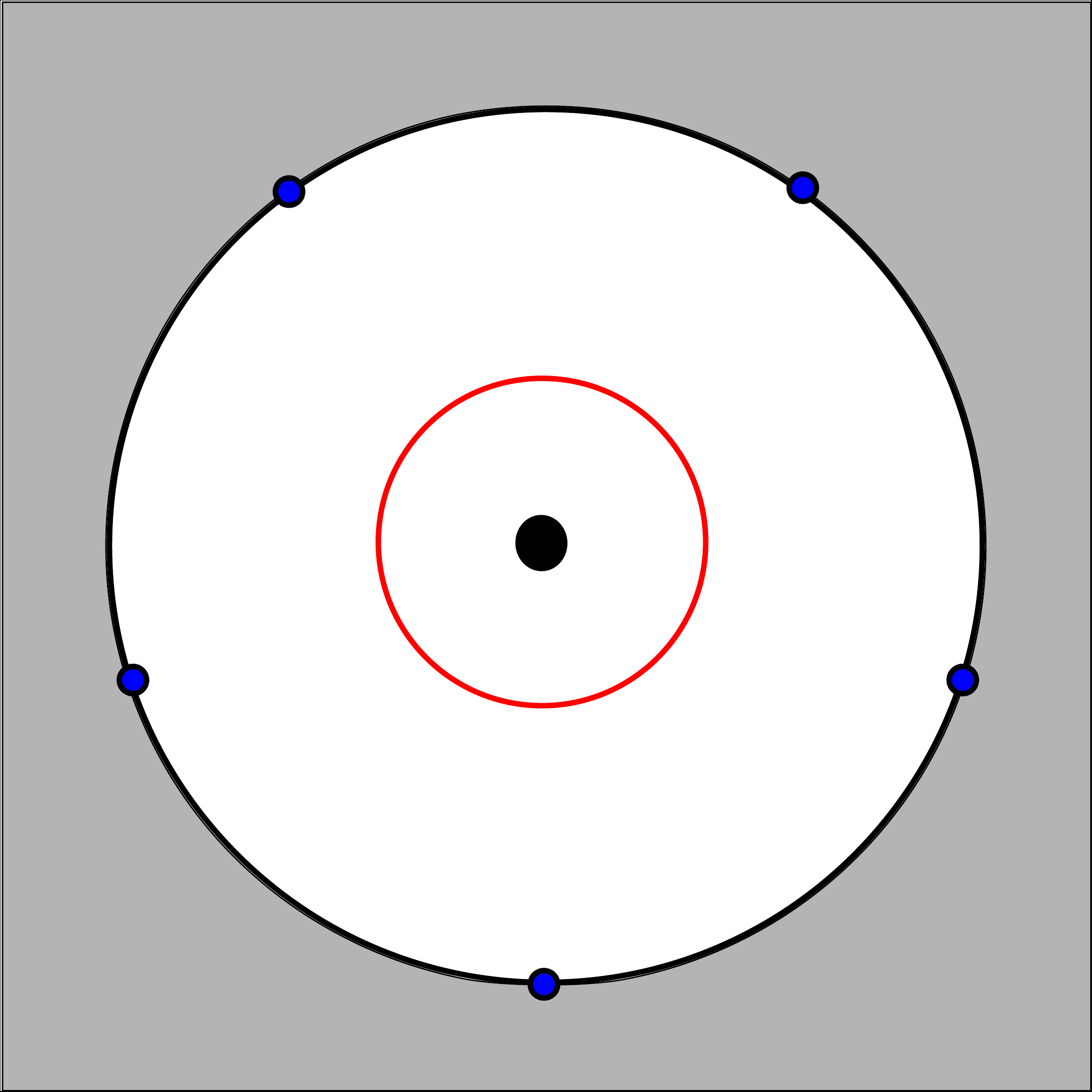}
			\caption{Type E}
			\label{D5lamE}
	\end{subfigure}
	\caption{Five types of laminations in $(A_1,D_5)$ Argyres-Douglas theory.}
	\label{D5lam}
\end{figure}

The line defect generators correspond to laminations that can not be expressed as sum of other laminations. In this case there are 21 such laminations. The lamination ($E$) which is a loop around the regular singularity corresponds to the pure flavor line defect. The other 20 laminations come in four types $A,B,C$ and $D$. We label their corresponding generators as $A_i$, $B_i$, $C_i$ and $D_i$ ($i=1,\dots,5$) and list laminations corresponding to the generators $A_1,B_1,C_1,D_1$ and $E$ in Figure \ref{D5lam}. Laminations corresponding to e.g. generators $A_{i+1}$ are obtained by rotating laminations for $A_{i}$ clockwise by $4\pi/5$.
We define the flavor charge $\gamma_f$ and $\gamma'$ as follows:
\begin{equation}
\gamma_f=\frac{\gamma_4+\gamma_5}{2},\quad \gamma'=\frac{\gamma_4-\gamma_5}{2}.
\end{equation}
The $SU(2)$ flavor fugacity is $z:=\text{Tr}(X_{\gamma_f})$. The generating functions are computed using the method as reviewed in \S\ref{sec:defects-class-S}. In particular, the line defect generator $D_2$ has framed BPS states with charge $2\gamma_2$ in a 3-dimensional multiplet of $SO(3)$:
\begin{align*}
F(A_1) & = X_{-\gamma_1}+X_{-\gamma_1-\gamma_2},\\
F(A_2) & = X_{-\gamma_1}+X_{\gamma_2}+X_{-\gamma_1+\gamma_2}+X_{\gamma_2+\gamma_3}+X_{-\gamma_1+\gamma_2+\gamma_3}+zX_{\gamma_2+\gamma_3+\gamma'}+zX_{-\gamma_1+\gamma_2+\gamma_3+\gamma'},\\
F(A_3) & = X_{\gamma_2}+X_{\gamma_1+\gamma_2}+X_{-\gamma_3}+X_{\gamma_2-\gamma_3}+X_{\gamma_1+\gamma_2-\gamma_3}+z^{-1}X_{-\gamma_3+\gamma'}+z^{-1}X_{\gamma_2-\gamma_3+\gamma'}\\
& \ \ +z^{-1}X_{\gamma_1+\gamma_2-\gamma_3+\gamma'},\\
F(A_4) & = X_{\gamma_1},\\
F(A_5) & = (z+z^{-1})X_{-\gamma'}+z^{-1}X_{-\gamma_3-\gamma'}+z^{-1}X_{-\gamma_2-\gamma_3-\gamma'}+zX_{\gamma_3-\gamma'}+(q^{1/2}+q^{-1/2})X_{-2\gamma'}\\
& \ \ +X_{-\gamma_2-2\gamma'}+X_{-\gamma_3-2\gamma'}+X_{-\gamma_2-\gamma_3-2\gamma'}+X_{\gamma_3-2\gamma'},\\
F(B_1) & = X_{-\gamma_1-\gamma'}+X_{-\gamma_1-\gamma_2-\gamma'}+X_{-\gamma_1+\gamma_3-\gamma'}+zX_{-\gamma_1+\gamma_3},\\
F(B_2) & = X_{-\gamma_1+\gamma'}+X_{\gamma_2+\gamma'}+X_{-\gamma_1+\gamma_2+\gamma'},\\
F(B_3) & = X_{\gamma_2+\gamma'}+X_{\gamma_1+\gamma_2+\gamma'},\\
F(B_4) & = z^{-1}X_{\gamma_1-\gamma_3}+X_{\gamma_1-\gamma'}+X_{\gamma_1-\gamma_3-\gamma'},\\
F(B_5) & = X_{-\gamma_2-\gamma'},\\
F(C_1) & = X_{-\gamma'}+X_{\gamma_3-\gamma'}+z X_{\gamma_3},\\
F(C_2) & = (q^{1/2}+q^{-1/2})\big(X_{-\gamma_1-\gamma'}+X_{\gamma_2-\gamma'}+X_{-\gamma_1+\gamma_2-\gamma'}+X_{-\gamma_1-\gamma_3-\gamma'}+z^{-1}X_{-\gamma_1-\gamma_3}\big)\\
& \ \ +X_{-\gamma'}+X_{-\gamma_3-\gamma'}+X_{-\gamma_1-\gamma_2-\gamma_3-\gamma'}+X_{\gamma_2-\gamma_3-\gamma'}+X_{-\gamma_1+\gamma_2-\gamma_3-\gamma'}+X_{\gamma_2+\gamma_3-\gamma'}\\
& \ \ +X_{-\gamma_1+\gamma_2+\gamma_3-\gamma'}+(z+z^{-1})X_{-\gamma_1}+(z+z^{-1})X_{\gamma_2}+(z+z^{-1})X_{-\gamma_1+\gamma_2}+z^{-1}X_{-\gamma_3}\\
& \ \ +z^{-1}X_{-\gamma_1-\gamma_2-\gamma_3}+z^{-1}X_{\gamma_2-\gamma_3}+z^{-1}X_{-\gamma_1+\gamma_2-\gamma_3}+zX_{\gamma_2+\gamma_3}+z X_{-\gamma_1+\gamma_2+\gamma_3},\\
F(C_3) & = X_{\gamma'},\\
F(C_4) & = (q^{1/2}+q^{-1/2})\big(X_{\gamma_2-\gamma'}+X_{\gamma_1+\gamma_2-\gamma'}\big)+X_{-\gamma'}+X_{-\gamma_3-\gamma'}+X_{\gamma_2-\gamma_3-\gamma'}+X_{\gamma_1+\gamma_2-\gamma_3-\gamma'}\\
& \ \ +X_{\gamma_2+\gamma_3-\gamma'}+X_{\gamma_1+\gamma_2+\gamma_3-\gamma'}+z^{-1}X_{\gamma_2}+z^{-1}X_{\gamma_1+\gamma_2}+z^{-1}X_{-\gamma_3}+z^{-1}X_{\gamma_2-\gamma_3}\\
& \ \ +z^{-1}X_{\gamma_1+\gamma_2-\gamma_3}+zX_{\gamma_2}+zX_{\gamma_1+\gamma_2}+zX_{\gamma_2+\gamma_3}+zX_{\gamma_1+\gamma_2+\gamma_3},\\
F(C_5) & = X_{-\gamma'}+X_{-\gamma_3-\gamma'}+X_{-\gamma_2-\gamma_3-\gamma'}+z^{-1}X_{-\gamma_3}+z^{-1}X_{-\gamma_2-\gamma_3},\\
F(D_1) & = X_{-\gamma_1+\gamma_3}+zX_{-\gamma_1+\gamma_3+\gamma'},\\
F(D_2) & = (q^{1/2}+q^{-1/2})X_{\gamma_2}+(q^{1/2}+q^{-1/2})X_{-\gamma_1+\gamma_2}+(1+1+q+q^{-1})X_{2\gamma_2}\\
& \ \ +(q^{1/2}+q^{-1/2})X_{-\gamma_1+2\gamma_2}+(q^{1/2}+q^{-1/2})X_{\gamma_1+2\gamma_2}+X_{-\gamma_1-\gamma_3}+(q^{1/2}+q^{-1/2})X_{\gamma_2-\gamma_3}\\
& \ \  +(q^{1/2}+q^{-1/2})X_{-\gamma_1+\gamma_2-\gamma_3}+(q^{1/2}+q^{-1/2})X_{2\gamma_2-\gamma_3}+X_{-\gamma_1+2\gamma_2-\gamma_3}+X_{\gamma_1+2\gamma_2-\gamma_3}\\
& \ \ +(q^{1/2}+q^{-1/2})X_{2\gamma_2+\gamma_3}+X_{-\gamma_1+2\gamma_2+\gamma_3}+X_{\gamma_1+2\gamma_2+\gamma_3}+(z+z^{-1})X_{\gamma_2+\gamma'}\\
& \ \ +(z+z^{-1})X_{-\gamma_1+\gamma_2+\gamma'}+(z+z^{-1})(q^{1/2}+q^{-1/2})X_{2\gamma_2+\gamma'}+(z+z^{-1})X_{-\gamma_1+2\gamma_2+\gamma'}\\
& \ \ +(z+z^{-1})X_{\gamma_1+2\gamma_2+\gamma'}+z^{-1}X_{-\gamma_1-\gamma_3+\gamma'}+(q^{1/2}+q^{-1/2})z^{-1}X_{\gamma_2-\gamma_3+\gamma'}\\
& \ \ +(q^{1/2}+q^{-1/2})z^{-1}X_{-\gamma_1+\gamma_2-\gamma_3+\gamma'}+(q^{1/2}+q^{-1/2})z^{-1}X_{2\gamma_2-\gamma_3+\gamma'}+z^{-1}X_{-\gamma_1+2\gamma_2-\gamma_3+\gamma'}\\
& \ \ +z^{-1}X_{\gamma_1+2\gamma_2-\gamma_3+\gamma'}+(q^{1/2}+q^{-1/2})z X_{2\gamma_2+\gamma_3+\gamma'}+zX_{-\gamma_1+2\gamma_2+\gamma_3+\gamma'}+zX_{\gamma_1+2\gamma_2+\gamma_3+\gamma'},\\
F(D_3) & = X_{\gamma_1-\gamma_3}+z^{-1}X_{\gamma_1-\gamma_3+\gamma'},\\
F(D_4) & = (q^{1/2}+q^{-1/2})X_{\gamma_1-2\gamma'}+X_{\gamma_1-\gamma_3-2\gamma'}+X_{\gamma_1+\gamma_3-2\gamma'}+z^{-1}X_{\gamma_1-\gamma'}+z^{-1}X_{\gamma_1-\gamma_3-\gamma'}\\
& \ \ +zX_{\gamma_1-\gamma'}+zX_{\gamma_1+\gamma_3-\gamma'},\\
F(D_5) & = (q^{1/2}+q^{-1/2})\big(X_{-\gamma_1-2\gamma'}+X_{-\gamma_1-\gamma_2-2\gamma'}+X_{-\gamma_1-\gamma_2-\gamma_3-2\gamma'}+z^{-1}X_{-\gamma_1-\gamma_2-\gamma_3-\gamma'}\big)\\
& \ \ +X_{-\gamma_1-\gamma_3-2\gamma'}+X_{-\gamma_1-2\gamma_2-\gamma_3-2\gamma'}+X_{-\gamma_1+\gamma_3-2\gamma'}+(z+z^{-1})X_{-\gamma_1-\gamma'}\\
& \ \ +(z+z^{-1})X_{-\gamma_1-\gamma_2-\gamma'}+z^{-1}X_{-\gamma_1-\gamma_3-\gamma'}+z^{-1}X_{-\gamma_1-2\gamma_2-\gamma_3-\gamma'}+zX_{-\gamma_1+\gamma_3-\gamma'},\\
F(E) & = z+z^{-1}.
\end{align*}

The line defect Schur index is
\begin{eqnarray}
\begin{split}
\cI_L(q,z)&=(q)_\infty^4\text{Tr}[F(L)S(q)\overline{S}(q)], \quad \text{with}\\
S(q)&=E_q(X_{\gamma_1})E_q(X_{\gamma_4})E_q(X_{\gamma_3})E_q(X_{\gamma_2})E_q(X_{\gamma_5}).
\end{split}
\end{eqnarray}
After inserting generating functions the calculation boils down to computing the following:
\begin{align*}
& \ \ \ (q)_\infty^4\text{Tr}[X_{a\gamma_1+b\gamma_2+c\gamma_3+d\gamma'}S(q)\overline{S}(q)]\\
& = (q)_\infty^4\Sum_{l_i,k_i=0}^{\infty}\frac{(-1)^{a+b+c+d}q^{A/2}z^{l_4+l_5-k_4-k_5}}{(q)_{l_1}\dots(q)_{l_5}(q)_{k_1}\dots(q)_{k_5}}\delta_{k_1,l_1+a}\delta_{k_2,l_2+b}\delta_{k_3,l_3+c}\delta_{k_4,l_4-l_5+k_5+d},\quad\text{with}\\
& A=\frac{1}{2}\Big(a + b + a b + c + b c - c d + d (1 + 2 c + 2 l_3) +
2 \big(l_1 + l_2 + a l_2 + c l_2 + l_1 l_2 + l_3\\
&\quad\quad + l_2 l_3 + k_5 (1 + c + l_3) +
l_4 + l_3 l_4\big)\Big).
\end{align*}
Within the same class line defect Schur indices are the same. The coefficients in $q$ are again characters of certain $SU(2)$ representations:
\begin{eqnarray}
\begin{split}
\cI_{A}(q,z) & = -q^{\frac{1}{2}}(1+\chi_{\bf{3}}q+\chi_{\bf{1\oplus 3\oplus 5}}q^2+\chi_{\bf{1}\oplus 3^{\oplus\text{2}}\oplus 5\oplus 7}q^3+\cdots),\\
\cI_{B}(q,z) & = q(\chi_{\bf{2}}+\chi_{\bf{4}}q+\chi_{\bf{2\oplus 4\oplus 6}}q^2+\chi_{\bf{2}^{\oplus\text{2}}\oplus 4^{\oplus\text{2}}\oplus 6\oplus 8}q^3+\cdots),\\
\cI_{C}(q,z) & =-q^{\frac{1}{2}}(\chi_{\bf{2}}+\chi_{\bf{4}}q+\chi_{\bf{2}^{\oplus\text{2}}\oplus 4\oplus 6}q^2+\chi_{\bf{2}^{\oplus\text{2}}\oplus 4^{\oplus\text{3}}\oplus 6\oplus 8}q^3+\cdots),\\
\cI_{D}(q,z) & = q(1+\chi_{\bf{3}}q^2+\chi_{\bf{1\oplus 3}}q^3+\cdots).
\end{split}
\end{eqnarray}
The chiral algebra corresponding to the $(A_1,D_5)$ Argyres-Douglas theory is $\widehat{\mathfrak{sl}(2)}_{-\frac{8}{5}}$ \cite{Cordova:2015nma,Beem:2013sza,Beem:2014zpa,Xie:2016evu}, which has five admissible representations with the following highest weights:
\begin{eqnarray}\label{d5admiss}
\begin{split}
\Phi_0&=\left[-\frac{8}{5},0\right],\quad \Phi_1=\left[-\frac{6}{5},-\frac{2}{5}\right],\quad
\Phi_2=\left[-\frac{4}{5},-\frac{4}{5}\right],\\
\Phi_3&=\left[-\frac{2}{5},-\frac{6}{5}\right],\quad
\Phi_4=\left[0,-\frac{8}{5}\right],
\end{split}
\end{eqnarray}
where $\Phi_0$ is the highest weight for the vacuum module. The characters of these representations can be worked out using the Kac-Wakimoto formula \cite{V.G.Kac}, which is a special case of the Kazhdan-Lusztig formula \cite{DeVos:1995an} (see also \cite{D.Francesco} for expressions in terms of generalized theta functions):
\begin{eqnarray}
\begin{split}
\chi_0(q,z)&=\frac{\sum_{m=0}^\infty (-1)^m\frac{z^{2m+1}-z^{-(2m+1)}}{z-z^{-1}}q^{\frac{5m(m+1)}{2}}}{\prod_{n=1}^\infty(1-q^n)(1-z^2q^n)(1-z^{-2}q^n)},\\
\chi_1(q,z)&=\frac{1+\sum_{m=1}^\infty(-1)^m(z^{-2m}q^{\frac{m(5m-3)}{2}}+z^{2m}q^{\frac{m(5m+3)}{2}})}{(1-z^{-2})\prod_{n=1}^\infty(1-q^n)(1-z^2q^n)(1-z^{-2}q^n)},\\
\chi_2(q,z)&=\frac{1+\sum_{m=1}^\infty(-1)^m(z^{-2m}q^{\frac{m(5m-1)}{2}}+z^{2m}q^{\frac{m(5m+1)}{2}})}{(1-z^{-2})\prod_{n=1}^\infty(1-q^n)(1-z^2q^n)(1-z^{-2}q^n)},\\
\chi_3(q,z)&=\frac{1+\sum_{m=1}^\infty(-1)^m(z^{2m}q^{\frac{m(5m-1)}{2}}+z^{-2m}q^{\frac{m(5m+1)}{2}})}{(1-z^{-2})\prod_{n=1}^\infty(1-q^n)(1-z^2q^n)(1-z^{-2}q^n)},\\
\chi_4(q,z)&=\frac{1+\sum_{m=1}^\infty(-1)^m(z^{2m}q^{\frac{m(5m-3)}{2}}+z^{-2m}q^{\frac{m(5m+3)}{2}})}{(1-z^{-2})\prod_{n=1}^\infty(1-q^n)(1-z^2q^n)(1-z^{-2}q^n)}.
\end{split}
\end{eqnarray}
The $\mathcal{S}$ matrix for these five admissible representations, in the order (\ref{d5admiss}), is \cite{D.Francesco}:
\begin{equation}
\mathcal{S}=\frac{1}{\sqrt{5}}\begin{pmatrix}
1 & -1 & 1 & -1 & 1 \\
-1 & -(-1)^{3/5} & (-1)^{1/5} & (-1)^{4/5} & -(-1)^{2/5} \\
1 & (-1)^{1/5} & (-1)^{2/5} & (-1)^{3/5} & (-1)^{4/5}\\
-1 & (-1)^{4/5} & (-1)^{3/5} & (-1)^{2/5} & (-1)^{1/5}\\
1 & -(-1)^{2/5} & (-1)^{4/5} & (-1)^{1/5} & -(-1)^{3/5}
\end{pmatrix}.
\end{equation}
Working out the conjugation matrix $\mathcal{C}=\mathcal{S}^2$ it's clear that $\Phi_1$ and $\Phi_4$ are conjugate to each other, $\Phi_2$ and $\Phi_3$ are conjugate to each other.
Using the Verlinde formula \cite{Verlinde:1988sn} the modular fusion rules for $\widehat{\mathfrak{sl}(2)}_{-\frac{8}{5}}$ are given by:
\begin{eqnarray}
\begin{split}
[\Phi_1]\times[\Phi_1]&=[\Phi_2],\quad [\Phi_1]\times[\Phi_2]=[\Phi_3],\quad [\Phi_1]\times[\Phi_3]=[\Phi_4],\\
[\Phi_1]\times[\Phi_4]&=-[\Phi_0],\quad [\Phi_2]\times[\Phi_2]=[\Phi_4],\quad
[\Phi_2]\times[\Phi_3]=-[\Phi_0],\\
[\Phi_2]\times[\Phi_4]&=-[\Phi_1],\quad [\Phi_3]\times[\Phi_3]=-[\Phi_1],\quad [\Phi_3]\times[\Phi_4]=-[\Phi_2],\\
[\Phi_4]\times[\Phi_4]&=-[\Phi_3].
\end{split}
\end{eqnarray}
As we will see shortly, multiplications in the deformed Verlinde-like algebra are again given by multiplying the $-1$ coefficients in the original modular fusion rules by a factor of $z^2$.

The line defect Schur indices for defect generators of type $A$, $B$, $C$ and $D$ admit the following character expansions:
\begin{eqnarray}
\begin{split}
\cI_{A}(q,z)&=q^{-1/2}\big(\chi_0(q,z)-\chi_1(q,z)+z^{-2}\chi_4(q,z)\big),\\
\cI_{B}(q,z)&=q^{-1}z^{-1}\big(\chi_2(q,z)-\chi_3(q,z)\big),\\
\cI_C(q,z)&=q^{-1/2}z^{-1}\big(-\chi_1(q,z)+\chi_2(q,z)-\chi_3(q,z)+\chi_4(q,z)\big),\\
\cI_D(q,z)&=\chi_0(q,z)-q^{-1}\big(\chi_1(q,z)-\chi_2(q,z)+z^{-2}\chi_3(q,z)-z^{-2}\chi_4(q,z)\big).
\end{split}
\end{eqnarray}
Now we again take the $q\to 1$ limit while keeping $z$ general,
giving the map
\begin{eqnarray}
\begin{split}
I&\xrightarrow{f}[\Phi_0],\\
A_i&\xrightarrow{f}[A]=[\Phi_0]-[\Phi_1]+z^{-2}[\Phi_4],\\
B_i&\xrightarrow{f}[B]=z^{-1}([\Phi_2]-[\Phi_3]),\\
C_i&\xrightarrow{f}[C]=z^{-1}(-[\Phi_1]+[\Phi_2]-[\Phi_3]+[\Phi_4]),\\
D_i&\xrightarrow{f}[D]=[\Phi_0]-[\Phi_1]+[\Phi_2]-z^{-2}[\Phi_3]+z^{-2}[\Phi_4].
\end{split}
\end{eqnarray}
This map is believed to be a homomorphism $f: \cL \to \cV_z$, when we
define the deformed Verlinde-like algebra $\cV_z$ by
the following $z$-deformed modular fusion rules:
\begin{eqnarray}\label{D5fus}
\begin{split}
[\Phi_1]\times[\Phi_1]&=[\Phi_2],\quad [\Phi_1]\times[\Phi_2]=[\Phi_3],\quad [\Phi_1]\times[\Phi_3]=[\Phi_4],\\
[\Phi_1]\times[\Phi_4]&=-z^2[\Phi_0],\quad [\Phi_2]\times[\Phi_2]=[\Phi_4],\quad
[\Phi_2]\times[\Phi_3]=-z^2[\Phi_0],\\
[\Phi_2]\times[\Phi_4]&=-z^2[\Phi_1],\quad [\Phi_3]\times[\Phi_3]=-z^2[\Phi_1],\quad [\Phi_3]\times[\Phi_4]=-z^2[\Phi_2],\\
[\Phi_4]\times[\Phi_4]&=-z^2[\Phi_3].
\end{split}
\end{eqnarray}
To check the homomorphism property we
consider Schur indices with insertion of two half line defects,
which can also be expanded in terms of characters of admissible representations. After setting $q\to 1$ the expansion coefficients do not depend on the $i$-index anymore:
\begin{align*}
A_iA_j&\xrightarrow{f}3[\Phi_0]-2[\Phi_1]+[\Phi_2]-z^{-2}[\Phi_3]+2z^{-2}[\Phi_4],\\
A_iB_j&\xrightarrow{f}z^{-1}(-[\Phi_1]+2[\Phi_2]-2[\Phi_3]+[\Phi_4]),\\
A_iC_j&\xrightarrow{f}(z+z^{-1})[\Phi_0]-2z^{-1}([\Phi_1]-[\Phi_4])+3z^{-1}([\Phi_2]-[\Phi_3]),\\
A_iD_j&\xrightarrow{f}3[\Phi_0]-3[\Phi_1]+(2+z^{-2})[\Phi_2]-(1+2z^{-2})[\Phi_3]+3z^{-2}[\Phi_4],\\
B_iB_j&\xrightarrow{f}2[\Phi_0]-[\Phi_1]+z^{-2}[\Phi_4],\\
B_iC_j&\xrightarrow{f}2[\Phi_0]-2[\Phi_1]+[\Phi_2]-z^{-2}[\Phi_3]+2z^{-2}[\Phi_4],\\
B_iD_j&\xrightarrow{f}(z+z^{-1})[\Phi_0]+2z^{-1}(-[\Phi_1]+[\Phi_2]-[\Phi_3]+[\Phi_4]),\\
C_iC_j&\xrightarrow{f}4[\Phi_0]-3[\Phi_1]+(2+z^{-2})[\Phi_2]-(1+2z^{-2})[\Phi_3]+3z^{-2}[\Phi_4],\\
C_iD_j&\xrightarrow{f}2(z+z^{-1})[\Phi_0]-(z+3z^{-1})[\Phi_1]+4z^{-1}([\Phi_2]-[\Phi_3])+(3z^{-1}+z^{-3})[\Phi_4],\\
D_iD_j&\xrightarrow{f}5[\Phi_0]-(4+z^{-2})[\Phi_1]+(3+2z^{-2})[\Phi_2]-(2+3z^{-2})[\Phi_3]+(1+4z^{-2})[\Phi_4].
\end{align*}
$f$ is a homomorphism if and only if the $z$-deformed fusion rules are as defined in (\ref{D5fus}).

\begin{figure}
	\centering
	\includegraphics[scale=0.25]{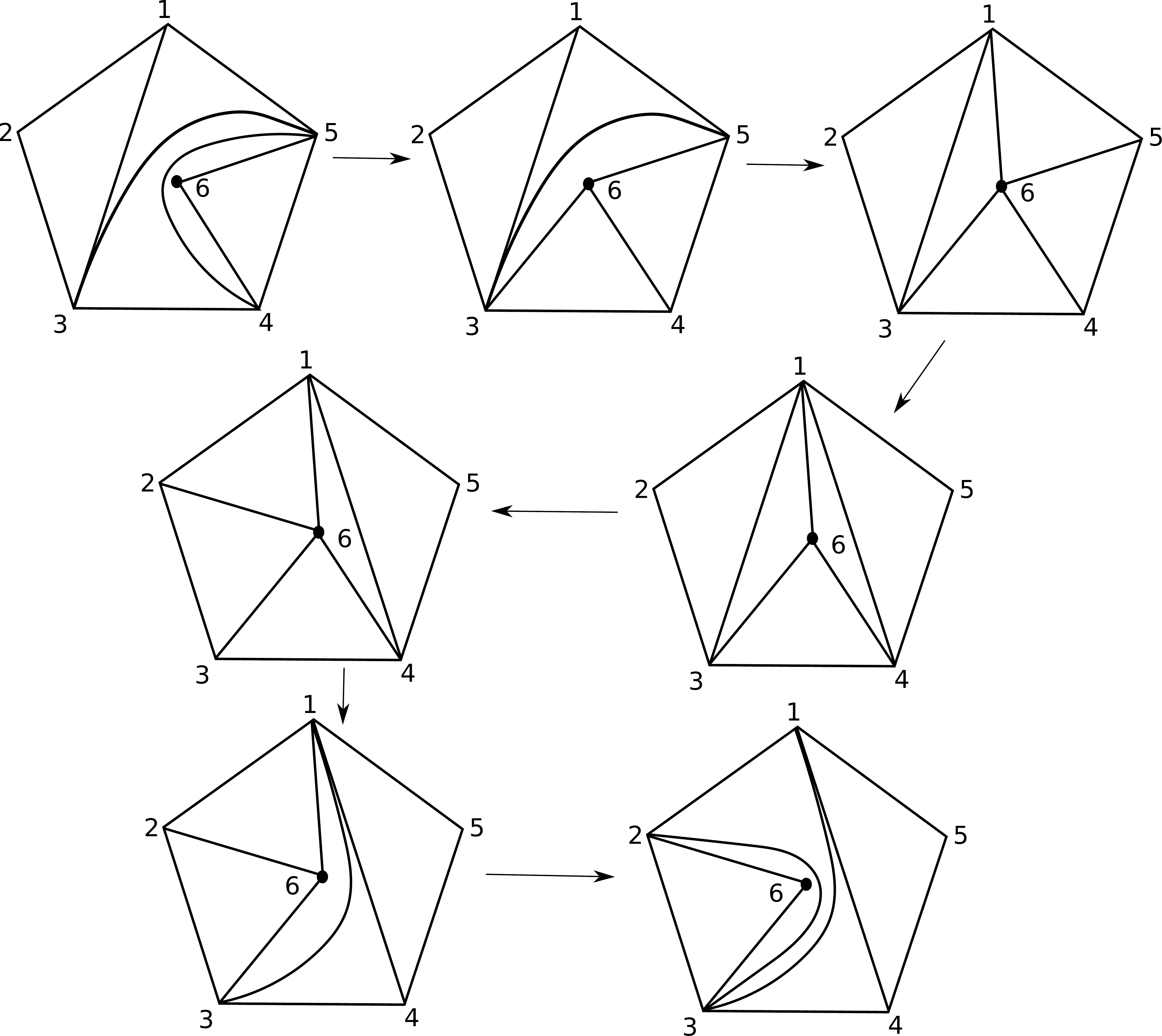}
	\caption{Monodromy action as a sequence of flips in the $(A_1,D_5)$ Argyres-Douglas theory.}
	\label{D5flip}
\end{figure}

The fusion matrices for non-vacuum modules are given as follows:
\begin{eqnarray}
\begin{split}
N_{\Phi_1}&=\begin{pmatrix}
0 & 1 & 0 & 0 & 0\\
0 & 0 & 1 & 0 & 0\\
0 & 0 & 0 & 1 & 0\\
0 & 0 & 0 & 0 & 1\\
-z^2 & 0 & 0 & 0 & 0
\end{pmatrix},\quad\quad\quad
N_{\Phi_2}=\begin{pmatrix}
0 & 0 & 1 & 0 & 0\\
0 & 0 & 0 & 1 & 0\\
0 & 0 & 0 & 0 & 1\\
-z^2 & 0 & 0 & 0 & 0\\
0 & -z^2 & 0 & 0 & 0
\end{pmatrix},\\
N_{\Phi_3}&=\begin{pmatrix}
0 & 0 & 0 & 1 & 0\\
0 & 0 & 0 & 0 & 1\\
-z^2 & 0 & 0 & 0 & 0\\
0 & -z^2 & 0 & 0 & 0\\
0 & 0 & -z^2 & 0 & 0
\end{pmatrix},\quad
N_{\Phi_4}=\begin{pmatrix}
0 & 0 & 0 & 0 & 1\\
-z^2 & 0 & 0 & 0 & 0\\
0 & -z^2 & 0 & 0 & 0\\
0 & 0 & -z^2 & 0 & 0\\
0 & 0 & 0 & -z^2 & 0
\end{pmatrix}.
\end{split}
\end{eqnarray}
For generic $z$ these four matrices are simultaneously diagonalizable with the following eigenvalues:
\begin{center}
	\begin{tabular}{ |c|c|c|c|c| }
		\hline
		eigenspace & $\lambda_{\Phi_1}$ & $\lambda_{\Phi_2}$ &$\lambda_{\Phi_3}$&$\lambda_{\Phi_4}$ \\ \hline \hline
		1 & $-z^{2/5}$ & $z^{4/5}$ &$-z^{6/5}$ & $z^{8/5}$   \\ \hline
		2 & $(-1)^{1/5}z^{2/5}$ & $(-1)^{2/5}z^{4/5}$&$(-1)^{3/5}z^{6/5}$ & $(-1)^{4/5}z^{8/5}$   \\ \hline
		3 & $-(-1)^{2/5}z^{2/5}$ & $(-z)^{4/5}$&$(-1)^{1/5}z^{6/5}$ & $-(-1)^{3/5}z^{8/5}$   \\ \hline
		4 & $(-1)^{3/5}z^{2/5}$ & $-(-1)^{1/5}z^{4/5}$&$-(-1)^{4/5}z^{6/5}$ & $(-1)^{2/5}z^{8/5}$   \\ \hline
		5 & $-(-1)^{4/5}z^{2/5}$ & $-(-1)^{3/5}z^{4/5}$&$-(-1)^{2/5}z^{6/5}$ & $-(-1)^{1/5}z^{8/5}$   \\ \hline
	\end{tabular}
\end{center}
The classical monodromy action in this chamber can be worked out as a composition of flips, as in Figure \ref{D5flip}:
\begin{align*}
\cX_{\gamma_1}&\to\frac{1+\cX_{\gamma_5}+\cX_{\gamma_3}\cX_{\gamma_5}+C}{\cX_{\gamma_2}\cX_{\gamma_3}\cX_{\gamma_4}},\\
\cX_{\gamma_2}&\to \frac{\cX_{\gamma_1}\cX_{\gamma_2}\cX_{\gamma_3}\cX_{\gamma_4}}{\big(1+\cX_{\gamma_2}(1+\cX_{\gamma_3}+\cX_{\gamma_3}\cX_{\gamma_4})\big)\big(1+\cX_{\gamma_5}+\cX_{\gamma_3}\cX_{\gamma_5}+(1+\cX_{\gamma_1})C\big)},\\
\cX_{\gamma_3}&\to\frac{\big(1+(1+\cX_{\gamma_1})\cX_{\gamma_2}(1+\cX_{\gamma_3})\big)(1+\cX_{\gamma_5}+\cX_{\gamma_3}\cX_{\gamma_5}+C)}{\cX_{\gamma_1}\cX_{\gamma_2}\cX_{\gamma_3}(1+\cX_{\gamma_3})\cX_{\gamma_4}\cX_{\gamma_5}},\\
\cX_{\gamma_4}&\to\frac{1+\cX_{\gamma_5}+\cX_{\gamma_3}\cX_{\gamma_5}+(1+\cX_{\gamma_1})C}{\cX_{\gamma_3}},\\
\cX_{\gamma_5}&\to\frac{\cX_{\gamma_3}\cX_{\gamma_4}\cX_{\gamma_5}}{1+\cX_{\gamma_5}+\cX_{\gamma_3}\cX_{\gamma_5}+(1+\cX_{\gamma_1})C},
\end{align*}
where
\begin{equation*}
C=\cX_{\gamma_2}(1+\cX_{\gamma_3})\big(1+\cX_{\gamma_5}(1+\cX_{\gamma_3}+\cX_{\gamma_3}\cX_{\gamma_4})\big).
\end{equation*}
For generic fixed $z\neq 0$, there are exactly five fixed points, matching the number of admissible representations of $\widehat{\mathfrak{sl}(2)}_{-\frac{8}{5}}$. Concretely, at the fixed locus $\cX_{\gamma_3}$ satisfies the following quintic equation:
\begin{equation} \label{eq:quintic-fixed}
z^6\cX_{\gamma_3}^5-5z^4\cX_{\gamma_3}^3-10z^4\cX_{\gamma_3}^2-5z^4\cX_{\gamma_3}-(z^4+z^2+1)=0,
\end{equation}
and $\cX_{\gamma_1},\cX_{\gamma_2},\cX_{\gamma'}$ are all determined by $\cX_{\gamma_3}$ and $z$ (by complicated algebraic expressions which we will not present here.)
As in previous examples, the values of line defect vevs at the fixed points do not depend on the index $i$.

The Galois group of the quintic \eqref{eq:quintic-fixed} is solvable according to {\tt sage}, so in principle one can give a solution
in radicals; we have not carried this out, however. Thus, here we cannot give a closed form for the values of the
$\cX_{\gamma}$ at the fixed points. Moreover, we also have the same problem as in \S\ref{sec:A1D3} above: we do not
know \ti{a priori} how to match the five fixed points and the five primaries.
Nevertheless we numerically sampled
various values of $z$ and confirmed that, for each $z$, there
does exist a matching between fixed points and primaries,
such that the corresponding $h$ makes the diagram commute.

\section{Verlinde algebra from Fixed Points Analysis}
Given the relations that we have discussed between the three algebras, one might ask whether we could say something about the Verlinde algebra through values of generating functions at the fixed points\footnote{We thank Shu-Heng Shao for mentioning this interesting perspective.}. The answer is that we can not determine Verlinde algebra from fixed points analysis alone, but we do obtain useful information about Verlinde algebra\footnote{More precisely we mean Verlinde-like algebra of the set of highest weight modules that correspond to the $U(1)_r$ fixed points, from direct application of the Verlinde formula.} and expansion of line defect Schur index in terms of characters.

First we would like to stress that, in principal one could obtain the (deformed) Verlinde algebra through computing Schur index with one half line and two half lines inserted and studying their images under the homomorphism $f$. In fact this is practically how we found the deformed Verlinde algebra in the $D_3$ and $D_5$ cases. However, in practice (at least for us) character expansions of line defect Schur index (especially Schur index with more than one line defect inserted) are not very easy to obtain. It would be nice if there is some way to simplify this procedure.

To begin with, suppose that we already know the image of $[\Phi_\alpha]$ under the isomorphism $h$, then the modular fusion rules among them are very easy to obtain since the corresponding multiplication in $\cO(F)$ is given directly by pointwise multiplication. Concretely, suppose that
\begin{equation*}
[\Phi_\alpha]\xrightarrow{h}\phi_\alpha:=(\lambda_\alpha^1,\dots,\lambda_\alpha^n),
\end{equation*}
then by expanding e.g.
\begin{equation*}
\phi_\alpha\phi_\beta=\Sum_\gamma c_{\alpha\beta}^\gamma \phi_\gamma,
\end{equation*}
the modular fusion coefficients are given by $c_{\alpha\beta}^\gamma$\footnote{Here to get the fusion coefficients we don't need to ``order" the fixed points. We don't need to know the exact correspondence between $U(1)_r$ fixed points and primaries.}. Now how do we determine $\phi_\alpha$? Since we know the values of $F_{L_{\alpha i}}$ at the $U(1)_r$ fixed points, if in addition we also know the image of $L_{\alpha i}$ under $f$, then $\phi_\alpha$ is given by taking the inverse of the linear relations. So we still need to work out the character expansions for single line defect Schur index. But this already saves the effort of working out the character expansions of two line defect Schur index. \par

Now suppose that the only data given are generating functions of line defect generators and their values at the $U(1)_r$ fixed points, what ``constraints" could we possibly put on the (deformed) Verlinde algebra? We illustrate this by looking at two simplest examples $A_2$ and $D_3$ Argyres-Douglas theories. Of course the Verlinde algebra in these cases were already known for a long time (see \cite{D.Francesco} and references therein), the hope is that this might shed light on unknown Verlinde algebras of certain 2d chiral algebras.

In $A_2$ case there are two fixed points, the values of $F_{L_i}$ don't depend on $i$ at the fixed points so we denote them as $F_L$.
Over the fixed points
\begin{equation}\label{eqn7.1}
F_L^2=I+F_L.
\end{equation}
This equation is understood in the context of values of line defects at fixed points. This could be obtained either by direct computation or through the relation
\begin{equation}
L_iL_{i+2}=1+q^{\frac{1}{2}}L_{i+1}.
\end{equation}
As discussed in \S \ref{sec:comments} in $(A_1,A_{2N})$ theories the vev of line defect generators themselves realize fusion rules over $U(1)_r$ fixed points. In particular (\ref{eqn7.1}) is the non-trivial fusion rule of the $(2,5)$ minimal model. However this is a special phenomenon only in $(A_1,A_{2N})$ theories. We would like to rediscover fusion rules in the basis of $[\Phi_\alpha]$ instead for the purpose of generalization.\par
We make the following ansatz for the image of $L_i$ under $f$:
\begin{equation}
L_i\xrightarrow{f}[L]:=a[\Phi_0]+b[\Phi_1],
\end{equation}
where $\Phi_0$ is the vacuum. We also make an ansatz for the fusion rule:
\begin{equation*}
[\Phi_1]\times[\Phi_1]=c[\Phi_0]+d[\Phi_1].
\end{equation*}
(\ref{eqn7.1}) would imply
\begin{equation}
[LL]=[L]\times[L]=(a+1)[\Phi_0]+b[\Phi_1],
\end{equation}
by comparing coefficients we get the following equations for $a,b,c,d$:
\begin{equation}
a^2+b^2c=a+1,\quad 2ab+b^2d=b.
\end{equation}
Now, $a$ and $b$ have to be integers. This was the observation made in \cite{Cordova:2016uwk}. We do not have an explanation but it is true in all the examples that we considered in this paper so we use this as an assumption. The fusion coefficients $c$ and $d$ have to be $0$ or $1$\footnote{We will discuss how this works for modular fusion rules with apparent $-1$ coefficients momentarily.}. Moreover given each candidate fusion rule one could check whether the solution is consistent with eigenvalues of the Verlinde matrix. These constraints pin down the only possible fusion rule to be the desired one in $(2,5)$ minimal model namely $c=1$ and $d=1$. There are two solutions for $a$ and $b$:
\begin{equation}
(a,b)=(1,-1) \quad\text{or} \quad (a,b)=(0,1).
\end{equation}
 The wrong answer could be easily ruled out by computing the single line defect Schur index. In more complicated cases the finite number of solutions of $(a,b)$ also offers ansatz for the character expansion of single line defect Schur index.

In the $D_3$ case we have more constraints due to the $z$-deformed Verlinde algebra. We take an assumption that the $z$-deformed Verlinde algebra always replaces the $-1$ coefficient by $-z^2$.\footnote{We conjecture this is true at least for $(A_1,D_{2N+1})$ Argyres-Douglas theories. For other theories one could first work out simple examples to find out patterns of deformed modular fusion rules.} In that case by taking $z=\I$ all the fusion coefficients are either $0$ or $1$. So this reduces to a similar case as in $A_2$. When $z=\I$,
\begin{equation}
[AB]=2[A],\quad [AA]=[\Phi_0]+[B],\quad
[BB]=2[\Phi_0]+[B].
\end{equation}
Again this was obtained either by directly looking at values of $F(L)$ at fixed points or through relations between generating functions. Similarly by making ansatz and comparing coefficients one could obtain the consistent fusion rules. Note that in this case there is one more constraint coming into play, namely the fusion matrices $N_{\Phi_1}$ and $N_{\Phi_2}$ have to be simultaneously diagonalizable. The only fusion rules passing these constraints are
\begin{eqnarray}\label{A3fusionI}
\begin{split}
[\Phi_1]\times[\Phi_1]&=[\Phi_2],\\
[\Phi_1]\times[\Phi_2]&=[\Phi_0],\\
[\Phi_2]\times[\Phi_2]&=[\Phi_1].
\end{split}
\end{eqnarray}
Note that here we can not physically distinguish $[\Phi_1]$ and $[\Phi_2]$, e.g. we can not compute their conformal weights etc in our setup. They only appear in our ansatz (for $z=\I$) for $[A]$ and $[B]$. This is the reason why we can't actually pin down the fusion rules.
Now in the deformed fusion rules each $+1$ coefficient in (\ref{A3fusionI}) could be either $+1$ or $-z^2$. We again make ansatz for $[A]$ and $[B]$, only now the coefficients are monomials in $z$ with integer coefficients. Again this is an assumption that we make through observations of known examples. For general $z$ the following holds:
\begin{eqnarray}
\begin{split}
[AB]&=(z+z^{-1})[\Phi_0]+2[A],\\
[AA]&=[\Phi_0]+[B],\\
[BB]&=2[\Phi_0]+(z+z^{-1})[A]+[B].
\end{split}
\end{eqnarray}
Imposing constraints and comparing coefficients gives us two possibilities. One of them, which is also the correct one, is
\begin{align*}
[\Phi_1]\times[\Phi_1]&=[\Phi_2],\\
[\Phi_1]\times[\Phi_2]&=-z^2[\Phi_0],\\
[\Phi_2]\times[\Phi_2]&=-z^2[\Phi_1],
\end{align*}
with the following images of $A_i$ and $B_i$ under $f$:
\begin{align*}
[A]&=\frac{1}{z}([\Phi_2]-[\Phi_1]),\\
[B]&=[\Phi_0]-[\Phi_1]+z^{-2}[\Phi_2].
\end{align*}
The other solution is simply given by swapping $[\Phi_1]$ with $[\Phi_2]$. Note that this is reasonable since we can not physically distinguish $[\Phi_1]$ and $[\Phi_2]$. So this is the best we could do with the available ansatz. In reality given access to characters of admissible representations it would be easy to rule out the wrong answer.

\newpage

\bibliographystyle{utphys}

\bibliography{verlinde-fixed}

\end{document}